\documentclass[natbib,smallextended]{svjour3}       % onecolumn (second format)
\smartqed  % flush right qed marks, e.g. at end of proof
\usepackage{graphicx}
\usepackage{newtxtext}
\usepackage[varvw]{newtxmath}
\usepackage{amsmath}
\usepackage{bm}
\usepackage{aas_macros}
\usepackage{lscape}

\usepackage{threeparttable}
\usepackage{longtable}
\usepackage{threeparttablex}

\usepackage[colorlinks=true,linkcolor=blue,citecolor=blue,urlcolor=blue]{hyperref}

\usepackage{color}
\newcommand{\cla}[1]{#1}  % uncomment this for final version

\newcommand{\bi}{$\langle B \rangle$}
\newcommand{\bv}{$\langle B_V \rangle$}
\newcommand{\bz}{$\langle B_{\rm z} \rangle$}
\newcommand{\kms}{km\,s$^{-1}$}
\newcommand{\vs}{$v_{\rm e}\sin i$}
\newcommand*\centerstack[2]{%
  \ensuremath{%
    \mathrel{%
      \mathchoice{%
        \vcenter{%
          \offinterlineskip
          \halign{\hfil$\displaystyle##$\hfil\cr#1\cr#2\cr}%
        }%
      }
      {%
        \vcenter{%
          \offinterlineskip
          \halign{\hfil$\textstyle##$\hfil\cr#1\cr#2\cr}%
        }%
      }
      {%
        \vcenter{%
          \offinterlineskip
          \halign{\hfil$\scriptstyle##$\hfil\cr#1\cr#2\cr}%
        }%
      }
      {%
        \vcenter{%
          \offinterlineskip
          \halign{\hfil$\scriptscriptstyle##$\hfil\cr#1\cr#2\cr}%
        }%
      }%
    }%
  }%
}
\newcommand*\la{\centerstack{<}{\sim}}
\newcommand*\ga{\centerstack{>}{\sim}}
\journalname{The Astronomy and Astrophysics Review}
\begin{document}

\title{Magnetic fields of M dwarfs
%\thanks{Grants or other notes
%about the article that should go on the front page should be
%placed here. General acknowledgments should be placed at the end of the article.}
}
%\subtitle{Do you have a subtitle?\\ If so, write it here}

%\titlerunning{Short form of title}        % if too long for running head

\author{Oleg Kochukhov}

%\authorrunning{Short form of author list} % if too long for running head

\institute{O. Kochukhov \at
              Department of Physics and Astronomy, Uppsala University, \\ 
              Box 516, SE-751 20 Uppsala, Sweden \\
              %Tel.: +46-18-471-5993\\
              \email{oleg.kochukhov@physics.uu.se}          
}

\date{Received: date / Accepted: date}
% The correct dates will be entered by the editor

\maketitle

\begin{abstract}
Magnetic fields play a fundamental role for interior and atmospheric properties of M dwarfs and greatly influence terrestrial planets orbiting in the habitable zones of these low-mass stars. Determination of the strength and topology of magnetic fields, both on stellar surfaces and throughout the extended stellar magnetospheres, is a key ingredient for advancing stellar and planetary science. Here modern methods of magnetic field measurements applied to M-dwarf stars are reviewed, with an emphasis on direct diagnostics based on interpretation of the Zeeman effect signatures in high-resolution intensity and polarisation spectra. Results of the mean field strength measurements derived from Zeeman broadening analyses as well as information on the global magnetic geometries inferred by \cla{applying} tomographic mapping methods \cla{to spectropolarimetric observations} are summarised and critically evaluated. The emerging understanding of the complex, multi-scale nature of M-dwarf magnetic fields is discussed in the context of theoretical models of hydromagnetic dynamos and stellar interior structure altered by magnetic fields.
% \subclass{MSC code1 \and MSC code2 \and more}
\end{abstract}

\setcounter{tocdepth}{3}
\tableofcontents

\section{Introduction}
\label{sect:intro}

M dwarfs are the lowest-mass stars, occupying the bottom of the main sequence. These stars dominate the local stellar population, accounting for 70--75\% of all stars in the solar neighbourhood \citep{bochanski:2010,winters:2019}. M dwarfs have masses of 0.08--0.55 $M_\odot$ and effective temperatures of 2500--4000~K \citep{pecaut:2013}. Their atmospheric characteristics span a wide range, from conditions similar to the upper layers of GK-star atmospheres in early-M dwarfs to temperatures and pressures comparable to those found in brown dwarfs and giant planets in late-M stars. The optical and near-infrared spectra of M dwarfs are distinguished by prominent absorption bands of diatomic molecules. This molecular absorption becomes progressively more important towards later spectral types, to the extent that hardly any atomic line is free from molecular blends. The interior structure of M dwarfs undergoes a transition at $M\approx0.35M_\odot$ \citep{chabrier:1997} from being similar to that of solar-like stars, with a thick convective envelope overlaying a radiative zone, to a fully convective interior structure not found in any other type of main sequence stars.

M dwarfs exhibit conspicuous and abundant evidence of surface activity: flares, photometric rotational variability, enhanced chromospheric and coronal emission in X-rays, UV and radio \citep[e.g.][]{hawley:2014,newton:2016,newton:2017,astudillo-defru:2017,wright:2018,villadsen:2019}. In hotter stars, including the Sun, all these phenomena are invariably correlated with the presence of intense magnetic fields generated by a dynamo mechanism. It is believed that the dynamo process in solar-type stars is closely linked to the stellar differential rotation and is largely driven by shearing at the tachocline -- a narrow boundary layer separating the convective and radiative zones \citep{charbonneau:2014}. Details of this complex hydromagnetic process are far from being settled even for the Sun, and possibility of other dynamo effects operating elsewhere in the solar interior has been discussed \citep[e.g.][]{brandenburg:2005}. The tachocline disappears in mid-M dwarfs, offering a unique chance to explore cool-star dynamo action in a different environment compared to the Sun. In this context, investigation of the surface magnetism of M dwarfs straddling the boundary of fully convective interior is critically important for guiding development of the stellar dynamo theory.

Active M dwarfs is the only class of stars for which magnetic field alters global stellar parameters in an observable and systematic way. Interior structure of these stars is expected to be relatively simple, especially beyond the limit of full convection. Despite this, many studies demonstrated that measured radii of M dwarfs tend to be significantly larger than those predicted by the stellar evolution theory \citep[e.g.][]{ribas:2006,torres:2013} and that this discrepancy correlates with the magnetic activity indicators, such as the Ca H\&K and X-ray emission \citep{lopez-morales:2007,feiden:2012,stassun:2012}. The leading hypothesis explaining the apparent inflation of M-dwarf radii is a modification of the convective energy transport, governing the interior structure of low-mass stars, by strong magnetic fields \citep{mullan:2001,chabrier:2007,macdonald:2014,feiden:2013,feiden:2014}. Empirical determinations of the surface magnetic field strengths of M dwarfs are therefore instrumental for constraining and testing theoretical models of magnetised stellar interiors.

M dwarfs have been recently established as favourable targets for searches of small exoplanets and in-depth studies of their atmospheres. Due to their low mass, M dwarfs exhibit a higher amplitude reflex radial velocity variation compared to a Sun-like star orbited by the same planet. Moreover, a lower luminosity of M dwarfs means that habitable zones are located much closer to the central star. This translates to shorter orbital periods and higher radial velocity amplitudes for terrestrial planets residing in those zones \citep{kasting:2014}. All these factors facilitate discovery and analysis of rocky planets in M-dwarf exoplanetary systems. Both ground-based radial velocity searches \citep{bonfils:2013} and surveys of transiting exoplanets from space \citep{dressing:2015} confirm existence of a large population of small planets orbiting M dwarfs. The closest potentially habitable Earth-size planets all have M dwarfs as host stars \citep{anglada-escude:2016,gillon:2017,ribas:2018}. These high-profile exoplanetary systems are targeted by numerous multi-wavelength observational campaigns, dedicated instruments, space missions, and in-depth theoretical studies. To this end, understanding fundamental properties and magnetic activity behaviour of their M-dwarf hosts is often a limiting factor and a major source of uncertainty for many of these investigations.

Despite significant gains provided by M dwarfs for studies of small exoplanets, magnetic activity of these stars interferes with detection of planets using the radial velocity method and may have a major impact on planetary atmospheres and habitability. The fraction of active M dwarfs increases dramatically towards later spectral types \citep[e.g.][]{reiners:2012}, leading to substantial radial velocity jitter due to dark spots \citep{barnes:2014,andersen:2015} and Zeeman broadening \citep{reiners:2013}. For this reason, monitoring stellar magnetic field is considered essential for efficient modelling and filtering M-dwarf activity jitter \citep{hebrard:2016,moutou:2017}. The enhanced steady short-wavelength radiation as well as episodic energetic particle and photon emission \cla{events} associated with flares and coronal mass ejections \citep{moschou:2019} \cla{are believed to have} a large impact on the atmospheres of potentially habitable rocky planets \citep{khodachenko:2007,lammer:2007,penz:2008}, possibly even stripping thinner atmospheres not protected by planetary magnetic field \citep{luger:2015}. Particularly intense stellar magnetic fields may deposit enough energy via induction heating to melt interiors of close-in rocky planets, resulting in increased volcanic activity or a permanent molten mantle state \citep{kislyakova:2017,kislyakova:2018}. \cla{On the other hand, powerful stellar magnetospheres can also protect planetary atmospheres from erosion by retarding the stellar wind and restraining propagation of coronal mass ejections \citep{alvarado-gomez:2018a,alvarado-gomez:2019a}. All} these profound effects depend \cla{sensitively} on the strength and configuration of the global component of M-dwarf magnetic field \citep{lang:2012,vidotto:2013,vidotto:2014a,cohen:2014}. Thus, certain very specific characteristics of extended stellar magnetospheres, not the overall average surface magnetic field strength, are most important for the star-planet magnetic interactions and space weather environments in M-dwarf exoplanetary systems.

This discussion shows that a meaningful progress in the research topics mentioned above -- from understanding stellar dynamos, impact of magnetic fields on the stellar interior structure and fundamental parameters to star-planet magnetic interaction and the role of stellar activity in determining the atmospheric structure, composition and habitability of exoplanets -- is all but impossible without detailed information about the strength and topology of both small- and large-scale magnetic fields on the surfaces of M-dwarf stars. An analysis of high-resolution optical and near-infrared stellar intensity and polarisation spectra is currently the only approach allowing one to obtain such information in a systematic and direct manner.

This review aims to summarise the current state of observational knowledge about magnetism of M-dwarf stars. We focus on providing a comprehensive overview of the methodology and results of direct magnetic field diagnostic based on observations of the Zeeman effect in stellar spectra. In the first part of the review (Sect.~\ref{sect:methods}), the basic physics of the Zeeman effect is described (Sect.~\ref{sect:zeeman}), followed by a discussion of the formation of local and disk-integrated intensity and polarisation spectra in the presence of a magnetic field (Sects.~\ref{sect:prt} and \ref{sect:disk}). The resulting impact of magnetic field on the shape and strength of spectral lines is discussed in Sect.~\ref{sect:zb}. The multi-line polarisation diagnostic approach and techniques of mapping global magnetic fields using high-resolution \cla{spectropolarimetric observations} are described in Sect.~\ref{sect:lsd} and \ref{sect:zdi}, respectively. Sect.~\ref{sect:instruments} touches upon instrumentation best suited for M-dwarf magnetic field measurements. The second part of this review (Sect.~\ref{sect:obs}) is dedicated to presentation of the observational results, including magnetic field characteristics inferred from intensity (Sect.~\ref{sect:stokesI}) and polarisation (Sect.~\ref{sect:stokesV}) data. Sect.~\ref{sect:detailedI} discusses measurements of the total magnetic field strengths using detailed line profile modelling. An up-to-date compilation of all such M-dwarf magnetic field measurements is provided. We also critically evaluate approximate methods of deriving mean field strength (Sect.~\ref{sect:approxI}) and discuss the relation between magnetic field and stellar rotation (Sect.~\ref{sect:rotat}). This is followed in Sect.~\ref{sect:stokesV} by the discussion of detection of Zeeman polarisation in M-dwarf spectra and presentation of the properties of global stellar magnetic field topologies derived from these observations. A compilation of all tomographic magnetic field mapping results available for M dwarfs is provided and employed to explore dependence of \cla{the global} magnetic field properties on the stellar mass and rotation (Sect.~\ref{sect:zdiresults}). A relationship between the global and small-scale magnetic fields of M dwarfs is assessed in Sect.~\ref{sect:glob} and implications for the properties of extended stellar magnetospheres are discussed in Sect.~\ref{sect:extrapol}. The review ends with a discussion and outlook (Sect.~\ref{sect:discussion}), where we briefly summarise theoretical efforts to shed a light on the origin, topology and variability of M-dwarf magnetic fields and to understand the impact of these fields on the interior structure and fundamental parameters of low-mass stars. We conclude with an outline of the most promising directions of future theoretical and observational research necessary for advancing our knowledge about the nature of M-dwarf magnetism.

\section{Methods of magnetic field measurements}
\label{sect:methods}

\subsection{Zeeman effect in spectral lines}
\label{sect:zeeman}

All commonly used methods of detecting and measuring magnetic fields on the surfaces of late-type stars rely on manifestations of the Zeeman effect in spectral lines. Here we provide a brief account of the basic physics behind the Zeeman splitting and polarisation in spectral lines. A more comprehensive reviews can be found in e.g. \citet{polarization:2004} and \citet{kochukhov:2018d}.

In the presence of an external magnetic field, the atomic or molecular energy levels split into a number of magnetic sub-levels. For the field strengths $\la100$~kG, encountered in non-degenerate stars, the splitting occurs in the linear Zeeman regime for most atomic lines. An energy level with the total angular momentum quantum number $J$ splits into $2J+1$ equidistant sub-levels with the magnetic quantum numbers $M=-J,-J+1,...,J-1,J$. The shift in energy relative to the initial value is given by
\begin{equation}
\Delta E = g\dfrac{e \hbar}{2m_{\rm e} c} B M.
\end{equation}
Here $B$ is the magnetic field strength and $g$ is the so-called Land\'e factor, characterising magnetic sensitivity of each energy level. For the case when $LS$-coupling applies (light and most iron-peak elements), $g$ can be found from the $L$, $S$, and $J$ quantum numbers
\begin{equation}
g=\dfrac{3}{2} + \dfrac{S(S+1)-L(L+1)}{2J(J+1)}.
\end{equation}
More accurate $g$-factors are often provided by the same atomic structure calculations that supply transition probabilities and other line parameters. These data are available from several astrophysical line data bases, such as {\sc VALD} \citep{ryabchikova:2015}.

\begin{figure}[!t]
\centering
\includegraphics[width=0.33\textwidth,angle=90]{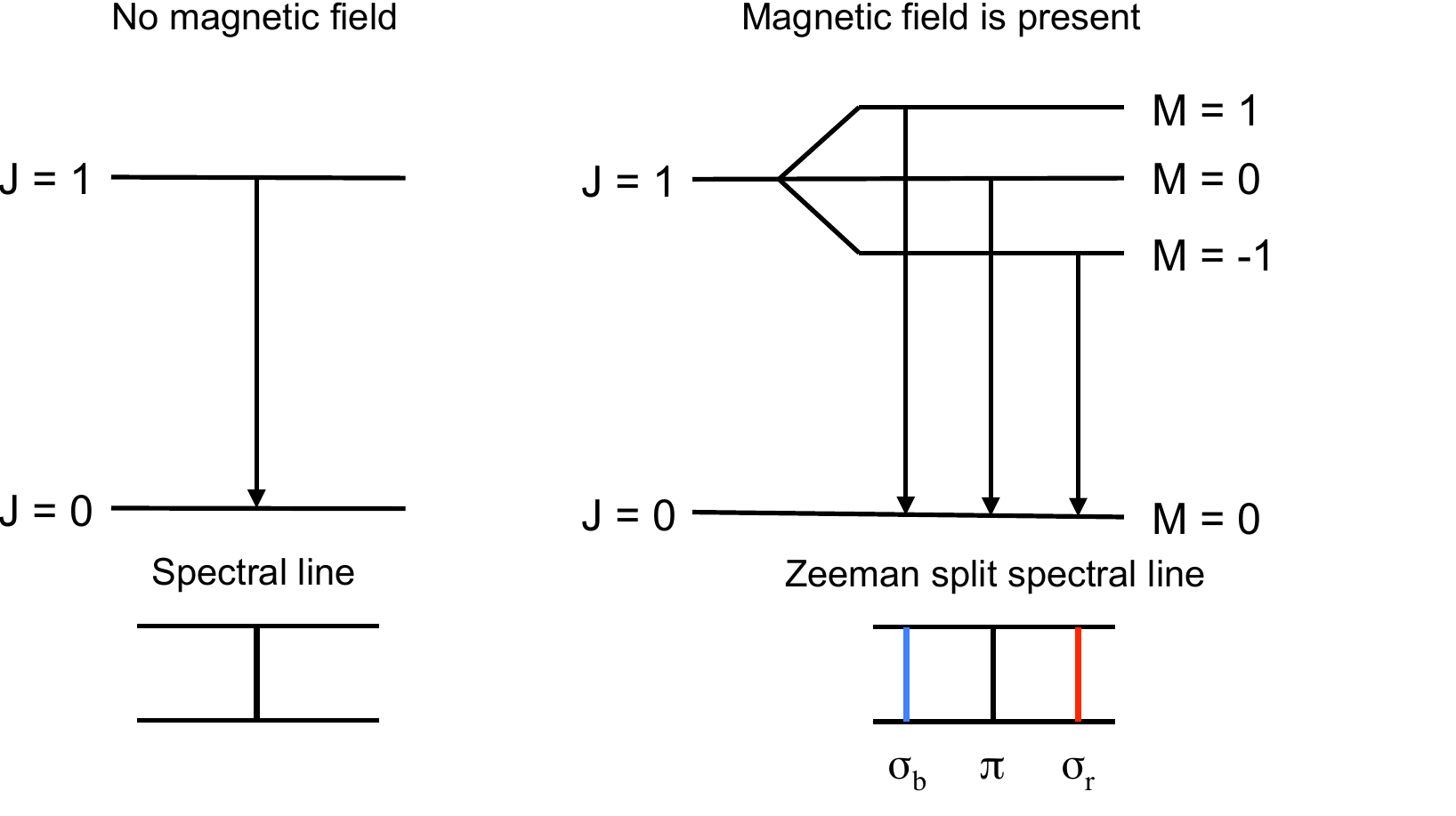}
\includegraphics[width=0.33\textwidth,angle=90]{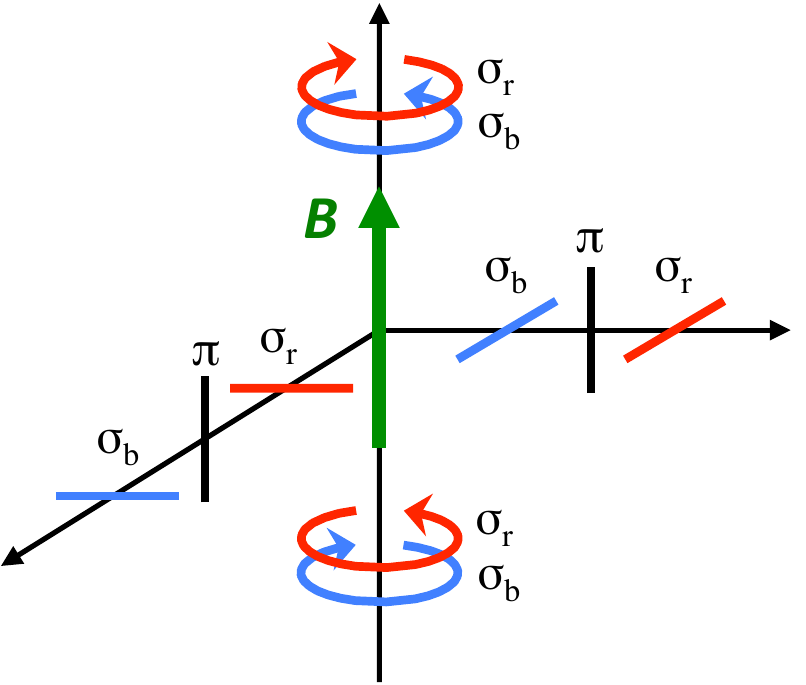}
\caption{Left: Zeeman splitting in a magnetic field. In the absence of the field, the transition between the upper and lower energy levels corresponds to a single spectral line. When an external field is present, the line splits into three ($\pi$, blue- and red-shifted $\sigma$) Zeeman components.
Right: Polarisation properties of the radiation emitted in the $\pi$ and $\sigma$ components for different orientations of the magnetic field vector relative to the line of sight. 
Image reproduced with permission from \citet{kochukhov:2018d}, copyright by CUP.}
%% Credit: \citeauthor{kochukhov:2018d}, 2018, in \textit{Cosmic Magnetic Fields}, Cambridge University Press, p.~47.}
\label{fig:zeeman}
\end{figure}

As a consequence of the Zeeman splitting of energy levels, the absorption or emission lines corresponding to the electron transitions between these levels split as well. This is illustrated in Fig.~\ref{fig:zeeman} (left panel) for a spectral line arising from the transition between the unsplit $J_{\rm l}=0$ lower level and the upper level with $J_{\rm u}=1$. The selection rules permit transitions with $\Delta M=0,\pm1$. This gives rise to the three groups of distinct Zeeman components. Those with $\Delta M=0$ are known as $\pi$ components, the ones with $\Delta M=\pm1$ are the blue- and red-shifted $\sigma$ components (denoted $\sigma_{\rm r}$ and $\sigma_{\rm b}$). For the simplest case of the so-called normal Zeeman triplet, illustrated in Fig.~\ref{fig:zeeman}, there is only one component of each type. In general (anomalous Zeeman splitting), there can be multiple components in each group.

The magnetic splitting of spectral lines in the linear Zeeman regime is symmetric with respect to the unperturbed wavelength $\lambda_0$. The wavelength displacement of the red $\sigma$ component (for a normal Zeeman triplet) or the centre-of-gravity of the group of $\sigma_{\rm r}$ components (for anomalous Zeeman splitting) is given by
\begin{equation}
\Delta\lambda_{\rm B}=g_{\rm eff}\dfrac{eB\lambda_0^2}{4\pi m_{\rm e}c^2}=4.67\times10^{-12}g_{\rm eff}B\lambda_0^2
\label{eq:aa}
\end{equation}
for the field strength in G and wavelength in nm. The parameter $g_{\rm eff}$ is known as the effective Land\'e factor. It provides a convenient measure of the magnetic sensitivity of a spectral line and can be calculated from the $g$ and $J$ values of the energy levels involved
\begin{equation}
g_{\rm eff}=\dfrac{1}{2}(g_{\rm l}+g_{\rm u})+\dfrac{1}{4}(g_{\rm l} - g_{\rm u})\left[ J_{\rm l} (J_{\rm l} +1) -  J_{\rm u} (J_{\rm u} +1) \right].
\end{equation}
The majority of spectral lines have $g_{\rm eff}\approx0.5$--1.5, with relatively uncommon but very useful magnetic null lines ($g_{\rm eff}=0$) and some very magnetically sensitive lines ($g_{\rm eff}=2.5$--3).

The basic picture of the Zeeman splitting, described here for atomic systems, also applies to many diatomic molecules \citep{herzberg:1950,berdyugina:2002}. In that case, however, interaction along the line joining the nuclei plays a central role, requiring different sets of quantum numbers depending on the coupling of the spin and angular momentum of electrons to the internuclear axis.

According to Eq.~(\ref{eq:aa}), separation of the Zeeman components grows linearly with the field strength and quadratically with wavelength. The field $B$ appearing in this equation is the absolute magnetic field strength value, independent of the field orientation. At the same time, the relative strengths as well as polarisation properties of the $\pi$ and $\sigma$ components depend on the orientation of magnetic field within the slab of gas where the absorption or emission line is produced. This dependence is illustrated by Fig.~\ref{fig:zeeman} (right panel). If the magnetic field vector is aligned with the line of sight, the $\sigma_{\rm b}$ and $\sigma_{\rm r}$ components are observed to have opposite circular polarisation and the $\pi$ components are absent. If the field vector is normal to the line of sight, the $\pi$ components are linearly polarised parallel to magnetic field and the $\sigma$ components are linearly polarised perpendicular to the field. For intermediate field vector orientations the $\pi$ components remain linearly polarised while the $\sigma$ components are polarised elliptically (i.e. exhibit both circular and linear polarisation).

In some situations the magnetic splitting may become comparable to the separation of neighbouring energy levels in the absence of the field. This can occur, even in moderate and weak fields, for certain atomic lines consisting of close multiplets (e.g. the Li~{\sc i} $\lambda$ 670.8~nm doublet), atomic lines with hyperfine structure and many molecular lines. In this situation, the Zeeman splitting can no longer be treated in the linear regime. A more complicated quantum mechanical calculation, sometimes referred to as incomplete Paschen-Back effect \citep{berdyugina:2005a,kochukhov:2008a}, has to be carried out leading to Zeeman splitting patterns that are no longer symmetric about the line centre $\lambda_0$, albeit retaining the same polarisation properties as described above.

\subsection{Local Stokes parameter spectra}
\label{sect:prt}

\begin{figure}[!t]
\centering
\includegraphics[height=\textwidth,angle=90]{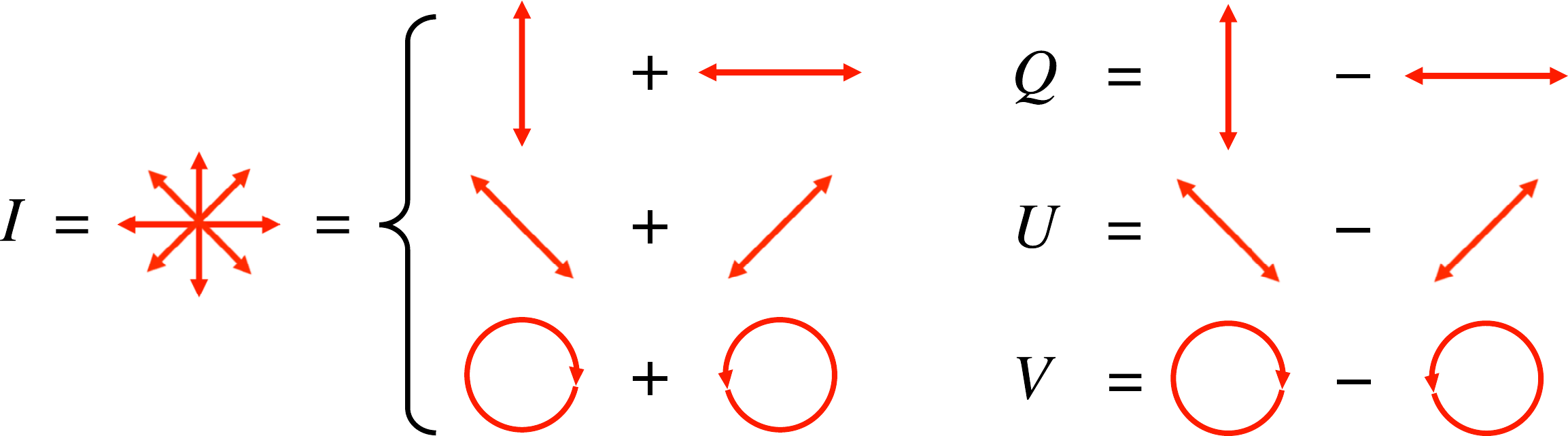}
\caption{\cla{Schematic representation of the Stokes parameter definitions.}}
\label{fig:stokes_def}
\end{figure}

The Stokes parameter formalism provides a convenient framework for quantifying polarisation of stellar radiation and relating theoretical predictions with observations. The components of the Stokes vector $\bm{I}=\{I,Q,U,V\}$ are defined as follows \citep{polarization:2004,bagnulo:2009}. The total radiation intensity is given by $I$. The Stokes $Q$ parameter measures the difference between the intensity of the radiation with the electric field oscillating along and perpendicular to the prescribed reference direction. Stokes $U$ is the difference between the intensity of the radiation with the electric field oscillating at 45$^{\rm o}$ and 135$^{\rm o}$ with respect to that direction. Together, the Stokes $Q$ and $U$ parameters fully describe the linear polarisation state of stellar radiation. Stokes $V$ is defined as the difference between the radiation intensity with the right-handed circular polarisation (the electric field vector rotates clockwise as seen by the observer looking at the radiation source) and with the left-handed circular polarisation (the electric field vector rotates counterclockwise). \cla{These definitions of the Stokes parameters are schematically illustrated in Fig.~\ref{fig:stokes_def}.}

An interaction between matter and radiation in the presence of a magnetic field is governed by the polarised radiative transfer (PRT) equation. This equation describes evolution of the Stokes vector $\bm{I}$ as it propagates outwards in the stellar surface layers. The most accurate and comprehensive treatment of this problem is a numerical solution of the PRT equation in a realistic stellar model atmosphere \citep[e.g.][]{landi-deglinnocenti:1976,piskunov:2002a,kochukhov:2018d}. In this approach, suitable for an arbitrary, possibly depth-dependent, magnetic field vector, one starts by specifying elemental abundances as well as temperature and pressure as a function of geometrical height in a stellar atmosphere. The {\sc Phoenix} \citep{hauschildt:1999} and {\sc MARCS} \citep{gustafsson:2008} model atmosphere grids are common choices for cool low-mass stars. Given this input, the system of equations describing ionisation and chemical balance between different molecular and atomic specifies is solved. This step requires a large amount of molecular and atomic data \citep[e.g.][]{piskunov:2017}, including molecular equilibrium constants, ionisation and dissociation potentials, partition functions, etc. The resulting concentrations of relevant species are then employed for calculation of the line and continuum opacities based on the lists of atomic and molecular transitions contributing to a given wavelength region. Relatively complete and accurate atomic line lists are readily available, e.g. from the {\sc VALD} database \citep{ryabchikova:2015}. On the other hand, the lists of molecular transitions relevant for M dwarfs are highly incomplete and often inaccurate. Only theoretical calculations are available for many molecules, leading to large offsets between the predicted and observed line positions. Given the continuum and line opacity coefficients, \cla{a radiative transfer problem is solved as a system of four coupled differential equations for the Stokes $I$, $Q$, $U$, and $V$ parameters through atmospheric layers using dedicated numerical algorithms}  \citep{rees:1989,piskunov:2002a,de-la-cruz-rodriguez:2013}. 
\cla{This provides} the emergent local \cla{Stokes parameter} spectra as a function of wavelength and the angle between the surface normal and the observer's line of sight.

\begin{figure}[!t]
\centering
\includegraphics[width=\textwidth]{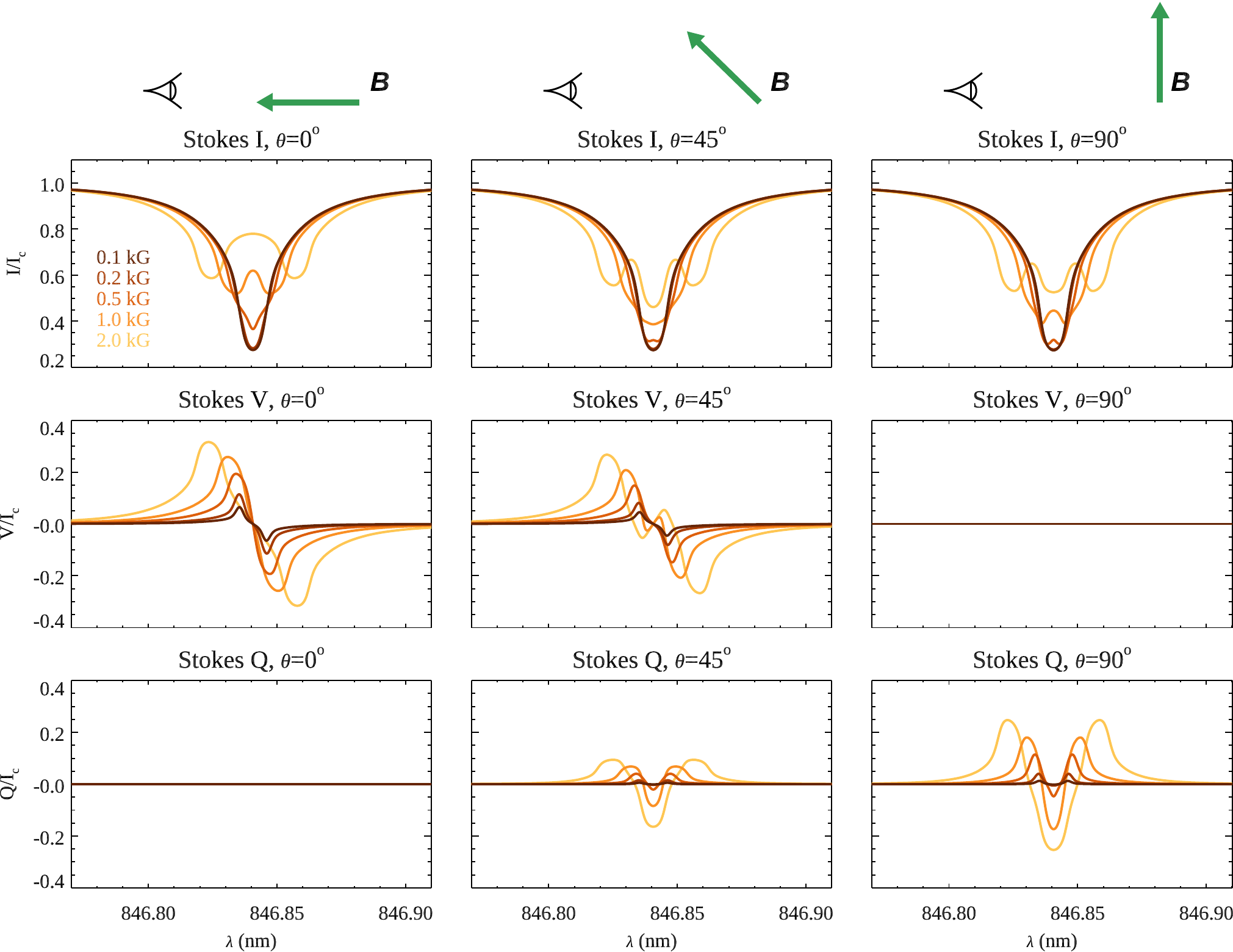}
\caption{Local Stokes $I$ (top row), $V$ (middle row), and $Q$ (bottom row) profiles of the Fe~{\sc i} $\lambda$ 846.84~nm line for the magnetic field strengths from 0.1 to 2.0 kG. The three columns show profiles for different inclinations ($\theta=0^{\rm o}$, 45$^{\rm o}$, and 90$^{\rm o}$) of the magnetic field vector relative to the line of sight, as illustrated schematically above each column. Theoretical profiles are computed at the disk centre using $T_{\rm eff}=3800$~K, $\log g=5.0$ model atmosphere.}
\label{fig:stokes_local}
\end{figure}

Figure~\ref{fig:stokes_local} shows an example of numerical calculation of the local Stokes $I$, $V$, and $Q$ profiles for the Fe~{\sc i} 846.84~nm line for three orientations of the magnetic field vector and the field strengths of 0.1--2~kG. These computations were carried out with the {\sc Synmast} code \citep{kochukhov:2010a} using $T_{\rm eff}=3800$~K, $\log g=5.0$ model atmosphere from the {\sc MARCS} grid, and assuming local thermodynamic equilibrium (LTE). The Stokes parameter profiles in Fig.~\ref{fig:stokes_local} follow the qualitative behaviour outlined in Sect.~\ref{sect:zeeman}. As the magnetic field strength increases, the Zeeman splitting in Stokes $I$ becomes wider. This causes, first, a deformation of the intensity profile for $B\la0.5$~kG and then appearance of the resolved Zeeman-split line components for $B$\,=\,1--2~kG, when the magnetic splitting exceeds the non-magnetic line broadening (in this case dominated by the thermal Doppler broadening in the line core and the van der Waals pressure damping in the outer wings). The Stokes $V$ profile exhibits the characteristic S-shape morphology with the positive and negative lobes at the positions of the $\sigma_{\rm b}$ and $\sigma_{\rm r}$ components, respectively. Stokes $V$ vanishes when the field vector is perpendicular to the line of sight. The Stokes $Q$ parameter shows the opposite behaviour with the highest linear polarisation amplitude corresponding to the transverse field orientation. The shape of the local Stokes $Q$ spectrum is more complex than that of Stokes $V$. For the line with a triplet Zeeman splitting pattern considered here, the $Q$ profile has three lobes: positive ones for the $\sigma_{\rm b}$ and $\sigma_{\rm r}$ components and a negative one for the central $\pi$ component. The Stokes $U$ parameter is essentially absent for the zero azimuthal field angle adopted for the calculations in Fig.~\ref{fig:stokes_local}. For other field orientations, its profile morphology is similar to that of Stokes $Q$. All local Stokes profiles exhibit distinct symmetry properties independently of the field orientation. Stokes $I$, $Q$ and $U$ are symmetric with respect to the line centre. Stokes $V$ is anti-symmetric. 

A detailed numerical treatment of the PRT problem is computationally demanding and requires massive amount of input laboratory data as well as the knowledge of stellar atmospheric parameters and chemical abundances. This level of detail is unattainable in many stellar magnetometry applications and may not be justified considering a limited quality of observed Stokes parameter spectra. In that case, it is appropriate to use an approximate analytical solution of the PRT equation. In particular, the Unno-Rachkovsky \citep[e.g.][]{polarization:2004} solution, obtained under the assumption of Milne-Eddington atmosphere, is frequently used for interpretation of the circular polarisation spectra of M dwarfs. This solution assumes a linear source function dependence on the optical depth, a constant magnetic field vector as well as depth-independent line and continuum opacities and line broadening parameters. It provides a set of closed analytical expressions for the Stokes $I$, $Q$, $U$, $V$ parameters for any magnetic field strength and an arbitrary Zeeman splitting pattern. However, because detailed physical treatment of line formation is replaced with parametrised formulas, several parameters of the Unno-Rachkovsky solution (the line strength, the Doppler and Lorentzian line broadening, the slope of the source function dependence on the optical depth) are not known \textit{ab initio} and have to be adjusted empirically.

Another, more restrictive, analytical PRT solution can be obtained in the weak-field limit. The latter is defined for a given line as the field strength that yields a Zeeman splitting which is much smaller than the intrinsic line width. If the latter is dominated by the thermal Doppler broadening, the weak-field approximation requires $\Delta\lambda_{\rm B}\ll \Delta\lambda_{\rm D}$. For a very magnetically sensitive line, such as the Fe~{\sc i} line illustrated in Fig.~\ref{fig:stokes_local} \cla{($g_{\rm eff}=2.5$)}, this condition is satisfied for the field strength $\la$\,0.5~kG. For lines with average magnetic sensitivity \cla{($g_{\rm eff}\approx1.0$)}, the weak-field approximation is valid up to $B\approx1$~kG. The assumption that $\Delta\lambda_{\rm B}$ is small allows one to apply the Taylor expansion to the PRT equation and establish that, to the first-order, magnetic field manifests itself in a Stokes $V$ \cla{signature} that has a simple relation to the derivative of the Stokes $I$ parameter in the absence of the field
\begin{equation}
V(v)=-\Delta\lambda_{\rm B} \cos{\theta} \dfrac{c}{\lambda_0} \dfrac{\partial I_0}{\partial v}= -1.4\times10^{-6}g_{\rm eff}\lambda_0 B_\parallel \dfrac{\partial I_0}{\partial v}.
\label{eq:ab}
\end{equation}
Here $\theta$ is the angle between the line of sight and the magnetic field vector, $B_\parallel \equiv B\cos{\theta}$ is the longitudinal component of the magnetic field, and $v=(\lambda-\lambda_0)/c$ is the velocity relative to the line centre. This relation provides an insight into how the Stokes $V$ profile shape relates to that of the intensity profile and how the amplitude of circular polarisation signature scales with wavelength and effective Land\'e factor. Eq.~(\ref{eq:ab}) is widely used for modelling $V$ profiles of active late-type stars. However, this formula is of limited usefulness for M dwarfs since their magnetic field strengths frequently exceed 1~kG.

\subsection{Disk-integrated Stokes parameters}
\label{sect:disk}

The methods of calculating local Stokes parameter spectra described in the previous section yield theoretical profiles suitable for direct comparison with observations of a resolved magnetic structure, such as local observations of magnetic regions on the solar surface. However, surfaces of stars other than the Sun are unresolved, meaning that their \cla{observed} Stokes spectra contain contributions from zones with different magnetic field strength and orientation. Additionally, these spectral contributions are modulated in time and Doppler shifted due to stellar rotation. Thus, a further step of disk integration is required to simulate observations of magnetic stars. In this procedure a certain surface distribution of magnetic field is assumed. The magnetic field vector map corresponding to this distribution is converted from the stellar to observer's reference frame for a given inclination angle of the stellar rotational axis and a given rotational phase. The visible stellar surface is divided into a number of elements. The Stokes parameter profiles are calculated for each of these zones with one of the methods described above, taking into account the local Doppler shifts and limb angles. Finally, these contributions are added together with a weight that incorporates the projected area of this surface element and the local continuum brightness (which varies across the disk due to limb darkening and, possibly, spots \cla{and plages}).

The impact of disk integration on the spectropolarimetric observables is profound. The Stokes parameter profiles loose their simple character and symmetry properties. The amplitude of polarisation signatures can greatly diminish due to a destructive addition of the spectral contributions with opposite signs of the Stokes $V$, $Q$, $U$ signals. Significant rotational modulation appears in some observables for magnetic field geometries dominated by a non-axisymmetric component. 

The effect of disk integration is qualitatively different for the intensity and polarisation profiles. The Stokes $I$ spectra are weakly sensitive to the magnetic field orientation but change significantly with the field strength. Consequently, it is often necessary to introduce a field strength distribution, i.e. combine spectra calculated with different field strength values, to adequately describe observations. The simplest form of such a distribution is a two-component model. It supposes that a fraction $f$ of the stellar surface is covered by the field strength $B$ and the rest of the surface is non-magnetic. A generalisation of this model is a multi-component field strength parameterisation, containing three or more spectral contributions corresponding to different field strengths. For the purpose of modelling Stokes $I$, each of these components is usually represented by a uniform surface magnetic field distribution. 

\begin{figure}[!t]
\centering
\includegraphics[width=0.75\textwidth]{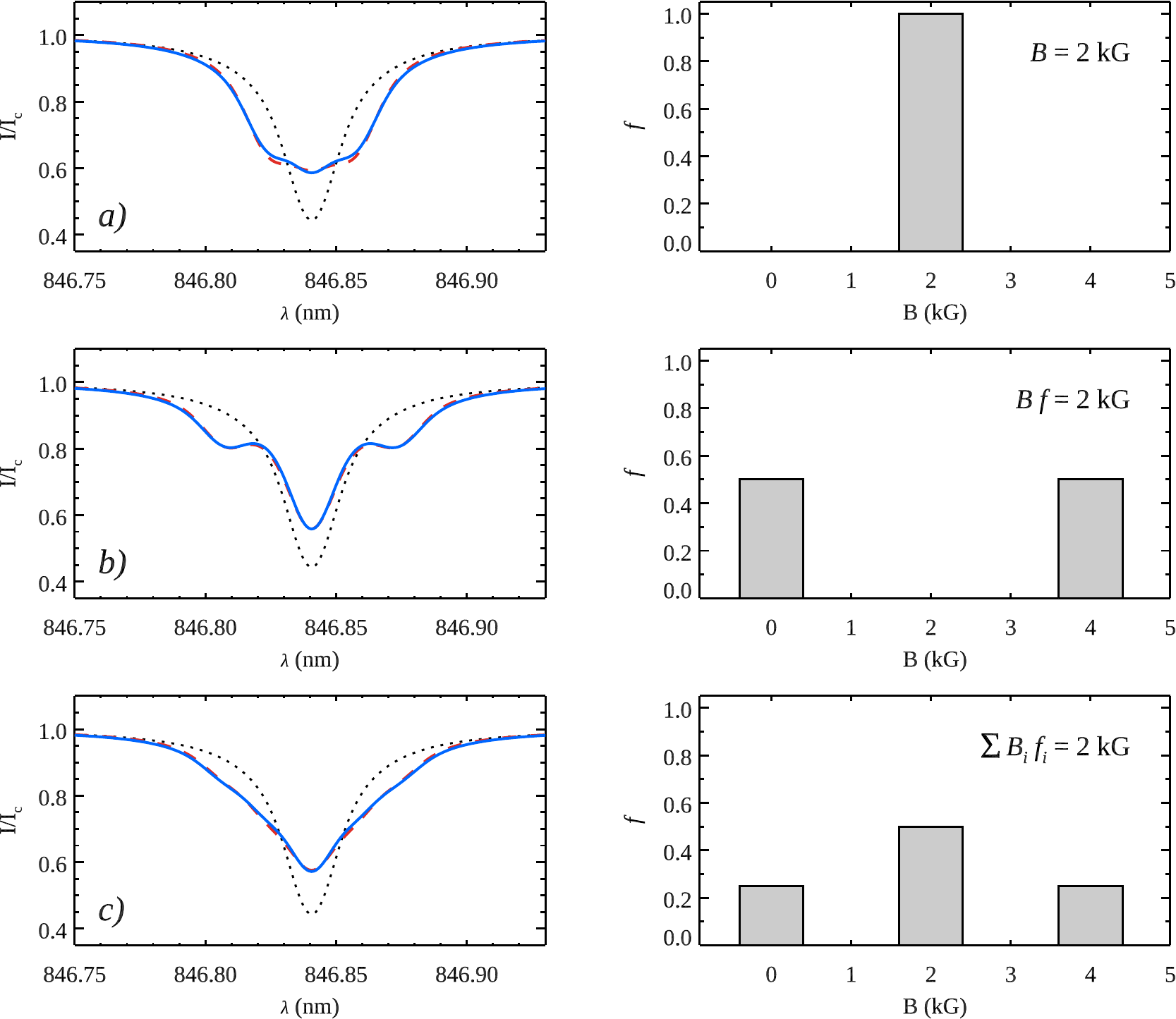}
\caption{Disk-integrated intensity profiles (left) of the Fe~{\sc i} $\lambda$ 846.84~nm line for a uniform magnetic field with different field strength distributions (right). (a) Homogeneous single-value field, (b) two-component field strength distribution, (c) three-component model. The mean field strength is 2~kG in all three cases. Solid lines correspond to the theoretical spectra calculated for a uniform radial field, dashed lines show calculations assuming a horizontal field, dotted lines illustrate non-magnetic profiles. Calculations are carried out for $T_{\rm eff}=3800$~K, $\log g=5.0$ model atmosphere. All spectra are convolved with a 5~\kms\ Gaussian kernel.}
\label{fig:bf_models}
\end{figure}

Fig.~\ref{fig:bf_models} shows Stokes $I$ profiles of the Fe~{\sc i} 846.84~nm line calculated with a single field strength $B=2$~kG (top), a 4~kG field covering 50\% of the stellar surface (middle), and a three-component model including contributions of 0, 2, and 4~kG fields (bottom). For each of these cases the average magnetic field, defined as \bi\,=\,$B$ for the first model, \bi\,=\,$Bf$ for the second, and \bi\,=\,$\sum B_i f_i$ for the third, is the same. Nevertheless, the profiles differ significantly, underscoring the necessity of applying a suitable field strength distribution in practical analyses of Stokes $I$ spectra of magnetic stars. At the same time, Fig.~\ref{fig:bf_models} also shows that changing from a purely radial homogenous field (parallel to the line of sight at the disk centre and perpendicular to the line of sight at the limb) to a uniform azimuthal field (perpendicular to the observer's line of sight and tangential to the stellar surface across the disk) has a very small impact on the disk-integrated Stokes $I$ profiles. For this reason, studies interpreting M-dwarf intensity spectra cannot determine the local field orientation and typically adopt a purely radial field.

\begin{figure}[!t]
\centering
\includegraphics[width=0.2\textwidth]{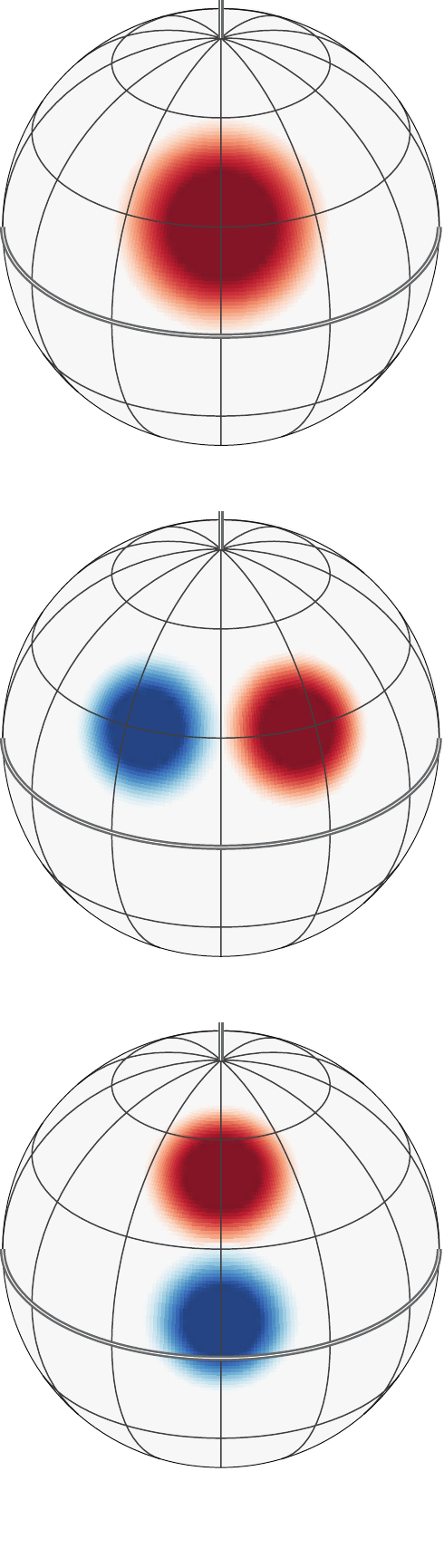}\hspace*{5mm}
\includegraphics[width=0.72\textwidth]{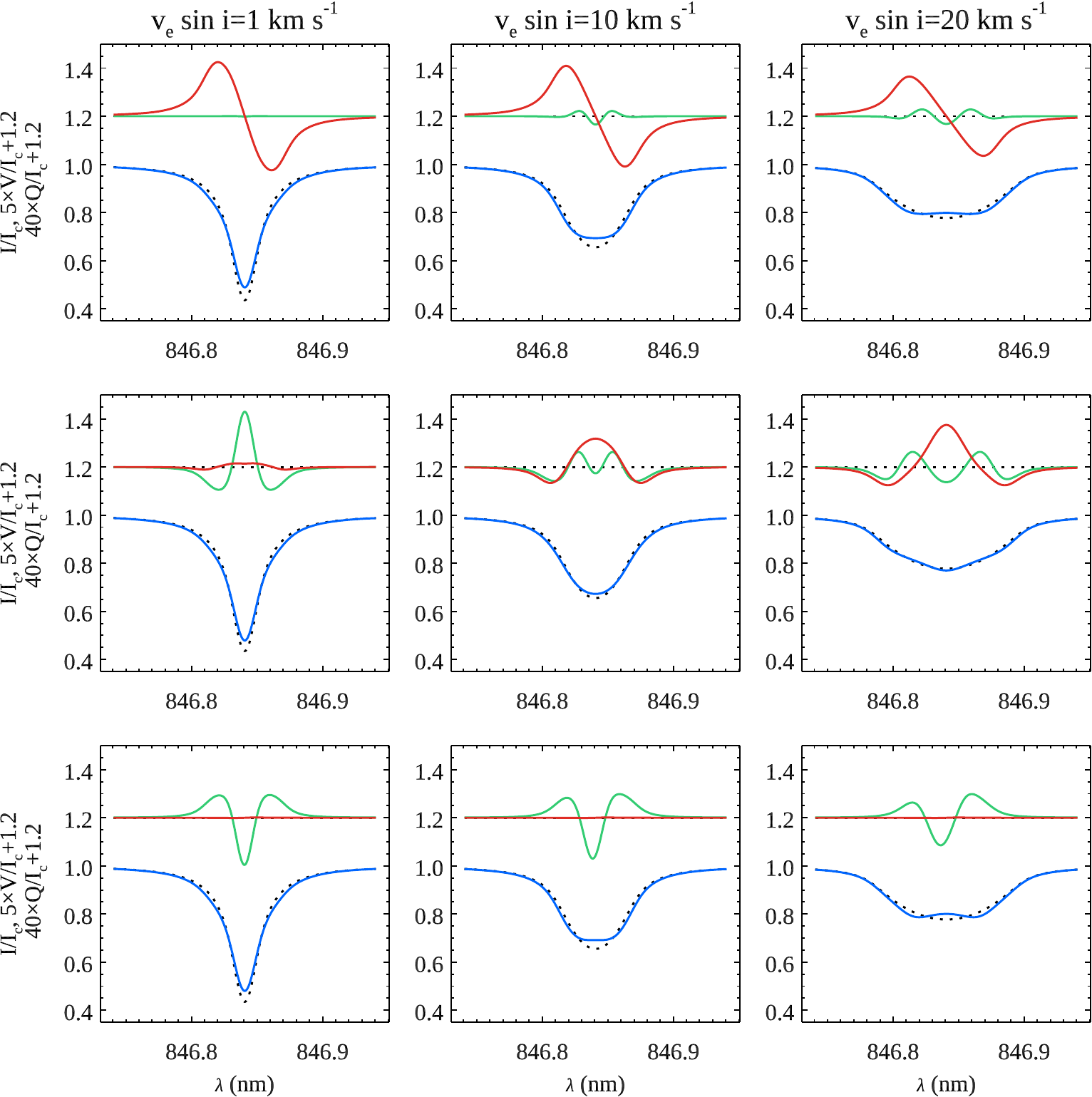}
\caption{\cla{Signatures} of simple magnetic field geometries in the disk-integrated Stokes $I$, $V$, and $Q$ profiles of the Fe~{\sc i} $\lambda$ 846.84~nm line. Top row: a single 3~kG radial field spot located at the disk centre. Middle row: two 3~kG spots with opposite field polarities offset in longitude. Bottom row: two 3~kG spots with opposite field polarities offset in latitude. Panels to the right of the spherical plots show theoretical Stokes parameter spectra calculated for $v_{\rm e}\sin i=\cla{1}$, 10, and 20~\kms. Each panel shows Stokes $I$ (blue solid line), $V$ (red solid line), and $Q$ (green solid line) profiles together with the calculation without magnetic field (black dotted line). Polarisation spectra are shifted vertically and amplified by a factor of 5 for Stokes $V$ and 40 for Stokes $Q$. Calculations are carried out for $T_{\rm eff}=3800$~K, $\log g=5.0$, and inclination angle $i=60^{\rm o}$.}
\label{fig:stokes_integ}
\end{figure}

The impact of disk integration on the Stokes $V$, $Q$, and $U$ profiles depends sensitively on the field orientation and degree of the field complexity. As one can see from Eq.~(\ref{eq:ab}), Stokes $V$ varies with the angle $\theta$ between the field vector and the line of sight as $\cos{\theta}$, thus changing sign for opposite field orientations and turning to zero at $\theta=90^{\rm o}$ and $270^{\rm o}$. Depending on the spatial scale of large changes in $\theta$, the disk-integrated circular polarisation profiles may duly reveal or entirely miss certain magnetic field configurations. Fig.~\ref{fig:stokes_integ} gives an example of these different outcomes of disk integration. The top row of this figure shows the Stokes profiles for a configuration with a single 3~kG radial field spot located at the disk centre. In this case, Stokes $V$ maintains its simple S-shape morphology independently of the projected rotational velocity $v_{\rm e}\sin i$. On the other hand, the linear polarisation amplitude is very low due to cancellation and lack of a substantial transverse field component. The middle row shows simulated Stokes spectra for a pair of spots with opposite polarities of 3~kG radially-oriented magnetic field. These spots are separated in longitude. The Stokes $V$ signal is almost fully cancelled out for small $v_{\rm e}\sin i$, but increases in amplitude as the Doppler effect separates profile contributions coming from the two spots. The disk-integrated Stokes $V$ spectrum for $v_{\rm e}\sin i=20$~\kms\ reaches almost the same amplitude as was obtained for the single-spot geometry and exhibits a symmetric W-shape, which could not be produced by the Zeeman effect in a local circular polarisation profile. If the same pair of magnetic spots is arranged along the central meridian (bottom row in Fig.~\ref{fig:stokes_integ}), their Stokes $V$ spectra remain undetectable at any $v_{\rm e}\sin i$ value. The Stokes $Q$ signals are noticeably stronger for the configurations with spot pairs compared to the single spot, but these linear polarisation signals remain about an order magnitude weaker than the Stokes $V$ signatures in the upper and middle rows of Fig.~\ref{fig:stokes_integ}. One can also notice that the presence of magnetic field is always recognisable in Stokes $I$, yielding slightly broader and shallower profiles compared to the case when the field is absent. However, the difference between the intensity profiles with and without magnetic field is small due to a low fraction of the stellar surface covered by the field in these calculations.

To summarise, the disk-integrated polarimetric observables provide a valuable information on the field geometry. However, they capture only some part of the actual stellar magnetic field. Depending on the degree of local field intermittency, this part can be significant or represent a minor fraction of the magnetic field present at the stellar surface. For this reason, it is important to distinguish the strength of the global magnetic field component inferred from polarisation profiles from the total magnetic field strength \bi\ found from Stokes $I$. Throughout this review we denote the surface-averaged global magnetic field as \bv\ since maps of large-scale fields are typically reconstructed from Stokes $V$ observations alone. For M dwarfs and other late-type active stars one always finds \bv\,$\ll$\,\bi.

\begin{figure}[!t]
\centering
\hspace*{5mm}\includegraphics[width=0.28\textwidth]{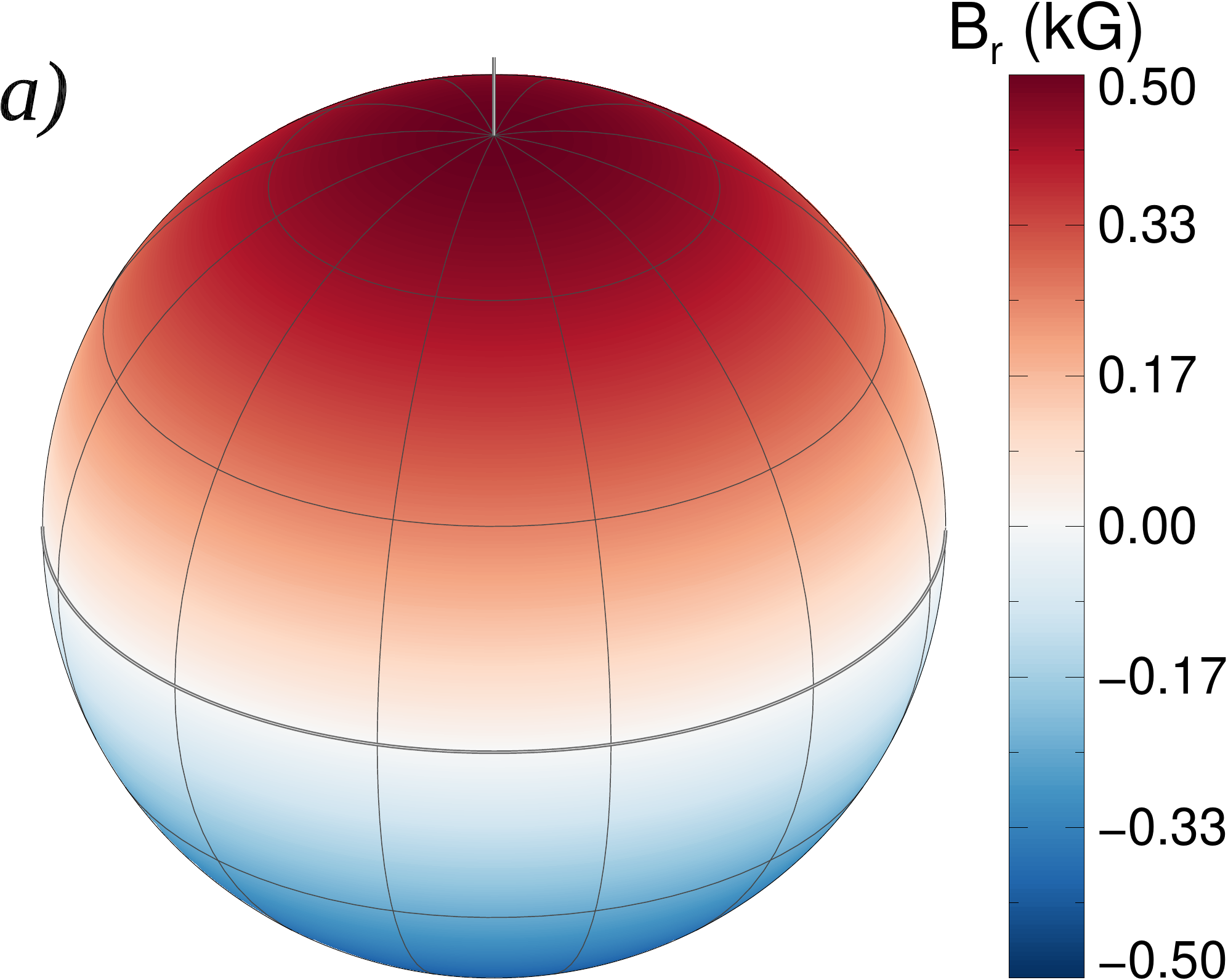}
\hspace*{5mm}\includegraphics[width=0.28\textwidth]{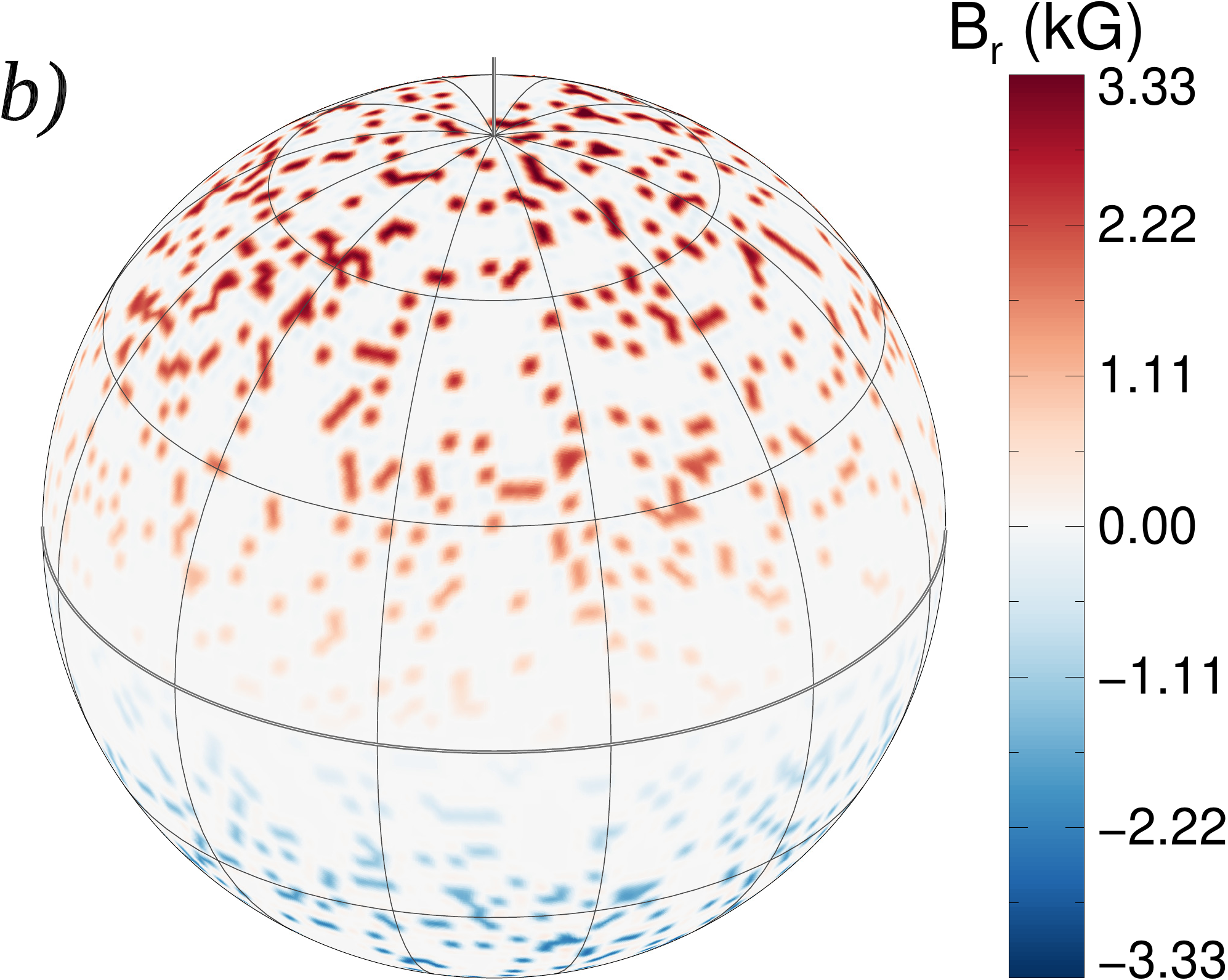}
\hspace*{5mm}\includegraphics[width=0.28\textwidth]{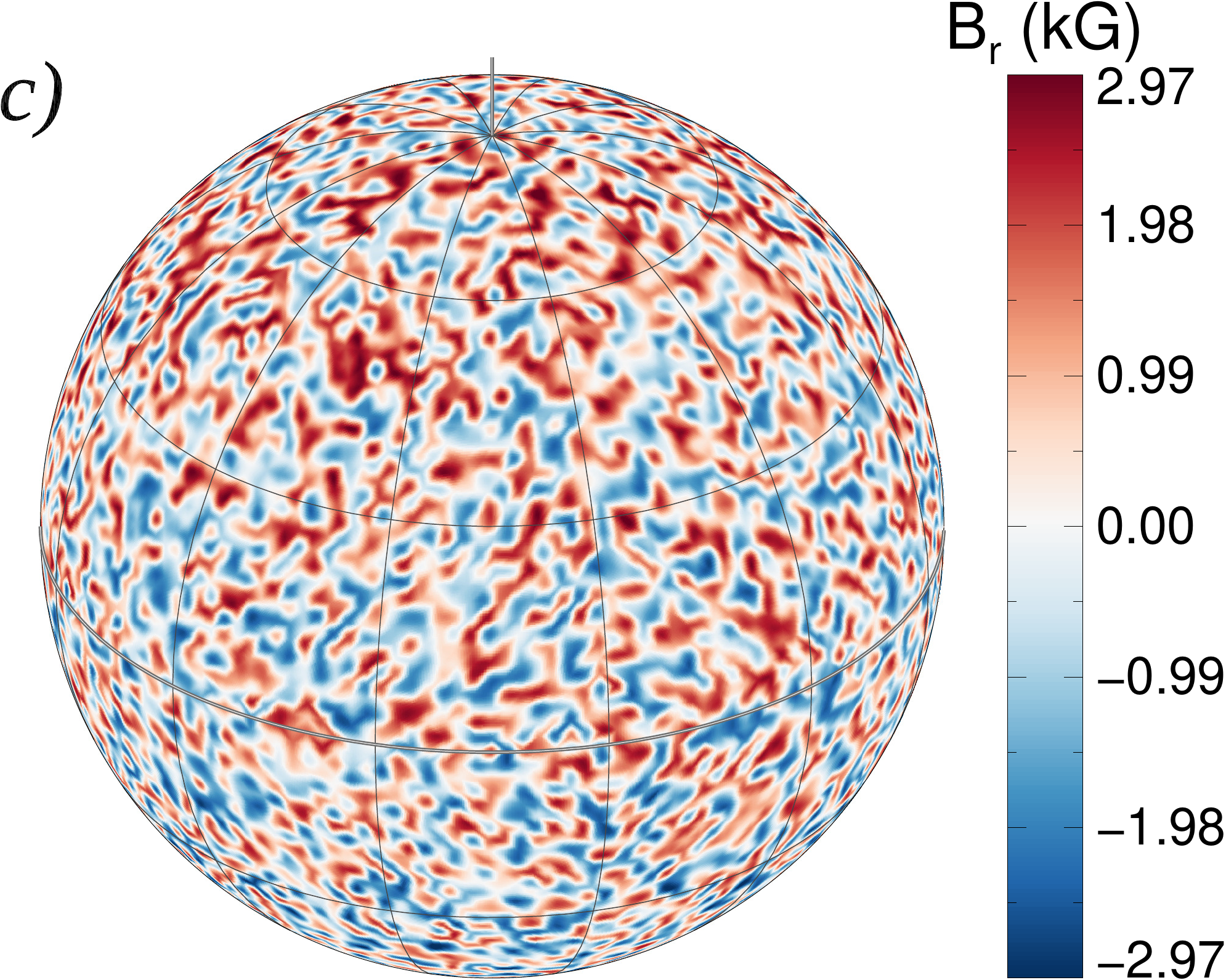}\\
\vspace*{5mm}
\includegraphics[width=0.95\textwidth]{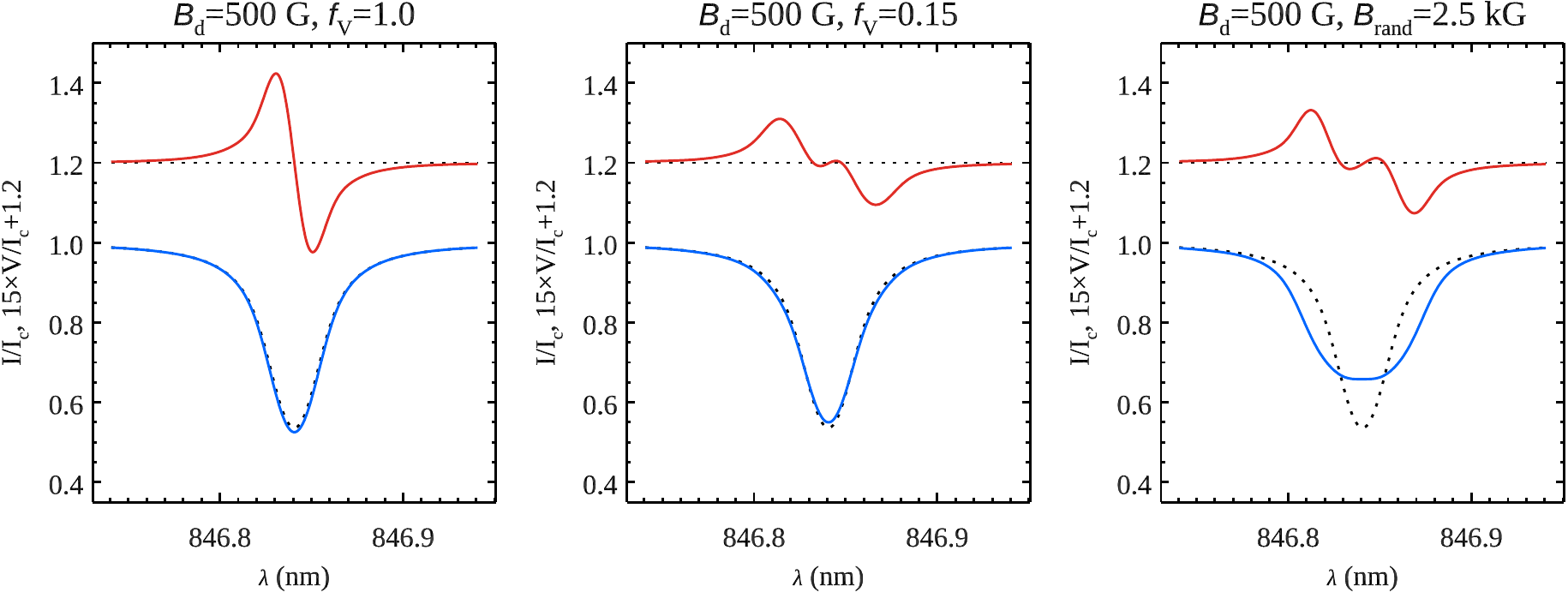}
\caption{Disk-integrated Stokes $I$ and $V$ profiles of the Fe~{\sc i} $\lambda$ 846.84~nm line computed with different treatment of the global magnetic field geometry. (a) Conventional axisymmetric dipolar field with 0.5~kG polar strength. (b) The same dipolar field, but occupying 15\% of the stellar surface. (c) Superposition of 0.5~kG dipolar field and 2.5~kG randomly oriented field. The spherical colour maps in the upper row illustrate the radial field distributions, with the side bar giving the field strength in kG. Plots in the lower row show the corresponding model Stokes $I$ (blue solid line) and Stokes $V$ (red solid line) profiles. The black dotted lines show calculations without magnetic field. The circular polarisation spectra are shifted vertically and amplified by a factor of 15 relative to Stokes $I$. Calculations are carried out for $T_{\rm eff}=3800$~K, $\log g=5.0$, and $v_{\rm e}\sin i=5$~\kms.}
\label{fig:v_models}
\end{figure}

An additional complication, specific to the interpretation of disk-integrated polarisation spectra of active M dwarfs, stems from the fact that formation of their Stokes parameter profiles cannot be treated in the weak-field limit. In the latter case, the local Stokes $V$ profile shape is independent of the field strength but scales in amplitude according to the magnitude of the line of sight field component. Then, the same Stokes $V$ profile is obtained for the surface element covered by a uniform field with a given strength $B_{\rm loc}$ and for $B_{\rm loc}/f$ field occupying a fraction $f$ of this element. As evident from Fig.~\ref{fig:stokes_local}, this equivalence breaks down for fields exceeding $\sim$\,1~kG. For stronger fields, the Stokes $V$ profile shape depends on the local field modulus. For active M dwarfs this \cla{often} means that some superposition of global and much stronger local fields has to be introduced in order to reproduce both the amplitude and width of the Stokes $V$ profiles seen in highly-quality observations. This has been accomplished with the help of the global field filling factor $f_V$, usually assumed to be the same for the entire stellar surface \citep{morin:2008}. In this approach the local Stokes parameter profiles are calculated for $B_V/f_V$ and then polarisation profiles are downscaled by multiplying them by $f_V$. As demonstrated by Fig.~\ref{fig:v_models}, this yields Stokes $V$ profiles that are qualitatively different -- showing a lower amplitude and wider wings -- than calculation with $f_V=1$. The physical interpretation of this global field filling factor is that polarisation signal is produced not by a continuous, monolithic global field geometry (upper panel in Fig.~\ref{fig:v_models}a), but by a system of strong-field spots arranged according to some large-scale configuration (upper panel in Fig.~\ref{fig:v_models}b). Similar Stokes $V$ profile shapes can be also obtained by direct superposition of an organised global field and an intermittent local field comprised of magnetic spots with random field vector orientation (e.g. \citealt{lang:2014} and upper panel in Fig.~\ref{fig:v_models}c). However, such composite field structure model is yet to be applied for practical modelling of polarisation spectra of M dwarfs.

\subsection{Zeeman broadening and intensification}
\label{sect:zb}

The consequence of the Zeeman effect for the line profiles in stellar intensity spectra is twofold. First, as illustrated by the calculations in the previous section and by Fig.~\ref{fig:bf_models}, separation of the Zeeman components leads to broadening and, eventually, to splitting of spectral lines. The magnitude of this effect is quantified by Eq.~(\ref{eq:aa}). An equivalent, and more informative, relation for separation of the Zeeman components in velocity units is given by
\begin{equation}
\Delta v_{\rm B} = 1.4\times10^{-3} g_{\rm eff}\lambda_0 B
\end{equation}
with $\Delta v_{\rm B}$ in \kms, field strength in kG and wavelength in nm. 

For the Zeeman broadening to be reliably identified, $\Delta v_{\rm B}$ must be at least comparable to, or exceed, other broadening contributions. The turbulent velocities in the atmospheres of M dwarfs are believed to be small \citep{wende:2009}, so the line width in the absence of a magnetic field is dominated by the instrumental broadening (3--6~\kms\ for the resolving power $R=\lambda/\Delta\lambda=0.5$--1$\times10^5$) and by the rotational Doppler effect. Considering parameters of the Fe~{\sc i} 846.84 nm line, which offers one of the best possibilities for detecting magnetic broadening in the optical M-dwarf spectra, one can determine $\Delta v_{\rm B} =3.0$ \kms\,kG$^{-1}$. This shows that practical applications of the Zeeman broadening analysis based on this line are limited to very active M dwarfs with multi-kG fields observed with a high signal-to-noise ratio ($S/N$), $R\ga10^5$ spectra \citep[e.g.][]{johns-krull:1996}. If the projected rotational velocity exceeds $\sim$\,5~\kms, identification of the Zeeman broadening becomes ambiguous. These requirements can be partly relaxed with observations at near-infrared wavelengths \citep{saar:1985,johns-krull:1999}. For example, for the $g_{\rm eff}=2.5$ Ti~{\sc i} line at $\lambda$ 2231.06~nm one finds $\Delta v_{\rm B} =7.8$ \kms\,kG$^{-1}$, indicating that the Zeeman broadening caused by a $\sim$\,2~kG field is still recognisable in the spectra of stars rotating with $v_{\rm e}\sin i\sim20$~\kms.

A less commonly discussed consequence of the Zeeman effect is the differential magnetic intensification of spectral lines. This effect occurs due to a desaturation of strong spectral lines associated with the wavelength separation of their Zeeman components \citep[e.g.][]{basri:1992,basri:1994,kochukhov:2020}. This is similar to how the isotope or hyperfine splitting increases the equivalent width of some spectral lines in the absence of the field. The Zeeman intensification increases monotonically with the field strength, until the Zeeman components are fully resolved. The magnitude of this effect is a complex function of the line strength and the Zeeman splitting pattern parameters and cannot be expressed analytically. Lines with the strongest magnetic intensification are not necessarily the same features that have the largest $g_{\rm eff}$ values and are most affected by the Zeeman broadening. Instead, stronger lines with a large number of widely separated Zeeman components tend to exhibit a larger amplification in a magnetic field.

\begin{figure}[!t]
\centering
\includegraphics[width=0.7\textwidth]{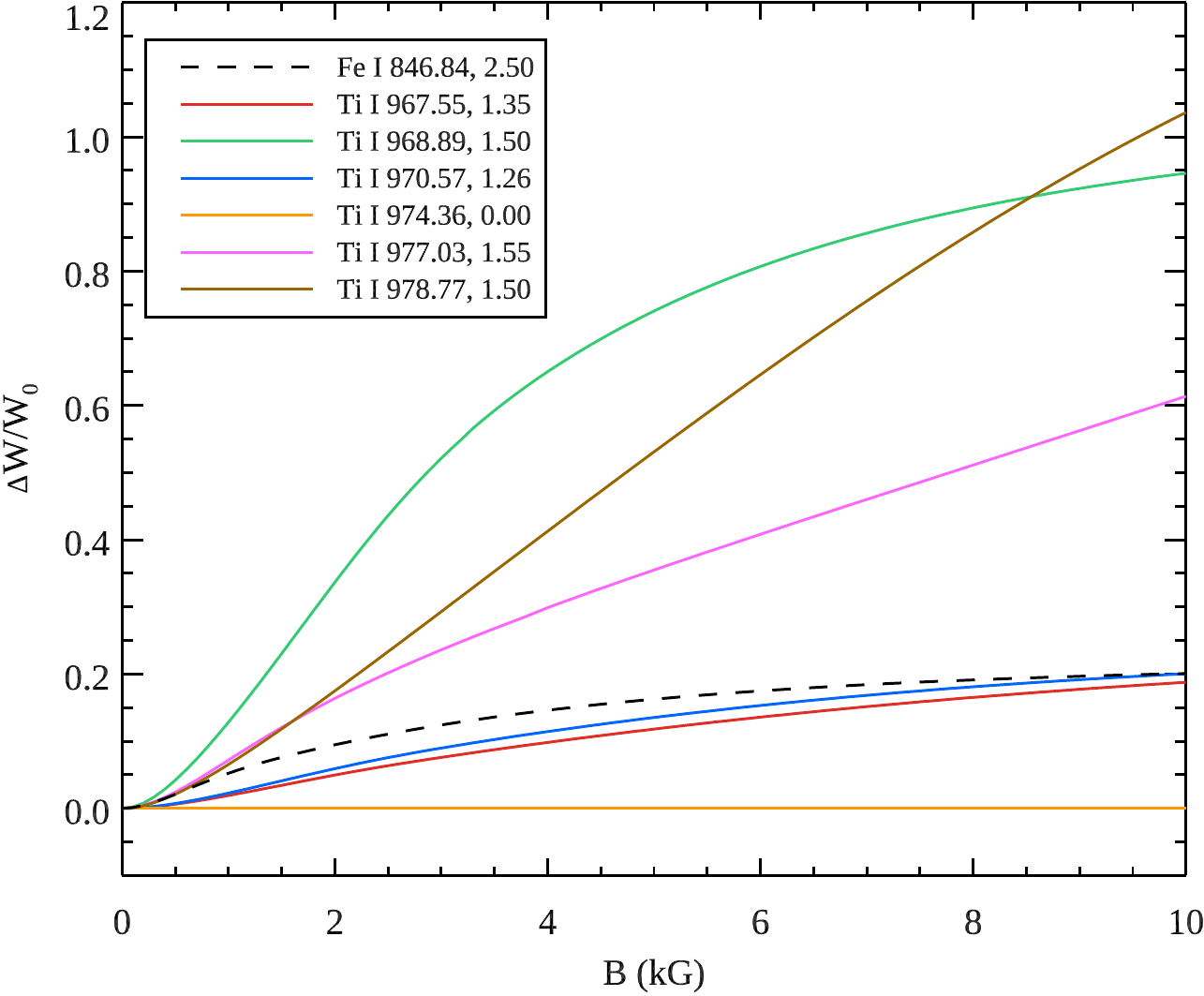}
\caption{Magnetic intensification of the Ti~{\sc i} lines from the $^5$F--$^5$F$^{\rm o}$ multiplet (solid curves) compared to the intensification of the Fe~{\sc i} 846.84~nm line (dashed curve). The legend lists the central wavelengths in nm and the effective Land\'e factors. Theoretical equivalent widths were obtained assuming a uniform radial magnetic field and using $T_{\rm eff}=3800$~K, $\log g=5.0$ model atmosphere.}
\label{fig:intens}
\end{figure}

The main advantage of using the Zeeman intensification for magnetic field measurements compared to analysing the Zeeman broadening is that the former method makes use of intensities or equivalent widths of spectral lines whereas the latter extracts information from detailed line profile shapes. Consequently, magnetic intensification analysis is far less demanding in terms of the quality of observational material and places no direct restrictions on the stellar $v_{\rm e}\sin i$. On the other hand, intensification analysis must rely on comparison of multiple lines with different responses to magnetic field in order to disentangle the magnetic line amplification from all other parameters influencing line strengths in stellar spectra. To this end, an analysis of the differential Zeeman intensification of lines belonging to the same multiplet is the optimal approach since it allows one to avoid errors related to uncertain line parameters, in particular the oscillator strengths, and ensure that the studied lines are \cla{formed} under similar conditions in the stellar atmosphere. 

\citet{kochukhov:2017c} and \citet{shulyak:2017} found that ten Ti~{\sc i} lines from the $^5$F--$^5$F$^{\rm o}$ multiplet at $\lambda$ 964.74--978.77~nm represent an exquisite diagnostic of M-dwarf magnetic fields owing to the fact that one of those lines, $\lambda$ 974.36~nm, has a zero effective Land\'e factor and thus can be employed for constraining titanium abundance and non-magnetic broadening. Fig.~\ref{fig:intens} illustrates a relative increase of the equivalent width as a function of the magnetic field strength for the seven Ti~{\sc i} lines from this multiplet and for the Fe~{\sc i} 846.84~nm line discussed earlier. One can see that the equivalent width of some of these Ti~{\sc i} lines increases by $\ge20$\% in a 2~kG field and that the magnetic response varies considerably from one line to another. Lines with nearly identical effective Land\'e factors (e.g. Ti~{\sc i} 968.89, 977.03, 978.77 nm) exhibit different intensification curves due to different Zeeman splitting patterns. The Fe~{\sc i} 846.84~nm, deemed to be very magnetically sensitive by Zeeman broadening applications, shows only a modest equivalent width increase in a magnetic field despite its large effective Land\'e factor.

\subsection{Least-squares deconvolution}
\label{sect:lsd}

The amplitude of circular polarisation signatures in the disk-integrated spectra of active late-type stars is usually too small for a reliable detection of these signatures in individual lines. This is also the case for the majority of M dwarfs, many of which are also too faint in the optical for high-quality ($S/N\gg100$), time-resolved spectra to be obtained at medium-size telescopes normally available for monitoring studies. The Zeeman linear polarisation signals are roughly one order magnitude weaker than Stokes $V$ and have never been detected in individual lines of any late-type star. These difficulties notwithstanding, one can still detect and model high-resolution polarisation signatures by combining information from many spectral lines \citep{semel:1996,donati:1997}. Eq.~(\ref{eq:ab}) shows that in the weak-field limit the Stokes $V$ profiles of different lines are self-similar and their amplitude scales with the Stokes $I$ line depth, effective Land\'e factor and \cla{central} wavelength. Therefore, a stellar spectrum can be represented as a convolution of a mean profile and a line mask composed of delta-functions at the \cla{wavelength positions} of considered lines with amplitudes equal to the expected line strengths. This coarse model assumes that the profiles of overlapping lines add up linearly. One can invert this model and determine the mean line profile for a given observed spectrum and a line mask. This can be accomplished with a set of matrix operations equivalent to solving a linear least-squares problem \citep{donati:1997,kochukhov:2010a}. This efficient line-addition algorithm is known as the least-squares deconvolution (LSD) and the resulting average line profiles are referred to as the LSD Stokes $IQUV$ profiles.

Only atomic lines are included in the LSD line masks since molecular features are often blended, lack accurate theoretical line lists and do not follow simple polarisation scaling relation given by Eq.~(\ref{eq:ab}). Moreover, any lines which deviate from the average behaviour (e.g. emission lines, very strong lines with broad wings) have to be excluded as well. Line strengths necessary for application of LSD are provided by the same theoretical spectrum synthesis calculations as required for the analysis of individual lines (e.g. Sect.~\ref{sect:prt}). However, the LSD profiles are only weakly sensitive to the adopted stellar parameters and abundances. Depending on the stellar spectral type and wavelength coverage of observations, one can find from a few hundred to $\sim$\,10$^4$ lines suitable for LSD and achieve a $S/N$ gain of 10--100 relative to the analysis of individual lines. A polarimetric sensitivity of $\sim$\,$10^{-5}$ has been achieved in the Stokes $V$ LSD profiles of bright solar-type stars \citep[e.g.][]{kochukhov:2011,metcalfe:2019}, while a precision of 1--$5\times10^{-4}$ is more typical of modern M-dwarf observations \citep{donati:2008,morin:2008}.

Although the original idea of LSD was based on the weak-field behaviour of Stokes $V$, this line-averaging technique is also routinely applied to the strongly magnetic Ap stars \citep{silvester:2012,kochukhov:2019} and was extended to the Stokes $Q$ and $U$ parameter spectra \citep{wade:2000b,kochukhov:2011}. Likewise, LSD has enabled detection of both circular and linear polarisation signatures in M-dwarf stars with multi-kG magnetic fields \citep{donati:2008,morin:2008,lavail:2018}. The question of interpretation of the LSD profiles for such strongly magnetic objects is not fully settled. Depending on what type of modelling is applied to these profiles, the basic assumptions and simplifications inherent to the LSD method might be largely irrelevant or give rise to major errors \citep{kochukhov:2010a}. For instance, it is understood that a measurement of the disk-integrated line of sight magnetic field -- the so-called mean longitudinal magnetic field \bz\ -- can be obtained from the normalised first moment of the Stokes $V$ LSD profile \cla{with the expression}
\begin{equation}
\langle B_{\rm z} \rangle = -7.145\times10^5 \dfrac{\int (v-v_0) Z_V \mathrm{d}v}{\langle\lambda_0\rangle \langle g_{\rm eff}\rangle\int (1-Z_I) \mathrm{d}v} 
\end{equation}
at any field strength that can be realistically expected on the surface of a late-type star. This formula gives \bz\ in G for wavelength in nm and velocity in \kms. $\langle\lambda_0\rangle$ and $\langle g_{\rm eff}\rangle$ correspond to the average wavelength and effective Land\'e factor of the set of lines employed for the LSD procedure. $Z_V$ and $Z_I$ are the circular polarisation and intensity LSD profiles, respectively, and $v_0$ is the centre-of-gravity of the Stokes $I$ profile. 

At the same time, detailed modelling of the LSD profile shapes, especially beyond the weak-field limit or when the surface magnetic field is accompanied by temperature spots, should be approached with care. The usual assumption made by many spectral modelling studies, including all spectropolarimetric investigations of M dwarfs published so far, is that the observed LSD spectra can be approximated by calculations for a single fiducial line that has $\lambda_0=\langle\lambda_0\rangle$, $g_{\rm eff}=\langle g_{\rm eff}\rangle$ and a triplet Zeeman splitting. \citet{kochukhov:2010a} demonstrated that this approximation becomes increasingly inaccurate for magnetic fields exceeding $\sim$\,2~kG and is not applicable to the Stokes $Q$ and $U$ LSD profiles at any field strength. Instead, it is possible to model the LSD profiles by comparing them with theoretical calculations in which detailed polarised radiative transfer computations for the entire stellar spectrum are processed by LSD using the same line mask as applied to observations \citep{kochukhov:2014,rosen:2015,strassmeier:2019}. This multi-line approach to the problem of interpretation of LSD spectra was applied to early-type magnetic stars and a few active solar-type stars but has not been adapted to M dwarfs. In the latter case, realistic PRT calculation of wide wavelength coverage spectra is greatly complicated by the presence of numerous molecular lines for which no complete lists are currently available and the Zeeman splitting requires a special treatment (see Sect.~\ref{sect:zeeman}).

\subsection{Zeeman Doppler imaging}
\label{sect:zdi}

Rotational modulation of the intensity and polarisation spectra of active stars can be exploited to reconstruct detailed maps of spots and magnetic fields on the stellar surfaces. For a Doppler-broadened line profile observed at a given rotational phase, there is a correspondence between the position of a spot relative to the central meridian on the stellar disk and location of the spectral contribution of this spot within the disk-integrated line profile. As the star rotates, the position of spot on the stellar disk changes in the observer's reference frame and thus a distortion, or a polarisation signature, associated with this spot moves across the line profile. Taking advantage of this behaviour, the techniques of Doppler and Zeeman (Magnetic) Doppler imaging (ZDI) invert high-resolution spectropolarimetric time-series observations into a two-dimensional surface distribution of star spots or magnetic field vector. These powerful tomographic imaging techniques have been applied to different types of active stars featuring diverse surface inhomogeneities (spots of temperature, element abundances, magnetic fields with both highly structured and simple, globally-organised topologies, non-radial pulsations, etc.). Comprehensive reviews of different \cla{applications} of indirect stellar surface imaging can be found elsewhere \citep[e.g.][]{kochukhov:2016}. Here the discussion will be restricted to a few methodological aspects of ZDI relevant for mapping M-dwarf magnetic fields.

The ZDI modelling of low-mass stars has so far relied exclusively on interpretation of the Stokes $V$ profile time-series. Information from linear polarisation was neglected due to difficulty of obtaining Stokes $QU$ observations of sufficient quality. Furthermore, no attempts were made to reproduce the Zeeman broadening of intensity spectra and the circular polarisation signatures with the same magnetic topology model. These methodological deficiencies have several important consequences. First, the lack of $Q$ and $U$ spectra leads to a certain ambiguity in recovering the local field inclination and cross-talks between maps of different magnetic field vector components \citep{donati:1997a,kochukhov:2002c,rosen:2012}. Furthermore, using Stokes $V$ alone for magnetic mapping inhibits reconstruction of small-scale components of magnetic geometries unless surface features are fully resolved by the stellar rotation \citep{kochukhov:2010,rosen:2015}. This exacerbates the signal cancellation problem existing for any disk-integrated polarimetric observable (Sect.~\ref{sect:disk}). Thus, ZDI is capable of recovering only a part, often a minor one in terms of the magnetic field energy fraction, of the total stellar magnetic field. The relationship between this global magnetic component and the total magnetic field present on a given M-dwarf star is generally unknown.

Although initial ZDI applications have focused on rapid rotators, useful magnetic field maps can also be obtained for narrow line stars with insignificant rotational Doppler broadening \citep[e.g.][]{donati:2006b,petit:2008a}. In this case, the information about spatial distribution of magnetic structures is extracted only from the temporal modulation of polarimetric signatures. For such low $v_{\rm e}\sin i$ targets the surface resolution of ZDI maps is reduced even further, independently of the quality of observations. Often, only the simplest, e.g. dipolar, component of the large-scale field can be constrained.

For M dwarfs, ZDI inversions are applied to the LSD Stokes $V$ profiles. Therefore, several caveats of theoretical interpretation of the LSD circular polarisation spectra discussed in the previous section are relevant. The ZDI codes which have been employed for these stars used either the Unno-Rachkovsky analytical solution of PRT \citep{morin:2008,kochukhov:2017c} or weak-field approximation \citep{donati:2006,morin:2008a} and treated LSD profile as a single line. These studies further assumed that the local surface features contributing to the disk-integrated polarisation signature are not correlated with cool spots, which may be present on late-type active stars and were indeed observed for a few M dwarfs \citep{morin:2008a,barnes:2017}. Only one recent ZDI study of the components of the early-M eclipsing binary YY~Gem \citep{kochukhov:2019a} took an inhomogeneous surface brightness distribution into account in modelling the Stokes $V$ profiles. It cannot be excluded that the global magnetic field strengths of M dwarfs are systematically underestimated if the circular polarisation signatures come preferentially from cool regions on the stellar surface.

\begin{figure}[!t]
\includegraphics[width=0.47\textwidth]{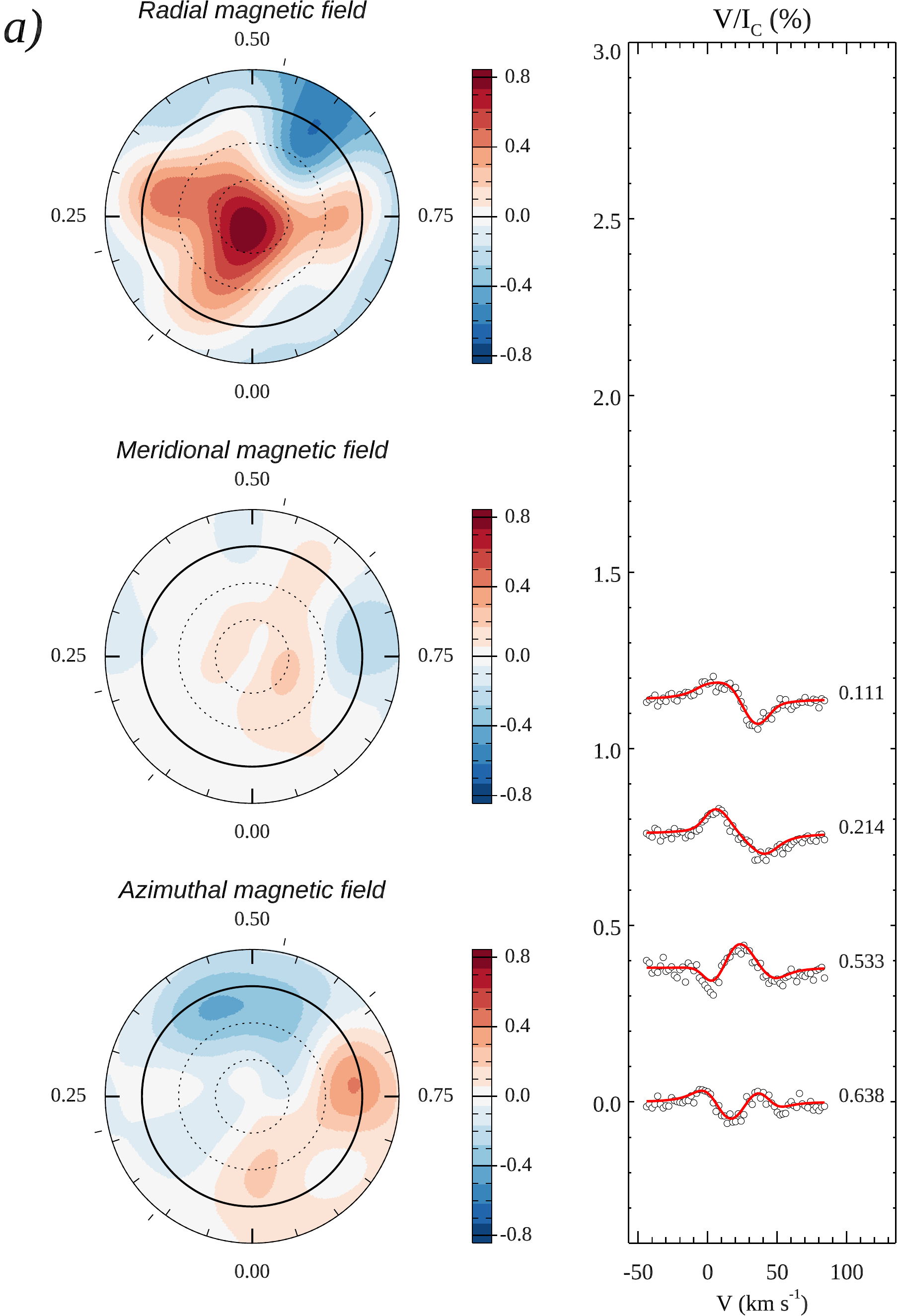}
\hspace*{5mm}
\includegraphics[width=0.47\textwidth]{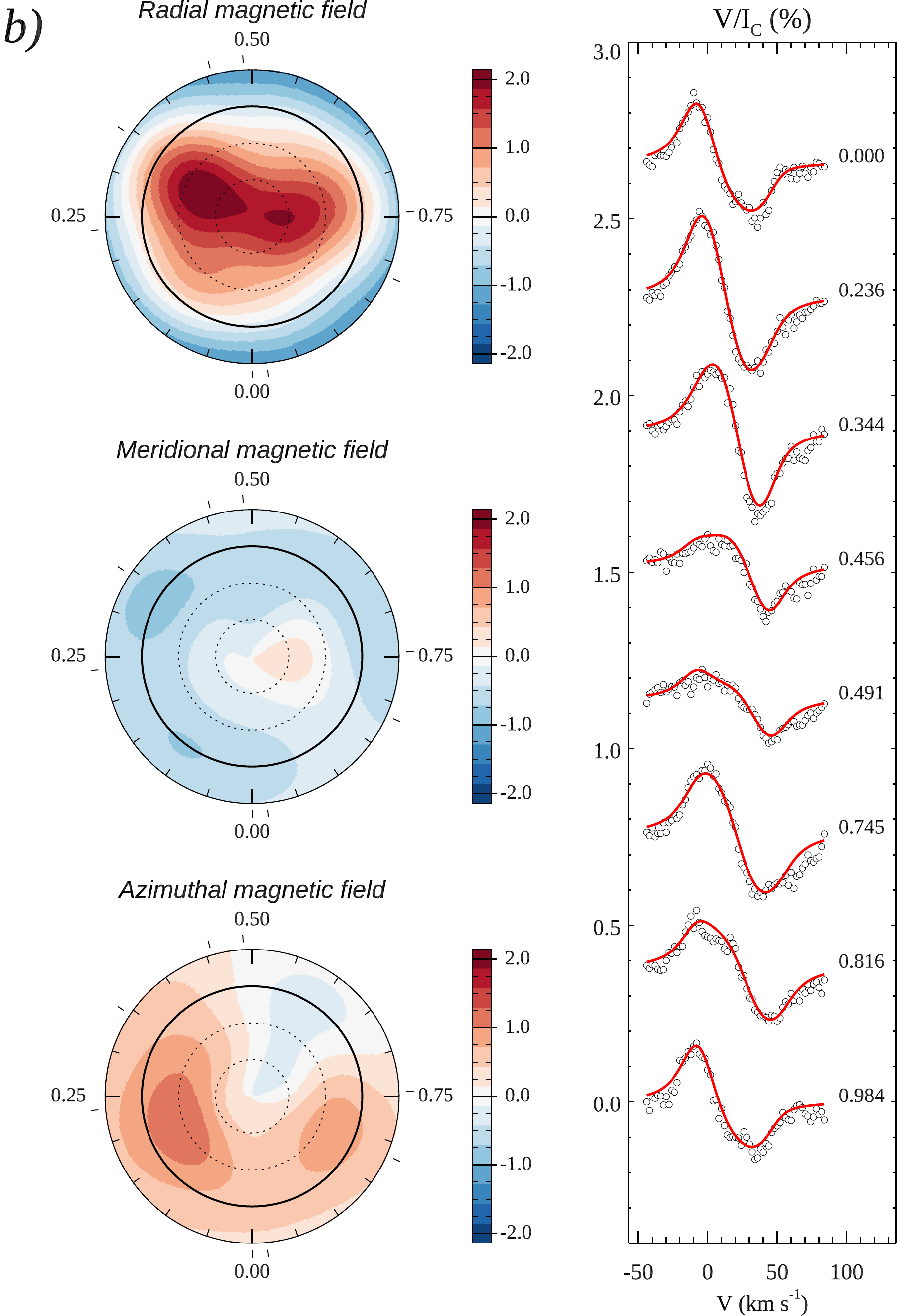}
\includegraphics[width=0.47\textwidth]{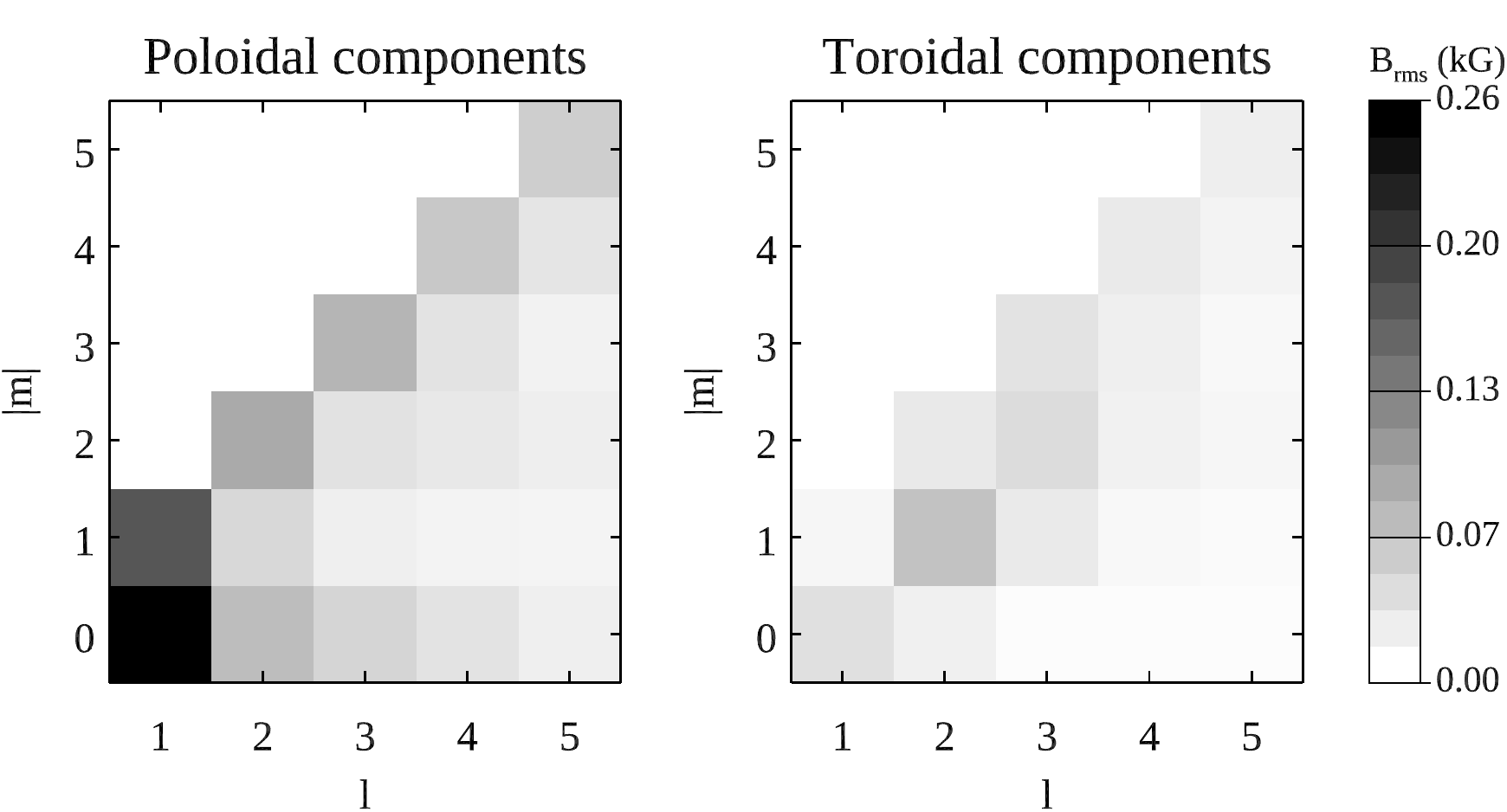}
\hspace*{5mm}
\includegraphics[width=0.47\textwidth]{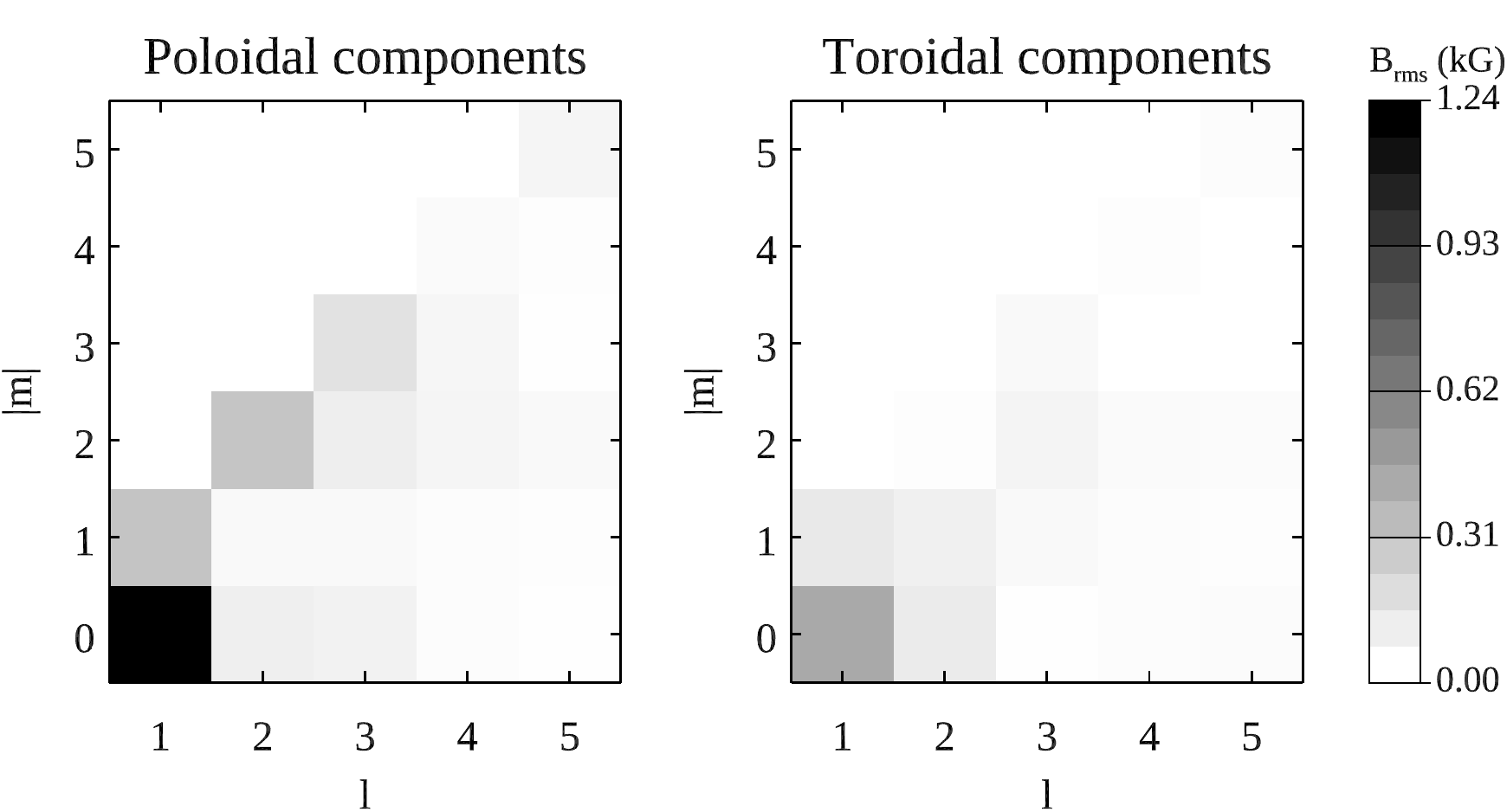}
\caption{Global magnetic field geometries of the rapidly rotating M dwarfs BL~Cet (GJ 65A, a) and UV~Cet (GJ~65B, b) derived with ZDI. The flattened polar projections of the radial, meridional, and azimuthal magnetic field vector components are shown on the left for each star. The thick solid line in these plots indicates the stellar equator. The dotted lines correspond to latitudes $+30^{\rm o}$ and $+60^{\rm o}$. The colour bar gives the field strength in kG. The plots on the right of the maps compare the observed (symbols) and model (solid red lines) Stokes $V$ profiles. The plots below illustrate the rms field strength of individual harmonic modes as a function of $\ell$ and $m$. Adapted from \citet{kochukhov:2017c}.}
\label{fig:zdi_gj65}
\end{figure}

Quantitative interpretation of M-dwarf ZDI results makes use of the expansion of their magnetic field maps in spherical harmonic series \citep{donati:2006b,kochukhov:2014}. This representation of a vector field provides a convenient way of characterising different morphological types and degree of complexity of the global magnetic geometries. For each spherical harmonic mode, defined by an angular degree $\ell$ and azimuthal number $m$, one can, in the most general case, expect contributions from three types of fields. There are (independent) radial and horizontal fields corresponding to the poloidal (potential) harmonic modes and horizontal fields associated with the toroidal (non-potential) harmonic components. The relative contribution of each field type is quantified by the magnetic energy (surface integral of $B^2$) corresponding to this component. In this way, ZDI studies distinguish stars with predominantly poloidal and mostly toroidal fields. Considering the strength of harmonic modes as a function of $\ell$, one can also separate the stars with geometrically simple global fields (most of the magnetic energy is in $\ell\le3$ or even $\ell=1$ modes) from the objects with more complex field geometries (spherical harmonic modes with $\ell>3$ are required to fit observations). Finally, depending on the strength of the low-$m$ ($|m|<\ell/2$) modes relative to the high-$m$ ($|m|\ge\ell/2$) ones, it is possible to identify the stars with predominantly axisymmetric and mostly non-axisymmetric global fields, respectively. 

An example of M-dwarf ZDI analysis results is shown in Fig.~\ref{fig:zdi_gj65}. This figure presents the flattened polar projections of the surface maps of the radial, meridional, and azimuthal magnetic field vector components and the fits of ZDI model spectra to the observed Stokes $V$ profiles for two rapidly rotating late-M dwarfs, GJ~65A (BL~Cet) and GJ~65B (UV~Cet). The distribution of the rms field strength as a function of the $\ell$ and $m$ numbers of spherical harmonic modes is also shown. According to this ZDI analysis, UV~Cet exhibits morphologically simple, strong and weakly variable Stokes $V$ profiles, indicating that its surface field geometry is dominated a positive axisymmetric \cla{dipolar} component. In contrast, BL~Cet shows much weaker and more complex Stokes $V$ signatures (at least judging by the few available observations). This is interpreted by ZDI as a weaker global field with a larger non-axisymmetric contribution compared to UV~Cet. The fields of both stars are predominantly poloidal, with the largest contribution coming from dipolar components.

\subsection{Instrumentation for magnetic field measurements}
\label{sect:instruments}

Measurements of the Zeeman broadening in the optical spectra of M dwarfs can be carried out by any high-resolution spectrometer covering the 500--1000~nm wavelength region. A spectral resolution of $R\ge50000$, and ideally $\sim$\,$10^5$, is necessary to study line profile shapes. A somewhat lower resolution of $\sim$\,30000 can be employed for analyses of the magnetic line intensification. Spectroscopic studies of M dwarfs greatly benefit from redder wavelength coverage due to intrinsic brightness of low-mass stars at near-infrared wavelengths and the $\lambda^2$ dependence of the Zeeman effect. To this end, several spectrographs working in the Y, J, H bands, such as the red arm of CARMENES at the 3.5-m telescope of Calar Alto observatory \citep{quirrenbach:2014}, the IRD instrument at Subaru \citep{kotani:2018}, and the upcoming ESO's NIRPS facility \citep{wildi:2017}, can be employed to provide observations of the well-established Ti~{\sc i} and FeH magnetic diagnostic lines blue-wards of 1~${\mu}$m with the added benefit of covering many atomic and molecular features in the J and H bands suitable for stellar parameter determination and abundance analysis \citep{lindgren:2017,passegger:2019}. Much fewer spectrographs are capable of obtaining high-resolution, broad bandwidth spectroscopic observations in the K band. Currently, such data can be collected at the 3-m class telescopes with iSHELL \citep[$R=75000$,][]{rayner:2016}, IGRINS \citep[$R=40000$,][]{park:2014}, SPIRou \citep[$R=75000$,][]{artigau:2014}, and GIANO \citep[$R=50000$,][]{oliva:2013}. The upgraded CRIRES instrument at the ESO 8-m VLT \citep{dorn:2016}, capable of delivering $R=10^5$ K-band spectra, will \cla{become operational} in 2020.

Spectropolarimetric studies of low-mass stars require highly specialised instrumentation -- a combination of high-resolution, dual-beam spectrometer and a polarimetric unit -- that is not commonly available at many observatories. Furthermore, another critical ingredient is the possibility to carry out monitoring or service mode observations in order to secure time series data with an appropriate rotational and activity cycle coverage. These considerations make ESPaDOnS at the 3.6-m CFHT \citep[$R=68000$,][]{manset:2003} the most efficient optical spectropolarimeter for M-dwarf studies. Its recently refurbished twin instrument NARVAL at the smaller 2-m TBL telescope has also been used for observations of brighter M dwarfs. Another widely used high-resolution spectropolarimeter, HARPSpol at the ESO 3.6-m telescope \citep[$R=110000$,][]{piskunov:2011}, is less suitable for polarisation measurements of M-dwarf stars due to its wavelength cutoff at $\lambda\approx700$~nm. The PEPSI instrument at the dual 8.4-m LBT \citep[$R=130000$,][]{strassmeier:2018} is the only high-resolution optical spectropolarimeter currently operating at a large telescope. It has a potential of making a substantial contribution to studies of low-mass stars despite its limited availability for stellar research and lack of service observing mode. Major advancements in magnetic diagnostics of M dwarfs are also expected from high-resolution polarisation measurements at near-infrared wavelengths. Both SPIRou and the upgraded CRIRES will be capable of providing such data, with the latter instrument being particularly promising for investigations of less active and fainter late-M dwarfs.

\section{Observations of M dwarf magnetic fields}
\label{sect:obs}

\subsection{Total magnetic fields from intensity spectra}
\label{sect:stokesI}

\subsubsection{Results from detailed line profile modelling}
\label{sect:detailedI}

The first direct magnetic field measurement for an M dwarf was presented by \citet{saar:1985}. Their study, based on $R\approx45000$ near-infrared Fourier transform spectrometer observations of GJ~388 (AD~Leo), demonstrated a clear resolved Zeeman splitting in the four Ti~{\sc i} lines at $\lambda$ 2221.1--2231.1~nm. Analysis of the profiles of these $g_{\rm eff}$\,=\,1.2--2.5 lines with a two-component spectrum synthesis model suggested that 73\% of the stellar surface is covered by a 3.8~kG field, corresponding to the average field strength \bi\,=\,2.8~kG\footnote{Although this magnetic field strength is reasonably consistent with later measurements for AD~Leo, more recent high-resolution spectra of the same K-band Ti~{\sc i} lines \citep{kochukhov:2009b} reveal no clear Zeeman splitting of their profiles. Such splitting is also not expected with the multi-component magnetic field strength distribution models derived for this M dwarf from optical lines \citep{shulyak:2017}. It is not known if this discrepancy originates from a systematic problem in the reduction of Fourier transform spectra in the study by \citet{saar:1985} or reflects a real change of the stellar magnetic field.}. Later, \citet{saar:1994a} reported a 2.3~kG field for GJ~803 (AU~Mic) and 3.7~kG field for GJ~873 (EV~Lac), derived using similar near-infrared spectroscopic data and analysis methodology. 

Based on high-quality optical spectra ($R\approx120000$, $S/N\approx200$), \citet{johns-krull:1996} identified significant excess broadening of the magnetically sensitive ($g_{\rm eff}$\,=\,2.5) Fe~{\sc i} $\lambda$ 846.84~nm line in several active M dwarfs compared to inactive stars. The authors measured 2.6--3.8~kG fields for GJ~729 and GJ~873 using a two-component spectral fitting approach. This analysis was revised by \citet{johns-krull:2000}, who fitted the same observations with theoretical spectra incorporating multiple magnetic components ranging from 0 to 9 kG in strength. That study also reported a relatively strong, \bi\,=\,3.3~kG, field for GJ~285 (YZ~CMi). Another magnetic field measurement using the Fe~{\sc i} 846.84~nm line, \bi\,=\,2.5~kG, was reported by \citet{kochukhov:2001} for the early-M dwarf GJ~1049. High-quality K-band $R\approx10^5$ spectra of several M dwarfs recorded with the CRIRES instrument at the ESO VLT were investigated by \citet{kochukhov:2009b}. They obtained new \bi\ estimates for GJ~285, GJ~388, GJ~1049 and measured \bi\,=\,4.3~kG field for GJ~398 (RY~Sex) using multi-component spectrum synthesis modelling of the Na~{\sc i} 2208.4~nm line.

These early studies of M-dwarf magnetic fields relying on a few, often just one, atomic lines were extremely challenging in several respects. The requirements for the resolution and $S/N$ ratio of observed spectra necessary for detecting subtle signatures of the Zeeman broadening were excessive and could be satisfied for only a small number of brightest active M dwarfs. Interpretation of these signatures became ambiguous as soon as \cla{the stellar} $v_{\rm e}\sin i$ exceeded $\approx$\,5~\kms\ and line profile details were washed out by the rotational Doppler broadening. This made it impossible to probe magnetic fields in faster rotating and, presumably, most magnetically active M dwarfs. Moreover, the blending of atomic lines by molecular absorption (TiO for the Fe~{\sc i} 846.84~nm line, H$_2$O for the Ti~{\sc i} lines in the K-band) becomes increasingly severe for later spectral types. This problem prevented analysis of M dwarfs cooler than about M4.5 and required fitting the ratio of active to inactive stellar spectra \cla{for warmer stars} \citep{johns-krull:1996} to partially offset the impact of molecular lines that could not be properly reproduced by theoretical spectra due to inaccurate and incomplete molecular line lists. In these circumstances, another magnetic field diagnostic in the form of the FeH lines in the Wing-Ford band (F$^4\Delta$--X$^4\Delta$ transitions) showed a great promise \citep{valenti:2001,berdyugina:2002}. These lines were known to be reasonably well reproduced by theoretical calculations for inactive stars and showed a significant Zeeman splitting in sunspots \citep{wallace:1999}. Developments in the theory of the molecular Zeeman effect \citep{berdyugina:2003b,asensio-ramos:2006} combined with a semi-empirical adjustment of missing molecular constants through the comparison of calculations with sunspot spectra \citep{afram:2008,shulyak:2010c} enabled practical application of Wing-Ford FeH lines to the problem of low-mass star magnetometry. In the context of FeH magnetic measurements utilising detailed radiative transfer calculations, \citet{afram:2009} reported magnetic fields for ten M dwarfs including objects as cool as M7.5. \citet{shulyak:2010c,shulyak:2011} enhanced this methodology by combining analysis of the FeH lines in the Wing-Ford band with the study of Ti~{\sc i} lines in the 1039.8--1073.5~nm region observed with CRIRES. Further development of this approach was presented by \citet{shulyak:2014}. They applied multi-component magnetic spectral fits to a small number of M3.5--M5.5 dwarfs, exploring feasibility of constraining individual magnetic filling factors and sensitivity of the results to assumptions regarding magnetic field orientation.

A significant breakthrough was achieved by \citet{kochukhov:2017c} and \citet{shulyak:2017}, who recognised particular usefulness of the group of ten Ti~{\sc i} lines at $\lambda$ 964.74--978.77~nm for M-dwarf magnetic field measurements. These strong neutral titanium lines belong to the same multiplet $^5$F--$^5$F$^{\rm o}$ and include a magnetically insensitive ($g_{\rm eff}=0$) line at $\lambda$ 974.36~nm. They are relatively free from molecular blends, even in late-M spectra. These Ti~{\sc i} lines were generally ignored by previous studies, likely due to blending by the telluric absorption. Their moderate Land\'e factors ($g_{\rm eff}\le1.55$) notwithstanding, these lines exhibit a strong response to the photospheric magnetic field not only in terms of the Zeeman distortion of their line profile shapes but also in the form of selective magnetic intensification (see Sect.~\ref{sect:zb}). Taking advantage of the latter, \citet{kochukhov:2017c} measured 5.2--6.7~kG fields in the extremely active M5.5--M6.0 components of the GJ~65 binary (BL Cet and UV Cet), rotating with $v_{\rm e}\sin i\approx30$~\kms\ and $P_{\rm rot}\approx0.2$~d. \citet{shulyak:2017} combined analysis of these Ti~{\sc i} lines with modelling of FeH lines, deriving magnetic fields for 20 M-dwarf stars. Their sample included several  rapid rotators. They also discovered a remarkable 7.3~kG field in the M6.0 star GJ~412B (WX~UMa) and demonstrated that the incidence of the strongest magnetic fields correlates with the properties of global magnetic field topologies obtained with ZDI (Sect.~\ref{sect:zdiresults}). \citet{shulyak:2019} have added further 29 field strength measurements obtained with the same approach. \citet{kochukhov:2019a} determined magnetic fields for both components of the short-period early-M eclipsing binary YY~Gem, thereby providing a key observational constraint for magnetohydrostatic stellar interior models of M dwarfs \citep[e.g.][]{feiden:2013,macdonald:2014}.

\begin{figure}[!t]
\centering
\includegraphics[width=\textwidth]{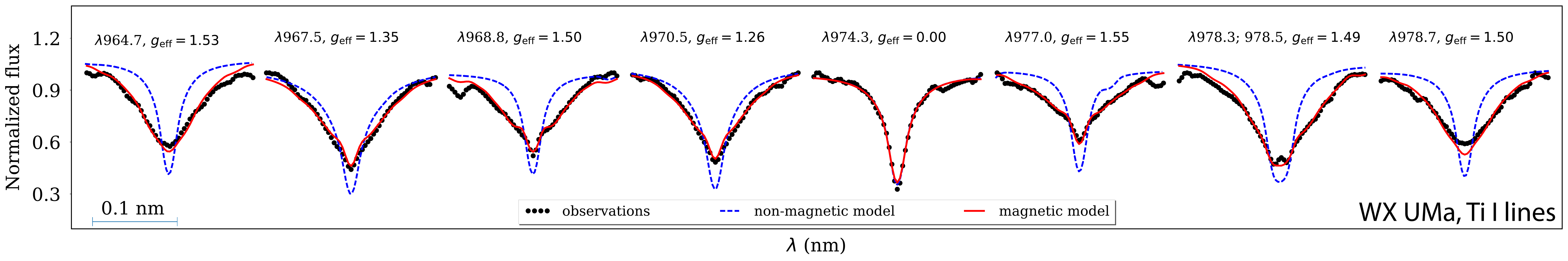}
\includegraphics[width=\textwidth]{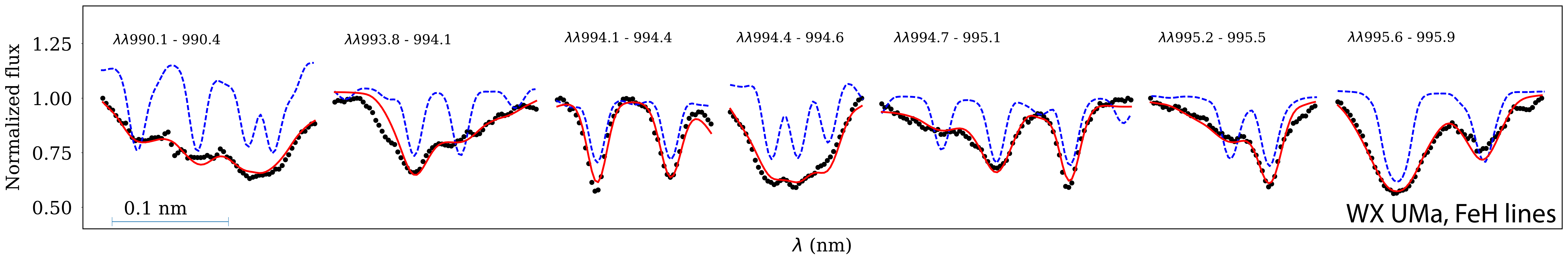}
\includegraphics[width=0.80\textwidth]{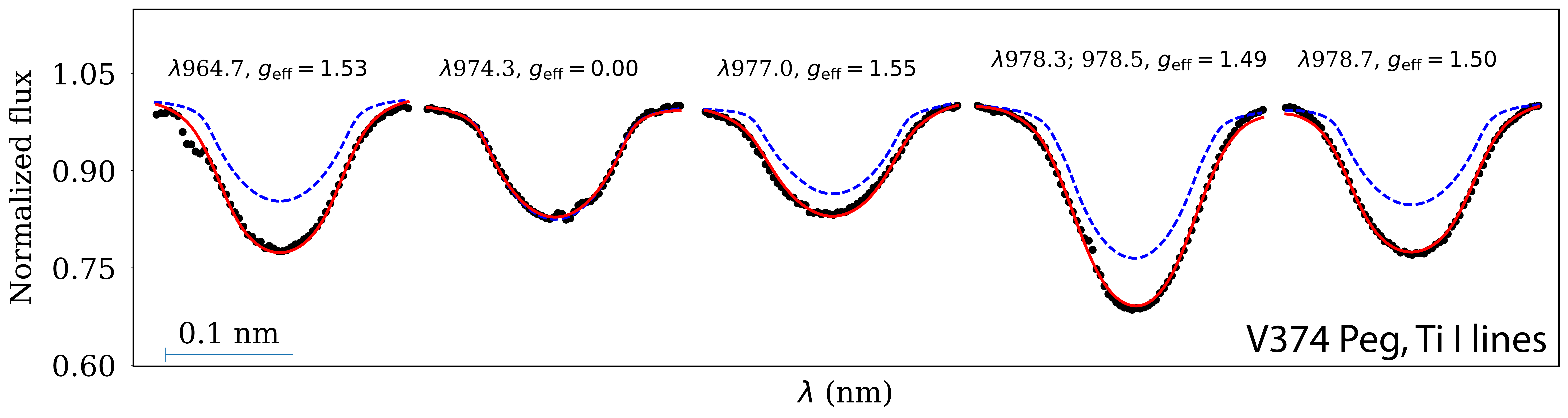}
\caption{Model fits to the Ti~{\sc i} and FeH line profiles in the spectra of active M dwarfs WX~UMa (top and middle panels) and V374~Peg (bottom panel). Black symbols correspond to observations, red solid line to best fit theoretical model spectrum, blue dashed line to spectrum computed assuming zero magnetic field. 
Image reproduced with permission from \citet{shulyak:2017}, copyright by Macmillan.}
 %% Credit: \citeauthor{shulyak:2017}, 2017, Nature Astronomy, 1, 184.}
\label{fig:fit_example}
\end{figure}

\begin{figure}[!t]
\centering
\includegraphics[width=0.45\textwidth]{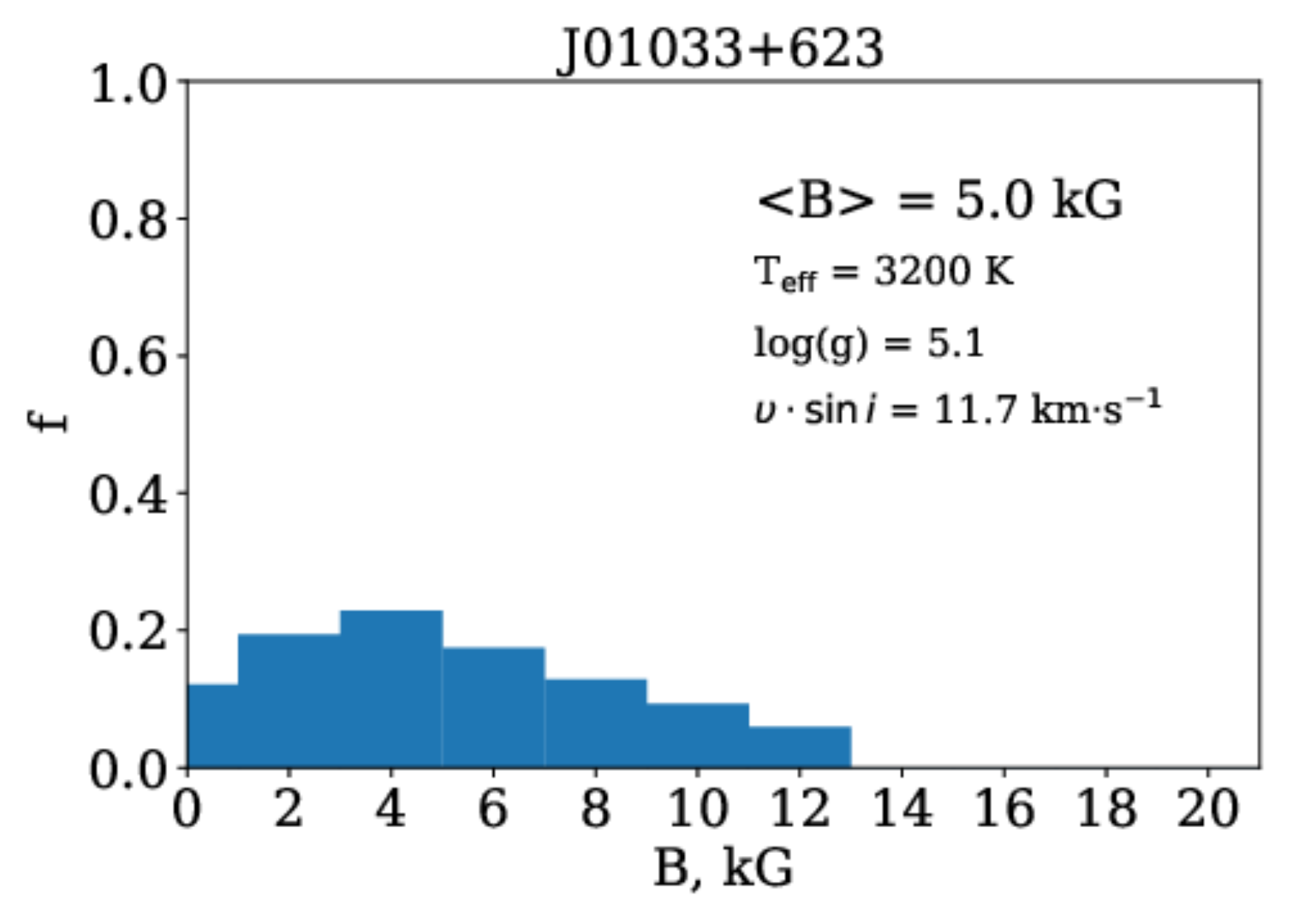}
\includegraphics[width=0.45\textwidth]{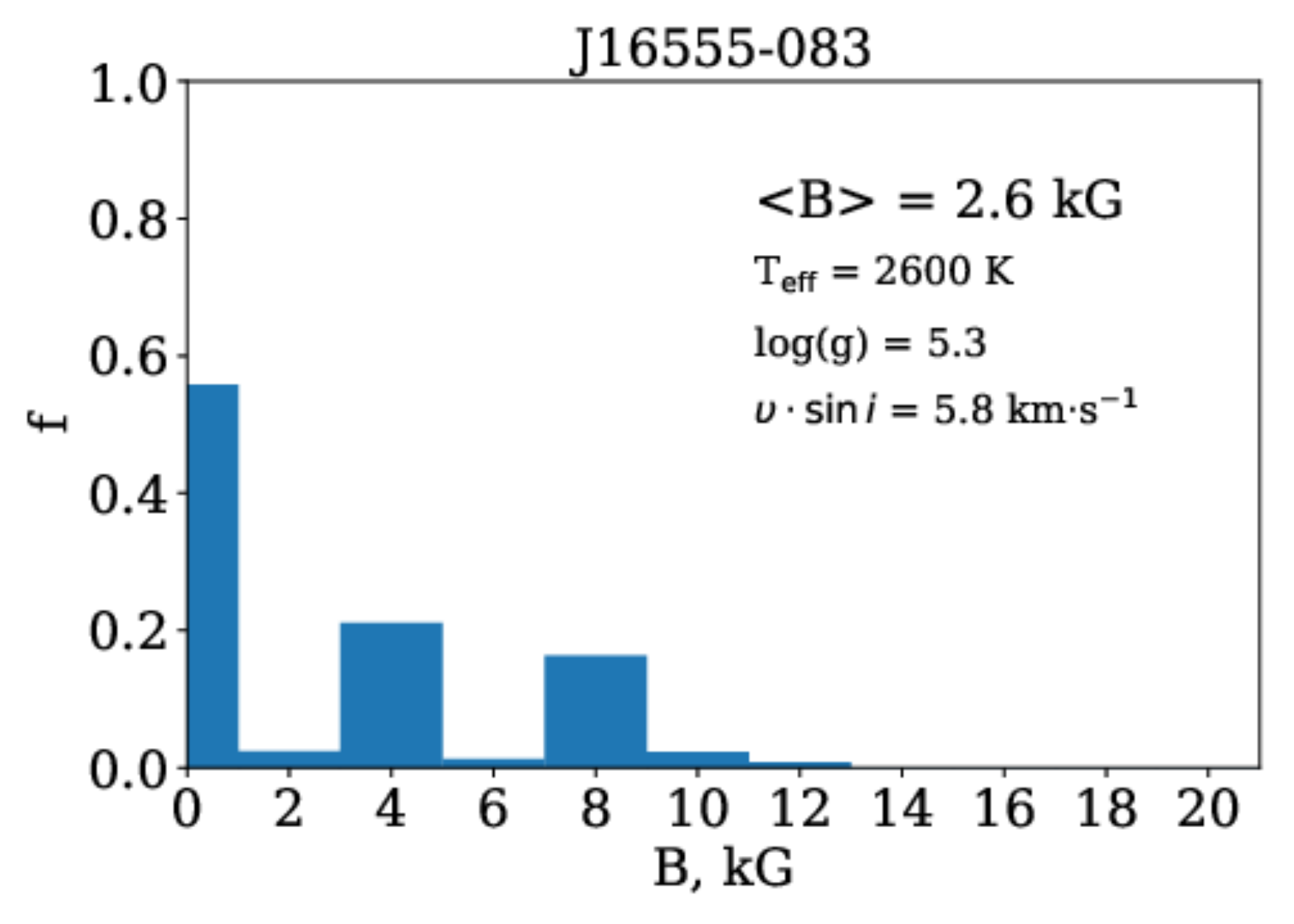}
\caption{Magnetic field strength distributions obtained with the spectrum synthesis analysis of Ti~{\sc i} lines in the spectra of M dwarfs GJ~51 (left) and GJ~644C (VB~8, right). 
Image reproduced with permission from \citet{shulyak:2019}, copyright by ESO.}
%% Credit: \citeauthor{shulyak:2019}, 2019, A\&A, 626, A86.}
\label{fig:bf_example}
\end{figure}

This series of investigations relied on $R=65000$--85000, $S/N\ge100$ ESPaDOnS and CARMENES spectra, illustrating that improvements of the modelling methodology have allowed one to relax the requirements on the quality of observational material compared to the single-line studies by \citet{johns-krull:1996} and \citet{kochukhov:2001}. Fig.~\ref{fig:fit_example} illustrates typical theoretical spectral fits to the Ti~{\sc i} and FeH lines from the study by \citet{shulyak:2017}. For slower rotating targets, in this case WX~UMa, both atomic and molecular diagnostics can be successfully reproduced with a model incorporating a wide distribution of magnetic field strengths, similar to what is illustrated in Fig.~\ref{fig:bf_example}. For such targets both the Zeeman broadening and magnetic intensification provide useful information about magnetic field parameters. For faster rotators, represented in Fig.~\ref{fig:fit_example} by GJ~4247 (V374~Peg), FeH lines are blended together and become more difficult to interpret. However, the intensification of the magnetically sensitive Ti~{\sc i} lines relative to the $\lambda$ 974.36~nm feature still unambiguously shows the presence of a 5.3~kG field.

As discussed by \citet{shulyak:2019}, the inferred fractional distributions of magnetic field strengths on M dwarfs are diverse and show no obvious correlation with stellar parameters or other magnetic field characteristics. There are examples of smooth distributions (left panel in Fig.~\ref{fig:bf_example}) as well as cases where a few magnetic components dominate (right panel in Fig.~\ref{fig:bf_example}). In the latter situation one often finds a significant field-free component. The field strengths exceeding 10~kG are often required to produce satisfactory fits to observations, although the reality of $>$\,15~kG components has been questioned \citep{shulyak:2017,shulyak:2019}. 

All studies described above derived average magnetic field strengths from snapshot spectra or observations averaged over rotation period. With the exception of GJ~65B \citep[UV Cet,][]{kochukhov:2017c}, no observational evidence of a non-uniform surface distribution of magnetic field diagnosed from Stokes $I$ was found. Consequently, these studies assumed that each magnetic field component is represented by a spatially uniform, depth-indepedent surface distribution of, typically radial, magnetic field.

An important limitation of these Zeeman analyses is that they employed model atmospheres from standard non-magnetic grids, such as {\sc Phoenix} \citep{hauschildt:1999} and {\sc Marcs} \citep{gustafsson:2008}, for radiative transfer modelling of atomic and molecular lines and ignored possible difference of thermodynamic structure of the surface regions with different magnetic field intensities. This assumption may be motivated by the observation of a general decline of the spot-to-photosphere temperature contrast with spectral type, from $>1500$~K for G dwarfs to a few hundred K for M dwarfs \citep{berdyugina:2005}, and the lack of high-contrast temperature inhomogeneities in the local magnetohydrodynamical simulations of M-dwarf atmospheres \citep{beeck:2015,beeck:2015a}. However, this view was challenged by \citet{afram:2019}, who modelled Fe and Ti atomic lines as well as TiO, FeH, MgH, and CaH molecular features in the spectra of nine M0--M7 stars using a two-component atmospheric approach. Their method allowed for different temperatures of the non-magnetic photosphere and magnetised regions (associated with ``starspots'' or ``magnetic network'', depending on the spectral line considered). These authors claimed extreme surface temperature contrasts, from 1700~K for early M dwarfs to about 1000 K for an M7 star. Unlike \citet{shulyak:2017,shulyak:2019}, who inferred mostly consistent average field strengths from atomic and molecular lines, \citet{afram:2019} required significantly different \bi\ values to fit different diagnostic features. These discrepancies were attributed to a complex height dependence of local magnetic field parameters. Results by \citet{afram:2019}, particularly the reality of a huge temperature difference between magnetic and non-magnetic regions, need to be independently verified in order to confirm reliability of their field strength measurements for different types of surface structures on the surfaces of active M dwarfs.

All measurements of the average magnetic field obtained for M dwarfs with the help of the spectral fitting techniques reviewed above are summarised in Table~\ref{tab:stokesI}. The results by \citet{shulyak:2010c} and \citet{afram:2019} are not included since those studies did not report specific average field strength values for individual stars. This compilation, complete as of the beginning of 2020, gives the common name of the star, the average field strength \bi\ with an uncertainty estimate whenever available, and the magnetic field distribution parameterisation adopted in the spectrum synthesis analysis. According to the recent detailed studies of this kind \citep[e.g.][]{shulyak:2017,shulyak:2019}, multi-component field distributions (method 2 and 3 in Table~\ref{tab:stokesI}) are compulsory for obtaining satisfactory fits of high $S/N$ ratio spectra of narrow-line active M dwarfs. Therefore, \bi\ measurements derived by some earlier studies \citep{afram:2009,shulyak:2011}, which relied on fitting observed spectra with theoretical calculations for a single field strength value only (method 1 in Table~\ref{tab:stokesI}), probably have a lower quality and might be affected by a systematic bias.

In total, 94 average magnetic field measurements are available for 62 stars, including binary components of YY~Gem. The field strengths resulting from multi-component analyses range from 0.8 to 7.3~kG, with WX~UMa (GJ~412B) possessing the strongest currently known average field for an M dwarf or any other late-type active star \citep{shulyak:2017}. So far, 17 stars were reported to have field strengths above the 4~kG threshold, which was previously thought to represent a saturation limit for the dynamo field intensity in rapidly rotating M dwarfs \citep{reiners:2009a}. Of these, nine stars have average fields above 5~kG and four above 6~kG.

For the four stars in Table~\ref{tab:stokesI} (GJ~285, GJ~388, GJ~729, GJ~873) five to seven average field strength measurements, spread over the period of up to $\approx$\,30 yr, are available. The standard deviation of \bi\ determinations reported for GJ~285 (YZ~CMi) is 0.67~kG, which is 2--3 times larger than the error bars of individual measurements. It is therefore possible that the average field intensity of this star varies by about 20\%. For the other three stars the standard deviation of independent \bi\ measurements is 0.21--0.26~kG, which is comparable to, or even lower than, the quoted uncertainties. Thus, repeated total field strength measurements from the Stokes $I$ spectra do not readily reveal a significant secular magnetic variability of M dwarfs.

\subsubsection{Approximate measurements of average magnetic fields}
\label{sect:approxI}

Determination of average magnetic fields and field strength distributions with the detailed spectrum synthesis modelling methods described in the previous section is time consuming and demanding in terms of the quality of observational material and robustness of theoretical spectra and model atmospheres. An approximate technique of magnetic field detection and coarse average field strength measurement without reliance on theoretical modelling is potentially very useful for analyses of large stellar samples and as a method to identify interesting targets for in-depth studies. Two such methods utilising lines in the Wing-Ford band of FeH at $\lambda$ 988--998~nm have been suggested. 

\citet{valenti:2001} showed that the Wing-Ford band spectrum of the active M dwarf AD Leo can \cla{be} well represented by a weighted superposition of the umbral solar contribution and the spectrum of an inactive M dwarf, allowing one estimate the field strength from the relative weights of the two components. This idea was further developed by \citet{reiners:2006a}. In their semi-empirical method the FeH spectrum of EV Lac was adopted as an active-star template with \bi\,=\,3.9~kG \citep{johns-krull:1996} and the spectrum of GJ~1227 was taken to be a zero-field reference. An average field of any other target was then derived by fitting its magnetically sensitive and insensitive FeH lines with a linear interpolation between these two templates, taking into account rotational Doppler broadening and scaling the line depth to track variation of FeH line intensities with temperature. \citet{reiners:2006a} argued that this methodology can yield a measurement of magnetic field for low-mass stars with \vs\ as large as 30~\kms. Subsequently, \citet{reiners:2007,reiners:2008,reiners:2010} and \citet{reiners:2009a} applied this approximate field strength measurement procedure to about 70 M0--M9 stars, mostly based on Keck spectra with the resolution of $R\approx30000$. These results were summarised by \citet{reiners:2012}. The average field strength for this sample is $1.9\pm0.8$~kG. Occasionally, field strengths exceeding the upper 3.9~kG limit set by the choice of the active M-dwarf template as well as fields below 1~kG were reported. The latter group of stars included the key rocky planet-hosting M dwarfs Proxima Cen and TRAPPIST-1, both of which were found to have 600~G mean fields \citep{reiners:2008,reiners:2010}.

Several error sources contribute to the uncertainties of this FeH template matching method. First, the choice of EV~Lac as an active M-dwarf template incurs a calibration error of $\sim$\,0.5~kG corresponding to the scatter of different independent field strength determinations (see Table~\ref{tab:stokesI}). Furthermore, this star possesses an unusual non-axisymmetric global magnetic field configuration (see Sect.~\ref{sect:zdiresults}), hinting that some of this scatter might be caused by the rotational modulation. Second, the methodology employed by Reiners and Basri relies on fairly crude assumptions that the spectral response induced by increasing magnetic field strength is linear with respect to \bi\ and that all targets share the same field strength distributions, the latter directly contradicting results of detailed spectrum synthesis modelling discussed in Sect.~\ref{sect:detailedI}. Considering some of these caveats, \citet{reiners:2006a} have initially estimated uncertainty of their method at a modest $\sim$\,1~kG level. However, later studies quoted an uncertainty of a few hundred G \citep[e.g.][]{reiners:2008,reiners:2012}.

\begin{figure}[!t]
\includegraphics[width=\textwidth]{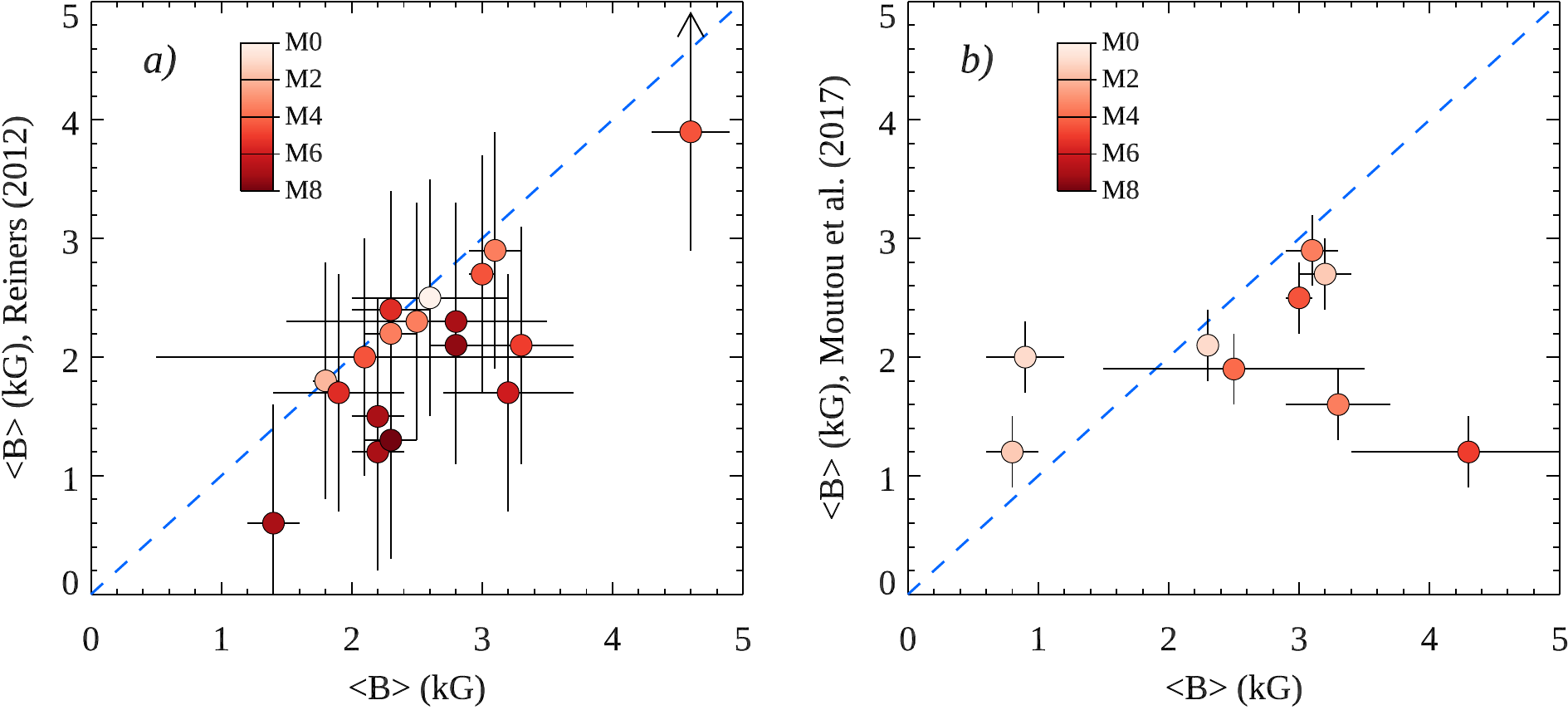}
\caption{Comparison between the average magnetic fields found with detailed spectrum synthesis modelling (see Table~\ref{tab:stokesI}) and approximate field strengths obtained by (a) interpolating between the FeH template spectra of active and inactive stars \citep[][and references therein]{reiners:2012} and (b) interpreting relative line widths of the FeH lines with different magnetic senstivity \citep{moutou:2017}. The symbol colour corresponds to the spectral type, as indicated by the colour bars.}
\label{fig:bi_approx}
\end{figure}

We can empirically assess the precision of Reiners and Basri's method by comparing their coarse \bi\ determinations with the results of detailed spectrum synthesis studies reported in Table~\ref{tab:stokesI}. A comparison of the field strengths found with both approaches is presented in Fig.~\ref{fig:bi_approx}a for 17 stars. It is evident that for some stars approximate \bi\ values agree well with the spectrum synthesis results, particularly for early M dwarfs. But for other stars, especially dwarfs with later spectral types, the field strengths determined by Reiners and Basri are underestimated by up to 1~kG. The mean absolute difference is 0.5~kG considering all the stars, but it increases to 0.8~kG for dwarfs with spectral types M5.5 and later. Judging from this comparison, the average fields of late-M dwarfs Proxima Cen and TRAPPIST-1 could be about a factor of two stronger than reported by Reiners and Basri.

Another version of an express field strength estimation procedure, also utilising the FeH lines from the Wing-Ford band, was proposed by \citet{moutou:2017}. They made use of only two FeH lines with different magnetic sensitivity and deduced field strengths assuming a linear relation between \bi\ and the line width difference of these two absorption features. Using this technique, \citet{moutou:2017} derived 0.4--2.9~kG fields for 136 M dwarfs based on $R\approx65000$ observations obtained with ESPaDOnS. The mean and scatter of their field strengths ($1.5\pm0.5$~kG) are both noticeably smaller than for the sample investigated by Reiners and Basri. A comparison of nine magnetic field measurements from \citet{moutou:2017} with the line profile fitting results from Table~\ref{tab:stokesI} is shown in Fig.~\ref{fig:bi_approx}b. No correlation is evident, with approximate \bi\ values offset from the literature values by 0.9~kG on average and up to 3~kG in extreme cases. This assessment suggests that the coarse field strength measurements by \citet{moutou:2017} are significantly more uncertain than other average field strengths found in the literature.

\subsubsection{Magnetic field and stellar rotation}
\label{sect:rotat}

Indirect magnetic activity proxies, such as chromospheric and coronal emission \cla{and photospheric absorption line strengths}, are known to exhibit a prominent dependence on stellar rotation \citep[e.g.][]{reiners:2014,newton:2017,wright:2011,wright:2018,muirhead:2020}. This dependence is best represented as a function of Rossby number (Ro), which is defined as the ratio of stellar rotation period $P_{\rm rot}$ and convective turnover time-scale $\tau$. The typical behaviour found for cool active stars, including M dwarfs, is an increase of activity proxies and magnetic field strength with decreasing Rossby number until a saturation  at Ro\,$\approx$\,0.1 \citep[e.g.][]{vidotto:2014,wright:2018,kochukhov:2020}.

To investigate a relationship between the average magnetic field strength and stellar rotation, we compiled in Table~\ref{tab:params} rotation periods for the M dwarfs with direct magnetic field measurements. This table also lists 2MASS cross-identifications and spectral types. For the majority of stars, $P_{\rm rot}$ values were taken from the same studies that measured magnetic fields. Additionally, for several targets $P_{\rm rot}$ was derived from ground-based photometry \citep{newton:2016} and Kepler light curves \citep{doyle:2018}. For six more stars (marked with asterisks in Table~\ref{tab:params}) new rotation periods could be determined from public 2-min cadence light curves obtained by the TESS satellite \citep{ricker:2015}. Eleven stars still lack any $P_{\rm rot}$ information. For these objects we derived upper limits of rotation periods from \vs\ reported by spectroscopic studies and stellar radii obtained from the spectral type-radius relation \citep{pecaut:2013}.

\begin{figure}[!t]
\includegraphics[width=\textwidth]{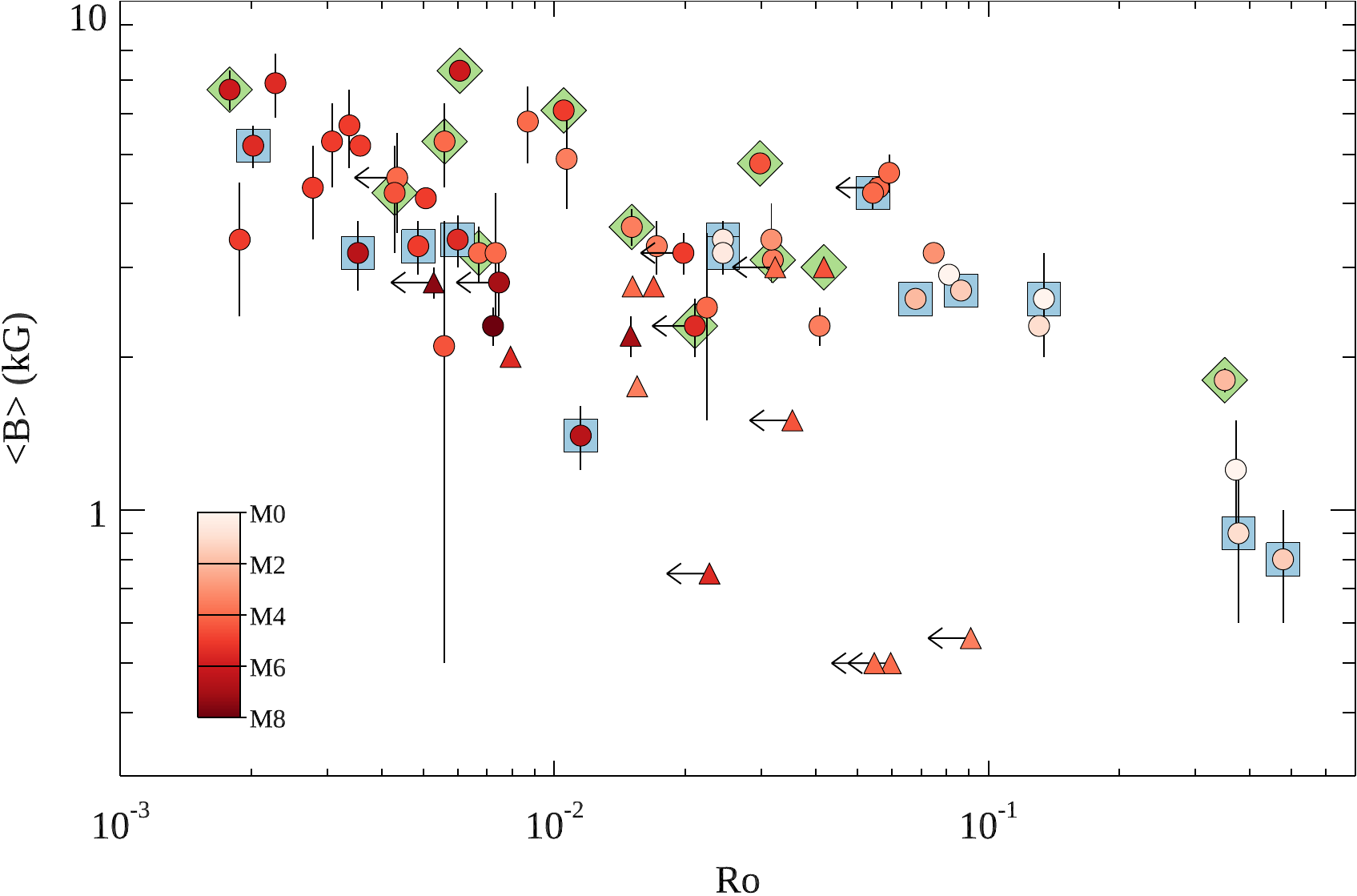}
\caption{Average magnetic field strengths obtained with the radiative transfer modelling of intensity spectra as a function of Rossby number. Circles represent the average field strengths determined with a multi-component approach; triangles correspond to less reliable results obtained by fitting a single field strength value. The symbol colour corresponds to the spectral type, as indicated by the colour bar. For stars studied with ZDI, the background symbols indicate the type of global magnetic field topology: green rhombs for stars with predominantly dipolar, axisymmetric fields and blue squares for stars with multipolar and/or non-axisymmetric dipolar fields.}
\label{fig:bi_summary}
\end{figure}

The average magnetic field for the sample of M dwarfs listed in Table~\ref{tab:stokesI} is shown as a function of Ro in Fig.~\ref{fig:bi_summary}. The rotation periods were taken from Table~\ref{tab:params} while the convective turnover times were calculated by determining the stellar mass from spectral type \citep{pecaut:2013} and applying the $\tau$ vs. mass calibration by \citet{wright:2011}. Fig.~\ref{fig:bi_summary} demonstrates that \bi\ initially increases with decreasing Ro, similar to indirect magnetic activity measures. This trend is observed in the unsaturated activity branch, which is traced mainly by early-M dwarfs. A number of magnetic measurements obtained by fitting stellar spectra with single magnetic field strength values (triangles in Fig.~\ref{fig:bi_summary}), originating primarily from the study by \citet{shulyak:2011}, fall significantly below the general relation. These results may be unreliable.

The scatter of \bi\ measurements in Fig.~\ref{fig:bi_summary} prevents determination of an accurate saturation threshold. Nevertheless, it appears that the field strength increases systematically until Ro drops to at least 2--3\,$\times10^{-2}$. A slower increase of \bi\ with decreasing rotation period cannot be excluded at even smaller Ro values. Considering 24 stars with Ro\,$\le10^{-2}$, the observed magnetic field strength ranges from 2.0 to 7.3 kG and is equal to $4.3\pm1.5$ kG on average. This group of rapidly rotating M dwarfs represents the most magnetised late-type stars currently known.

Additional symbols in Fig.~\ref{fig:bi_summary} illustrate relation between the total magnetic field strengths measured from Stokes $I$ and the properties of global field topologies studied for a smaller number of M dwarfs with polarimetry (see Sect.~\ref{sect:stokesV}). This comparison shows, in agreement with the results by \citet{shulyak:2017}, that the strongest average fields are typically found in stars with predominantly dipolar, axisymmetric large-scale magnetic geometries (shown with green rhombs in Fig.~\ref{fig:bi_summary}) rather than in M dwarfs with more complex multipolar or/and non-axisymmetric global fields (blue squares in Fig.~\ref{fig:bi_summary}). The star with the strongest average field belonging to the latter type is GJ~65A (BL~Cet) with \bi\,=\,5.2~kG. On the other hand, at least four stars with global fields of the former type (GJ~51, GJ~65B, GJ~412B, GJ~4247) possess 5.3--7.3~kG fields.

The distribution of measurements in Fig.~\ref{fig:bi_summary} suggests a substantial scatter of \bi\ for M dwarfs with similar Ro. This scatter is significantly larger than typical uncertainties of the average field strength measurements and is most certainly real. Comparison of GJ~51 (M5.0, Ro\,=\,0.011, \bi\,=\,6.1~kG) and GJ~3622 (M6.5, Ro\,=\,0.012, \bi\,=\,1.4~kG) provides one of the more extreme examples of this scatter. It is therefore very likely that parameters other than the stellar mass and rotation \cla{period} influence dynamo efficiency in M-dwarf stars.

\subsection{Large-scale magnetic fields from spectropolarimetry}
\label{sect:stokesV}

\subsubsection{Polarisation in M-dwarf spectra}
\label{sect:polaris}

Active M dwarfs exhibit higher amplitude circular polarisation signals in spectral lines compared to more massive late-type active stars. This suggests the presence of intense global magnetic fields. The first detection of Zeeman circular polarisation in the optical spectrum of an M dwarf was reported by \citet{donati:2006} for the rapidly rotating M4 star GJ~4247 (V374~Peg) using then newly commissioned ESPaDOnS spectropolarimeter. These authors applied the least-squares deconvolution technique (see Sect.~\ref{sect:lsd}) to about 5000 metal lines, obtaining repeated detection of a morphologically simple, rotationally-modulated Stokes $V$ profile with an amplitude of $\pm$\,0.3\% relative to the unpolarised continuum. Subsequent observations with ESPaDOnS and its twin instrument Narval \citep{donati:2008,morin:2008a,morin:2008,morin:2010} showed that LSD Stokes $V$ signals with an amplitude of up to $\sim$\,1\% are common for those M dwarfs in which multi-kG average magnetic fields were previously found by the Stokes $I$ profile analyses. For the brightest of those stars \cla{circular} polarisation signatures can be observed even in individual spectral lines \citep[e.g.][]{lavail:2018}.

\begin{figure}[!t]
\includegraphics[width=\textwidth]{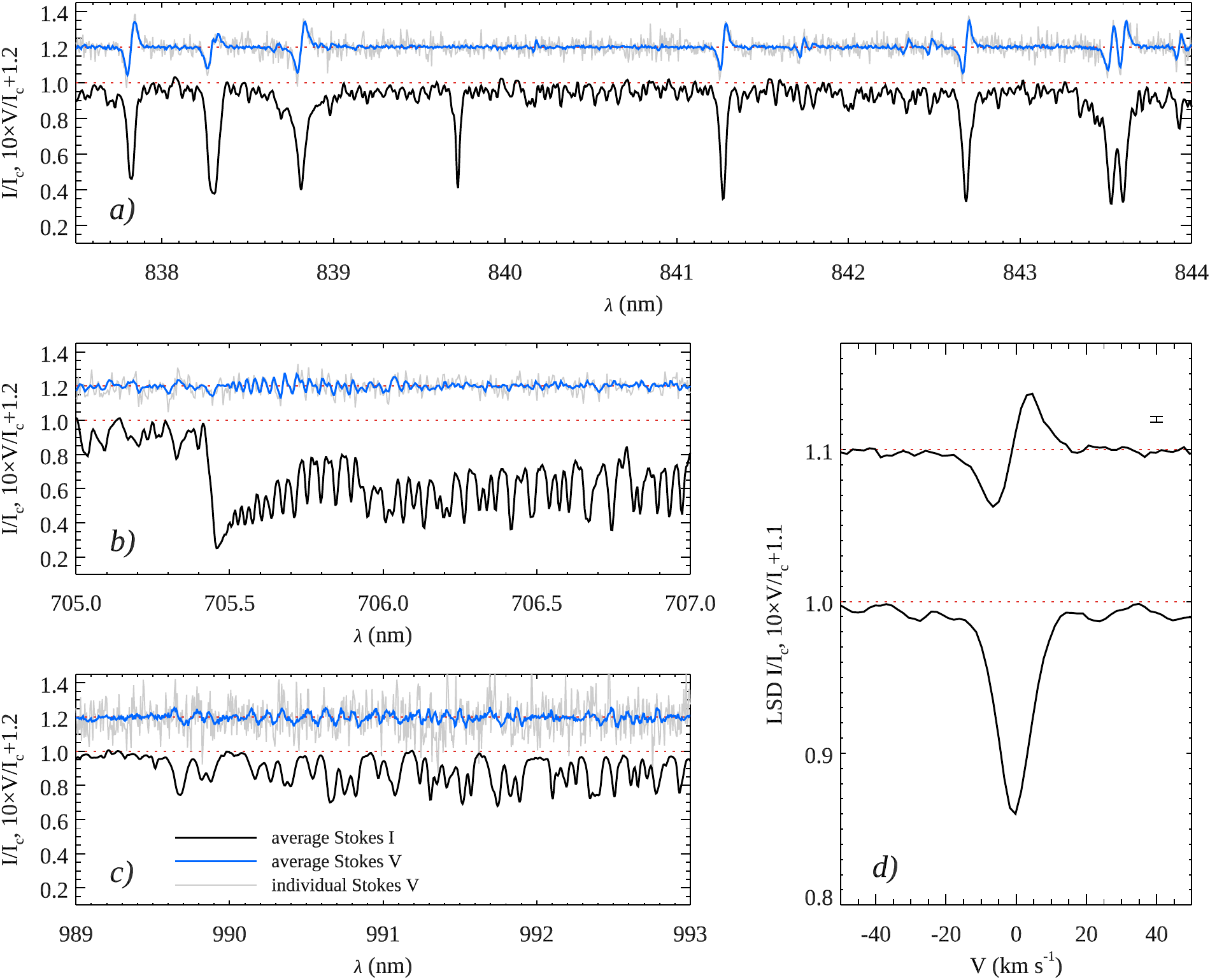}
\caption{Circular polarisation signatures in the spectrum of bright, active M dwarf AD Leo (GJ 388). (a) Fe~{\sc i} and Ti~{\sc i} lines in the 837--844~nm wavelength interval, (b) the TiO 705.5 nm bandhead, (c) the Wing-Ford FeH band at $\lambda\approx990$~nm. In each of these plots the black line shows the Stokes $I$ spectrum, the thick blue line corresponds to the average Stokes $V$ spectrum ($S/N\approx1800$) and the thin grey line illustrates typical high-quality individual Stokes $V$ observation ($S/N\approx270$). The circular polarisation spectra are shifted vertically and amplified by a factor of 10 relative to Stokes $I$. Panel (d) presents the LSD Stokes $I$ and $V$ profiles ($S/N\approx5000$) derived from atomic lines in the same individual observation as shown with the grey curve in panels (a)--(c).}
\label{fig:adleo_stokesv}
\end{figure}

Several examples of Zeeman Stokes $V$ profiles for the active M dwarf GJ~388 (AD~Leo) are illustrated in Fig.~\ref{fig:adleo_stokesv}. This star has a simple, nearly axisymmetric global magnetic field topology. Consequently, its Stokes $V$ profiles exhibit very little rotational modulation. This allows one to co-add many individual observations of AD~Leo, yielding a high $S/N$ Stokes $V$ spectral atlas. Figs.~\ref{fig:adleo_stokesv}a--c show several wavelength regions in this average Stokes $V$ spectrum in comparison to typical high-quality individual circular polarisation observation. The characteristic S-shaped circular polarisation profiles are seen in many atomic lines for both the average and individual Stokes $V$ spectrum. In the latter case, polarisation in individual lines reaches $\pm$1.5\% and is detected at, typically, 3--4$\sigma$ significance level. Circular polarisation signals can be also recognised in the average Stokes $V$ spectra of the most magnetically sensitive optical molecular features -- the TiO bandhead at $\lambda\approx 705.5$~nm (Fig.~\ref{fig:adleo_stokesv}b) and the Wing-Ford FeH lines (Fig.~\ref{fig:adleo_stokesv}c). However, molecular Zeeman signals are noticeably weaker than those observed in atomic lines and can hardly be seen in typical individual spectropolarimetric observations of M dwarfs. Fig.~\ref{fig:adleo_stokesv}d also displays the LSD Stokes $I$ and $V$ profiles derived from the same individual AD~Leo observation as shown in Figs.~\ref{fig:adleo_stokesv}a--c. In this particular case, the LSD circular polarisation signature exceeds the noise level by a factor of $\approx 20$, making this profile suitable for detailed modelling with ZDI.

So far, spectropolarimetry of active M dwarfs was limited to Stokes $V$ observations. An attempt to detect Stokes $QU$ signatures in the optical spectrum of AD~Leo was made by \citet{lavail:2018}. They succeeded in achieving a definite detection of linear polarisation signals with LSD in less than half of their very high $S/N$ ESPaDOnS observations. According to that study, the Stokes $QU$ amplitudes in spectral lines are, on average, 13 times smaller than the Stokes $V$ signals observed on the same nights. This is a somewhat larger Stokes $V$ to $QU$ amplitude ratio compared to the factor of 5--10 reported for other late-type stars studied with high-resolution linear spectropolarimetry \citep{kochukhov:2011,rosen:2013}. Due to this order of magnitude drop in polarisation signal amplitude from Stokes $V$ to Stokes $QU$, full four Stokes parameter investigations of active M dwarfs require considerable observing time investment and are unlikely to be feasible except for a few brightest stars.

The mean longitudinal magnetic field, \bz, can be readily determined from the first moment of the Stokes $V$ profile, as described in Sect.~\ref{sect:lsd}. This magnetic observable characterises the line-of-sight magnetic field component integrated over the stellar disk. This measurement can be employed to determine approximate strength of magnetic field at the stellar surface without performing detailed polarisation profile modelling. Time-resolved \bz\ measurements were reported for 26 active M dwarfs by \citet{donati:2008}, \citet{morin:2008,morin:2010}, \citet{phan-bao:2009}, \citet{kochukhov:2017c}. These studies found typical maximum \bz\ values in the 200--700~G range. The stars GJ~51 and GJ~412B (WX~UMa) show exceptionally strong circular polarisation signatures, corresponding to $\langle B_{\rm z} \rangle_{\rm max} = 2.1$--2.4~kG \citep{morin:2010}. The longitudinal magnetic fields of moderately active and inactive M dwarfs were assessed in the time-resolved study of five stars by \citet{hebrard:2016} and in the snapshot survey of about 100 targets by \citet{moutou:2017}. These authors measured \bz\ below 100~G for the majority of their targets. According to these results, the median maximum \bz\ of normal M dwarfs is about 30~G.

The maximum observed longitudinal magnetic field can be converted to a lower limit of the surface dipolar magnetic field strength, $B_{\rm d}$. For the case when a dipolar geometry is observed pole-on, we have $B_{\rm d}\approx 3 \langle B_{\rm z} \rangle_{\rm max}$. This indicates that, assuming purely dipolar field topologies, $B_{\rm d}\approx100$~G for inactive M dwarfs and up to several kG for active stars.

\subsubsection{Zeeman Doppler imaging results}
\label{sect:zdiresults}

Reconstruction of vector maps of the surface magnetic field with a tomographic imaging technique represents an ultimate method of interpreting time-series Stokes $V$ profile observations of active stars. The first ZDI study of an M dwarf was carried out by \citet{donati:2006}. They reconstructed the global magnetic field of GJ~4247 (V374 Peg, M4), demonstrating that this star hosts a nearly axisymmetric field geometry, dominated by a poloidal dipolar component, with the maximum local field strength of 2~kG. A follow-up study by \citet{morin:2008a} concluded that the global field of V374~Peg is stable on the time-scale of one year. The ZDI modelling by \citet{donati:2006} and \citet{morin:2008a} relied on the weak-field approximation of the Stokes $V$ profiles and included a reconstruction of the continuum brightness inhomogeneities from the Stokes $I$ LSD spectra alongside magnetic mapping. It was found that star spots on the surface of V374~Peg have a low contrast and a small fractional area coverage compared to more massive late-type stars with similar rotation rates. No connection between the spot distribution and global magnetic field topology was found.

\citet{donati:2008}, \citet{morin:2008} and \citet{phan-bao:2009} presented further ZDI studies of six early- and seven mid-M rapidly rotating dwarfs. These studies indicated that the global magnetic field topologies of low-mass stars undergo a gradual transition from relatively weak, complex, often non-axisymmetric configurations with a significant toroidal contribution in early-M dwarfs to much stronger, predominantly axisymmetric, poloidal fields of the type previously found for V374~Peg in mid-M dwarfs. The star GJ~873 (EV~Lac) appears to be an exception within the group of M3--M4.5 stars studied by \citet{morin:2008} due to its strong, poloidal, yet decidely non-axisymmetric large-scale field. The emergence of strong axisymmetric fields in mid-M dwarfs occurs close to the critical mass of $\approx0.35M_\odot$ where the stellar interior becomes fully convective \citep{chabrier:1997}.

The strength of magnetic fields in mid-M dwarfs reconstructed with ZDI reaches 0.8--3~kG locally according to the published maps, but does not exceed 0.8~kG when averaged over the entire stellar surface. It is thus clear that ZDI analyses of Stokes $V$ spectra miss significant part of the total 3--4~kG fields measured for the same stars from unpolarised spectra. Nevertheless, the Stokes $V$ LSD spectra of sharp-line M dwarfs GJ~285 (YZ~CMi), GJ~388 (AD~Leo), GJ~873 (EV~Lac), and GJ~896A (EQ~Peg~A) still reveal the presence of these tangled local fields through an excessive broadening of the circular polarisation profiles (see Fig.~\ref{fig:v_models}). This required \citet{morin:2008} to incorporate the large-scale field filling factor, $f_V$, as a free parameter in their modelling based on the Unno-Rachkovsky analytical solution of the polarised radiative transfer equation. $f_V$ values of 0.10--0.14 were deduced for the four M dwarfs mentioned above, meaning that the actual global magnetic field distributions suggested for these stars are not the smooth magnetic flux ($B_V f_V$) maps published by \citet{morin:2008}, but a highly intermittent, unresolved system of localised spots, similar to the one illustrated in Fig.~\ref{fig:v_models}b. This also implies that the true maximum local field strength -- given by the extreme values in ZDI maps divided by $f_V$ -- are in the range of 7.3--27.3~kG. It is not known if such ultra-strong magnetic fields are real or represent an unphysical artefact resulting from a simplified treatment of the Stokes $V$ LSD profiles.

Repeated ZDI inversions based on data sets obtained 1--2 years apart revealed no major systematic evolution of M-dwarf global field topologies. Analysis of the longest Stokes $V$ profile time-series data available for AD~Leo \citep{lavail:2018} provided an evidence of the reduction of the circular polarisation amplitude (both in the LSD profiles and in individual lines) for the data collected in 2016 compared to the observations \cla{made} in 2006--2012 period. This is, so far, the only report of possible secular change of the global magnetic field in an M dwarf. Applying the same ZDI methodology as \cla{used} in the earlier study of AD~Leo \citep{morin:2008}, \citet{lavail:2018} concluded that the change of the Stokes $V$ profile morphology can be understood in terms of a modification of the filling factor $f_V$, with the global field geometry remaining the same in 2016 as in previous epochs.

Stokes $V$ ZDI analysis was extended to late-M dwarfs by \citet{morin:2010}. This study investigated six M5--M6.5 stars, based on the ESPaDOnS observations obtained at 2--4 epochs during consecutive years. It was discovered that the global magnetic topologies of these stars fall into two distinct categories. Some stars have strong, axisymmetric, dipole-dominated fields reminiscent of the field configurations of mid-M dwarfs. For example, GJ~51 and GJ~412B (WX UMa) were found to host axisymmetric dipolar fields with the maximum magnetic flux of 4--5~kG and the surface-averaged field strength \bv\,=\,1.0--1.6~kG.  The global field filling factor $f_V=0.12$ was deduced for both stars, implying an improbable local field strength of up to $5/0.12\approx42$~kG. Several other late-M stars observed by \citet{morin:2010}, including GJ~406 (CN~Leo), GJ~1154A, GJ~1224, likely possess the same type of global magnetic field, but could not be analysed with ZDI due to the absence of any rotational modulation of their strong Stokes $V$ signatures. At the same time, other late-M stars, including GJ~1111, 1156, 1245B, 3622, turned out to have considerably weaker, \bv\,=\,50--200~G, global fields, often with a significant non-axisymmetric contribution. Low-mass dwarfs with both types of global magnetic topologies have similar fundamental parameters and rotation periods.

A ZDI study of the M5.5--M6 components of the GJ~65 binary system (BL~Cet and UV~Cet) by \citet{kochukhov:2017c} provided another illustration of this puzzling magnetic dichotomy of late-M dwarfs. These two stars with nearly identical masses and rotation rates form a physical pair and, thus, must have come from the same protostellar cloud and followed the same evolutionary path. Yet, UV~Cet is known to show a very different activity pattern in X-ray and radio compared to BL~Cet. \citet{kochukhov:2017c} were able to link this discrepancy to dissimilar large-scale fields of the two stars. They found that UV~Cet has a strong, axisymmetric field while BL~Cet hosts four times weaker non-axisymmetric magnetic topology.

\begin{figure}[!t]
\includegraphics[width=\textwidth]{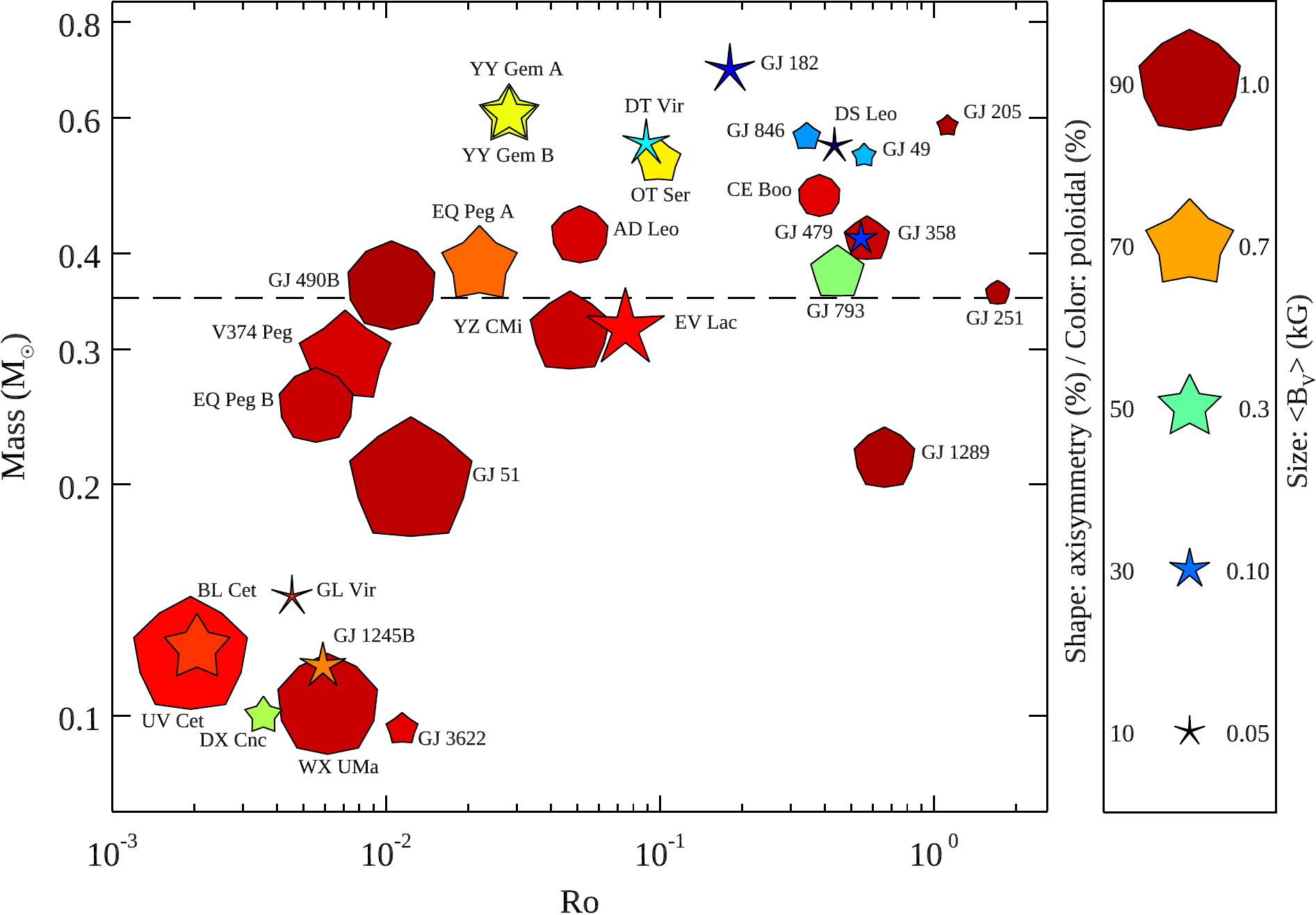}
\caption{Properties of the global magnetic field topologies of M dwarfs obtained with the ZDI modelling of Stokes $V$ spectra as a function of Rossby number and stellar mass. The symbol size is proportional to the mean global field strength \bv. \cla{The symbol} colour illustrates the energy fraction of the poloidal global field component (from dark red for purely poloidal fields to dark blue for purely toroidal fields). The symbol shapes correspond to the degree of global field axisymmetry (from decagons for purely axisymmetric fields to sharp stars for purely non-axisymmetric topologies). The horizontal dashed line marks the theoretical limit of fully convective interior $M_\star\approx0.35M_\odot$ \citep{chabrier:1997}.}
\label{fig:zdi_summary}
\end{figure}

All ZDI investigations summarised above targeted very active stars with rotation periods less than $\sim$\,10~d. Tomographic mapping of the global fields of seven slower rotating ($P_{\rm rot}$ up to 90~d), moderately active early- and mid-M dwarfs was presented by \citet{hebrard:2016} and \citet{moutou:2017}, respectively. These studies inferred topologically simple, often axisymmetric fields with the average strength \bv\,=\,20--130~G for M0--M2 stars and 30--270~G for M3--M4.5 dwarfs. On the other hand, \citet{kochukhov:2019a} reconstructed complex, non-axisymmetric fields with an average strength of 200--260~G and a maximum local field intensity of $\approx500$~G for the twin M0.5 components of the short-period (0.8~d) eclipsing binary YY~Gem.

A summary of the results of all ZDI studies of M-dwarf stars is compiled in Table~\ref{tab:stokesV}. This table reports the large-scale field characteristics of 30 stars based on 53 separate magnetic field maps. For each map, we list the average global field strength \bv\, as well as the fraction of the magnetic field energy contained in the poloidal component, dipolar ($\ell=1$) component, and in the axisymmetric ($|m|<\ell/2$) part of the global field.

The same magnetic mapping results are illustrated in Fig.~\ref{fig:zdi_summary}, which shows selected global field parameters as a function of Rossby number and stellar mass. The latter was calculated for single stars according to the $M_{K_S}$--$M_\star$ calibration by \citet{mann:2019}, using the $K_S$ magnitudes and parallaxes found in the SIMBAD data base. Fig.~\ref{fig:zdi_summary} demonstrates several main features and trends of the global magnetic field geometries of M dwarfs discussed above. Early-M dwarfs tend to have diverse, weak fields. Mid-M stars transition to simple, strong, predominantly axisymmetric fields when stellar interior becomes fully convective. Late-M dwarfs exhibit magnetic dichotomy, with two very different types of magnetic geometries found irrespective of stellar mass or rotation rate. This plot also shows that significant areas of the mass-rotation parameter space remain unexplored. In particular, there is a distinct lack of ZDI studies of fully convective stars with long rotation periods.

\subsubsection{Comparison of global and total magnetic fields}
\label{sect:glob}

\begin{figure}[!t]
\includegraphics[width=\textwidth]{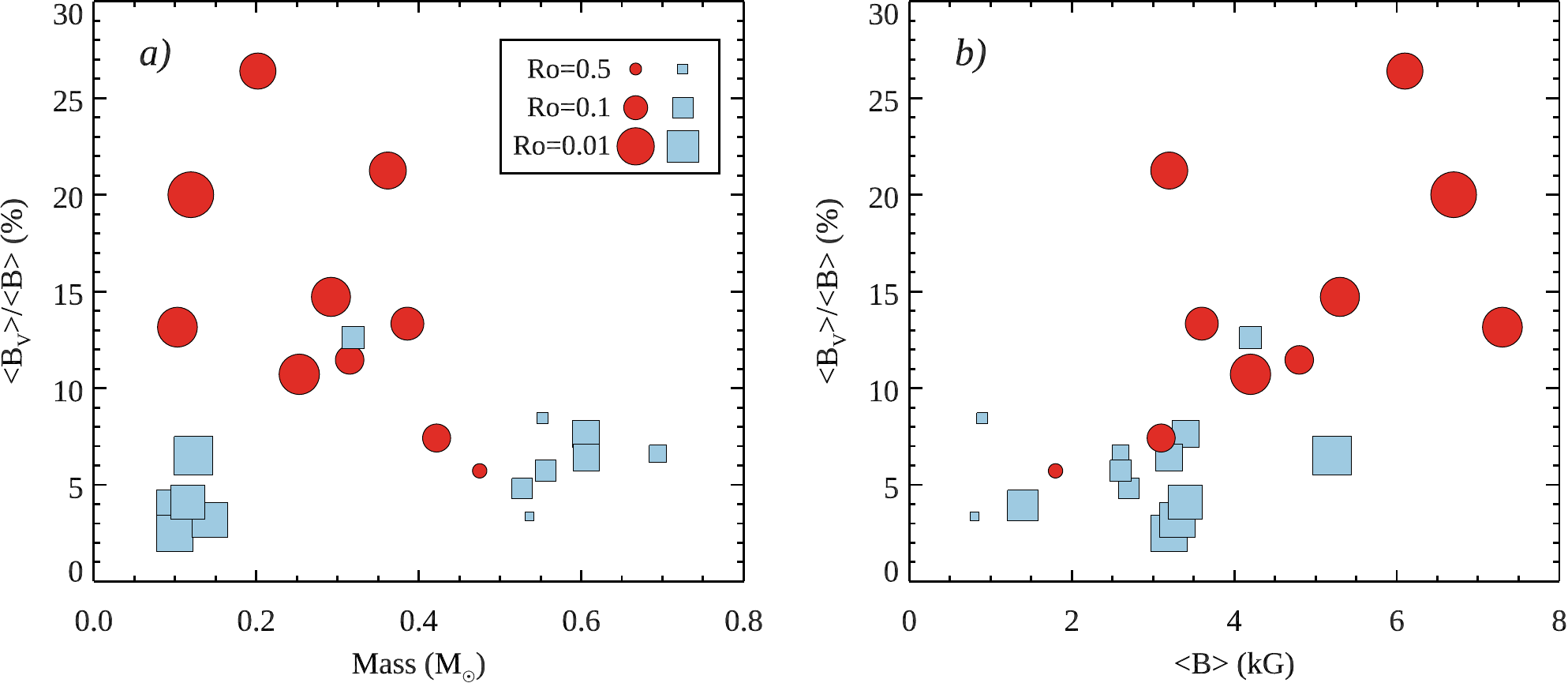}
\caption{Ratio of the average global magnetic field strength \bv\ derived by ZDI studies to the total field strength \bi\ obtained from Stokes $I$ as a function of stellar mass (a) and total field strength (b). The red circles show stars with predominantly dipolar, axisymmetric global fields. The blue squares correspond to stars with multipolar or non-axisymmetric dipolar fields. The symbol size reflects the Rossby number, as indicated by the legend in panel (a).}
\label{fig:bi_bv}
\end{figure}

It is instructive to compare magnetic fields of M dwarfs derived with the two direct diagnostic methods based on the Zeeman effect. As discussed in Sect.~\ref{sect:disk}, the total magnetic field \bi\ inferred from the Zeeman broadening and magnetic intensification of lines in intensity spectrum is likely to exceed the average large-scale magnetic field \bv\ obtained with ZDI from polarimetric observations. The degree to which the two magnetic measurements diverge characterises complexity of the underlying M-dwarf magnetic topology. Initial comparisons of this kind \citep{morin:2008,morin:2010,reiners:2009} were limited to a small number of stars and relied on coarse \bi\ measured with approximate techniques (see Sect.~\ref{sect:approxI}). More recently, \citet{kochukhov:2019a} extended this assessment to many more low-mass stars, taking advantage of the increased sample of M dwarfs with accurate total field strength determinations. Fig.~\ref{fig:bi_bv} presents an updated version of the plot from \citet{kochukhov:2019a}, using the data from Tables~\ref{tab:stokesI}--\ref{tab:params}. There are 23 stars with both \bi\ and \bv\ measurements. Ten of them have global magnetic configurations which are dominated by an axisymmetric poloidal dipolar field. The remaining 13 stars exhibit other types of large-scale fields, including mixtures of poloidal and toroidal components, oblique poloidal fields, etc.

When plotted as a function of stellar mass (Fig.~\ref{fig:bi_bv}a), the ratio of the global to total mean field, \bv$/$\bi, does not exceed 9\% for partially convective stars. For fully convective M dwarfs, this ratio appears to depend on the type of global magnetic field topology. The stars with axisymmetric dipolar fields tend to have a larger fraction of the total field strength recovered by ZDI, \bv$/$\bi\,=\,11--21\% and up to 26\% for GJ~51. On the other hand, stars with other types of large-scale fields have \bv$/$\bi\,$<$\,7\%.

The ratio \bv$/$\bi\ correlates with the total magnetic field strength (Fig.~\ref{fig:bi_bv}b). One can conclude that the stronger is the average field in an M dwarf, the more likely that this star has a strong, dipole-dominated global field. There are exceptions to these trends. The star GJ~873 (EV~Lac) has a decidedly non-axisymmetric field, but exhibits a large \bv$/$\bi\ ratio. Its total field also appears to be too strong for the 4.37~d rotation period. The primary component of the GJ~65 system (BL~Cet) also has an unusually strong total field given its non-axisymmetric configuration. A couple of stars, GJ~388 (AD~Leo) and GJ~569A (CE~Boo), seem to have axisymmetric, dipole-dominated global fields, but a relatively weak total field and a low \bv$/$\bi\ ratio.

\subsubsection{Extended stellar magnetospheres}
\label{sect:extrapol}

\begin{figure}[!t]
\centering
\includegraphics[width=0.6\textwidth]{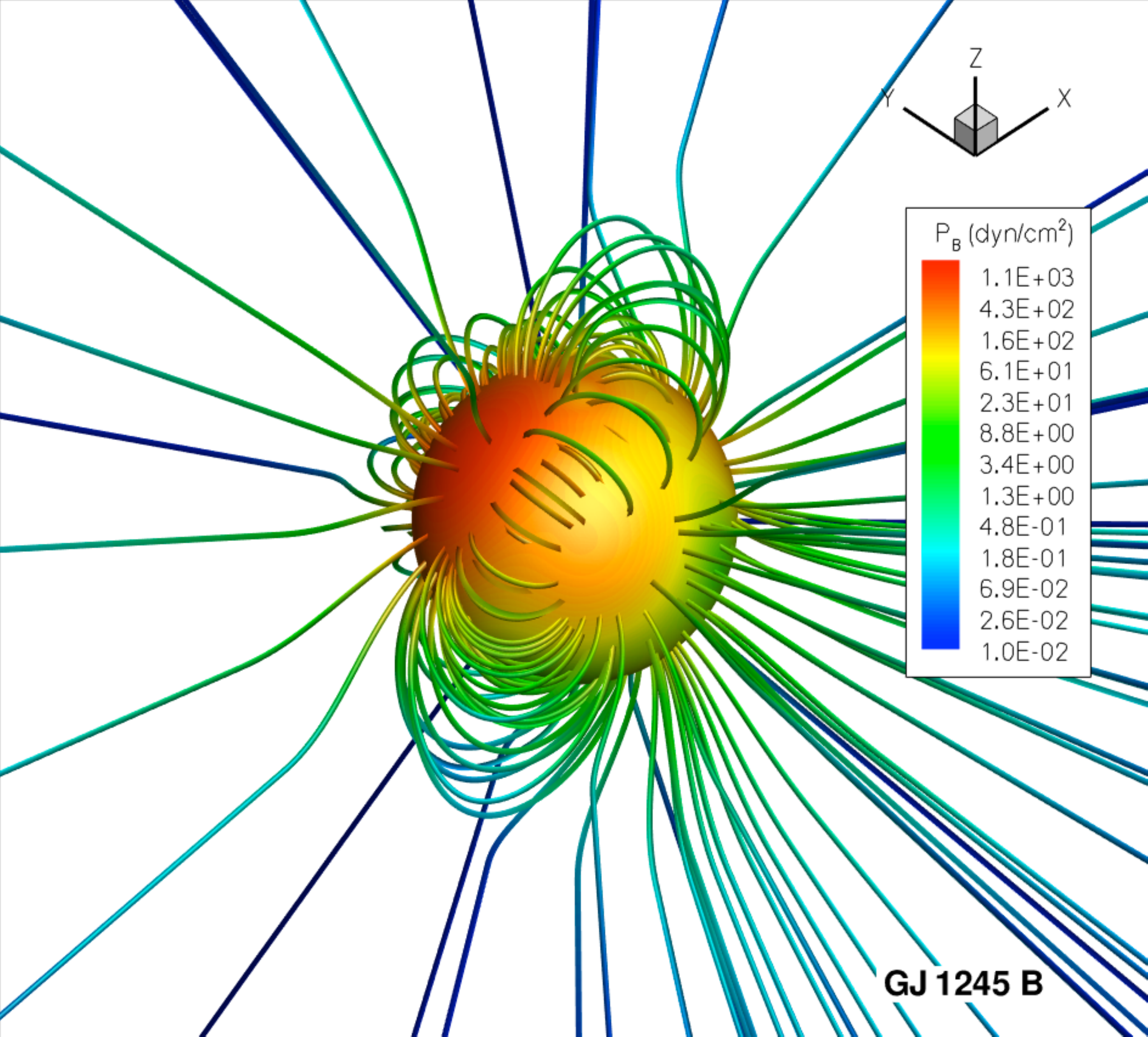}
\caption{Coronal magnetic field of GJ~1245B obtained by extrapolating the ZDI surface magnetic field map with the potential field source surface technique. The colour of the magnetic field lines corresponds to the local magnetic pressure that would be exerted on a planet orbiting this star. 
Image reproduced with permission from \citet{vidotto:2013}, copyright by ESO.}
%% Credit: \citeauthor{vidotto:2013}, 2013, A\&A, 557, A67.}

\label{fig:pfss}
\end{figure}

Maps of large-scale surface magnetic fields of M dwarfs reconstructed with ZDI form the basis of our knowledge about extended magnetospheres of these stars. Different methods of vector field extrapolation can be applied to ZDI maps with the goal to assess the three-dimensional structure of stellar wind, determine the mass and angular momentum loss, and investigate the impact of stellar magnetic field on planets orbiting M dwarfs. For example, \citet{lang:2012} and \citet{vidotto:2013} applied the potential field source surface extrapolation technique \citep[PFSS, e.g.][]{jardine:2002a} to the M-dwarf ZDI maps calculated by \citet{donati:2008} and \citet{morin:2008,morin:2010}. In this method, the three-dimensional potential field structure is established analytically by taking the observed stellar radial magnetic field as one boundary condition and supposing that the magnetic field lines become open at a certain distance from the star, called the source surface radius, $R_{s}$. An example of the extended M-dwarf magnetosphere calculated in this manner for GJ~1245B by \citet{vidotto:2013} using the ZDI results by \citet{morin:2010} and assuming $R_{s}=2.5R_\star$ is shown in Fig.~\ref{fig:pfss}.

With the help of the PFSS approach, \citet{lang:2012} predicted the X-ray emission of M dwarfs and were able to reproduce the observed saturation of coronal emission at Ro\,$\le$\,0.1. \citet{vidotto:2013} estimated magnetic pressure experienced by hypothetical planets in the habitable zones of M dwarfs with different global field geometries, finding that these planets must possess much stronger magnetic fields than the Earth to sustain their magnetospheres. Conversely, for an Earth-like planetary magnetic field, an M-dwarf host star must rotate much slower than the Sun to provide low enough magnetic pressure to ensure atmospheric retention and hence planetary habitability.

A more sophisticated \cla{magnetohydrodynamic (MHD)} modelling procedure was used by \citet{vidotto:2011,vidotto:2014a} to establish magnetospheric structure of six early-M dwarfs and the M4 star V374~Peg, again starting from the published ZDI maps of these stars. These studies indicated that detailed structure of the stellar surface magnetic field has a large impact on the planetary system environment, with non-axisymmetric global field configurations offering a better shielding of terrestrial-size planets from cosmic rays. On the other hand, \citet{cohen:2014} concluded that a strongly non-axisymmetric M-dwarf magnetic field, such as the one found for GJ~873 (EV~Lac), leads to a large, time-dependent Joule heating of the upper layers of planetary atmospheres. This contribution to the energy budget has to be taken into account in models of planetary atmospheres.

The small-scale magnetic fields, missing in ZDI reconstructions, are unlikely to affect the structure of extended magnetospheres except very close to the stellar surface \citep{lang:2014}. The neglect of non-potential (toroidal) magnetic field components, occasionally found by ZDI analyses of M dwarfs but omitted by magnetospheric studies, is also not expected to have a major impact on the inferences about stellar wind and angular momentum loss \citep{jardine:2013}. Despite this apparent robustness of the field extrapolation techniques, the lack of self-consistency with the ZDI modelling of stellar surface fields remains a major source of uncertainty. Calculation of the Stokes $V$ profiles in ZDI has, so far, relied on unphysical global field filling factors and did not directly incorporate strong small-scale fields. The ZDI inverse problem solution also depends on the adopted harmonic field parameterisation. As shown for early-type stars by \citet{kochukhov:2016a}, restricting this parametrisation to the one consistent with the PFSS framework may yield a drastically different surface magnetic field structure than the more general potential field parametrisation usually employed in ZDI (see Sect.~\ref{sect:zdi}). Ideally, the extended magnetic field of M dwarfs has to be determined self-consistently with the interpretation of observed Stokes $I$ and $V$ spectra of these stars.

Needless to say, the diversity of global magnetic fields of M dwarfs, especially among late-M stars, makes many conclusions and theoretical inferences based on ZDI maps of selected active stars highly questionable. Fully convective stars do not seem to possess a ``representative'' global magnetic field topology that changes systematically with the stellar mass or rotation rate. Magnetic fields of specific planet-hosting M dwarfs have to be detected and analysed, similar to the existing studies of active M dwarfs without confirmed planets, in order to reach robust conclusions about possible effects of these fields on exoplanet atmospheres, interiors, and space weather environments.

\section{Discussion and outlook}
\label{sect:discussion}

\subsection{Summary of observational results}

Recent progress in numerical methods of modelling spectra of low-mass magnetic stars combined with the availability of high-quality spectropolarimetric observations of these objects have ushered a higher level of clarity in understanding of their surface magnetic fields. We now know that active, rapidly rotating M dwarfs possess multi-kG average (total) fields. The strongest fields found on these stars reach 6--7~kG. This measure corresponds to the disk-integrated magnetic field modulus, suggesting that the field strength likely exceeds $\sim$\,10~kG within localised surface magnetic features. It appears that the average field strength generally grows towards later spectral types, but no prominent qualitative change of this field characteristic is observed at the transition from partially to fully convective stellar interiors. At the same time, M dwarfs possess very diverse large-scale magnetic field geometries. Early-M dwarfs tend to have complex, weak global fields, often with a significant toroidal contribution. Stronger poloidal global fields dominated by an axisymmetric dipolar component are common in mid-M dwarfs. These fields emerge roughly at the limit of full convection. Late-M dwarfs exhibit a puzzling global field dichotomy, with both early-M and mid-M field types present in stars of similar mass and rotation rate. The average strength of the global magnetic fields of M dwarfs seldom reaches 1~kG. A striking disparity with much stronger total fields implies that magnetic fields of M dwarfs have a complex, intermittent character, even in stars possessing seemingly simple dipolar large-scale magnetic topologies. A qualitative model consistent with all currently available observations envisages M-dwarf field structure comprising a carpet of strong, intermittent fields superimposed onto a weaker large-scale component (e.g. as depicted in Fig.~\ref{fig:v_models}c). It is not entirely clear if this global magnetic component is a separate physical entity, perhaps connected to a separate dynamo process, or is associated with one of the components (e.g. strong-field spots) of the continuous power spectrum describing the small-scale magnetic structures. Diversity of the relationships between the properties of total and large-scale fields observed for active M dwarfs (Sect.~\ref{sect:glob}) argues in favour of the former conjecture.

\subsection{Theoretical dynamo models}

Significant theoretical effort has been expended on understanding the physical processes responsible for magnetic field generation in low-mass stars and comparison of predicted and observed magnetic field geometries. So far, this work was focused on fully convective stars since these objects represent particularly useful laboratories for studying distinctly non-solar dynamo action. Magnetohydrodynamic models applied to M dwarfs can be divided into two broad categories. Mean-field dynamo models \citep{chabrier:2006,kitchatinov:2014,shulyak:2015a,pipin:2017,pipin:2018} explore large-scale magnetic configurations with a particular focus on probing long time scales. By construction, these models average out small-scale magnetic structures and thus are not capable of assessing the total M-dwarf field strengths. With the exception of recent non-linear mean-field calculations by \citet{pipin:2017} and \citet{pipin:2018}, this class of models also provides no information on the global field strength, leaving only the large-scale field geometry and its temporal evolution as testable predictions. Generally, the outcome of mean-field dynamo studies for M dwarfs has been inconclusive. Different published calculations found diverse dynamo solutions, depending on physical assumptions, model parameters and assumed magnitude of differential rotation, which is poorly known for M dwarfs. However, a common feature of most mean-field dynamo calculations is the prediction of a solar-like cyclic behaviour of M-dwarf global fields, with reversals of the dominant field polarity every 10--20 yr. In the framework of this hypothesis, different types of global fields observed in late-M dwarfs are attributed to different phases of a long activity cycle \citep[e.g.][]{kitchatinov:2014}. This idea can be tested by a long-term magnetic monitoring of selected stars. One of the best studied active M dwarfs with an axisymmetric dipolar field, AD Leo (GJ~388), has been observed with spectropolarimetry over a period of about 14 years (2005--2019). These observations reveal no evidence of a reversing large-scale field \citep[][and private communication]{lavail:2018}. There is also an indirect evidence, from X-ray and radio activity proxies, that the current global field dichotomy of the twin components in the GJ~65 binary system has persisted for at least 30~yr \citep{kochukhov:2017c}. These observations are incompatible with a magnetic cycle length shorter than $\sim$\,50~yr. \cla{Continuation of monitoring of this system will allow one to probe possibility of even longer activity cycles.}

\begin{figure}[!t]
\centering
\includegraphics[width=0.4\textwidth]{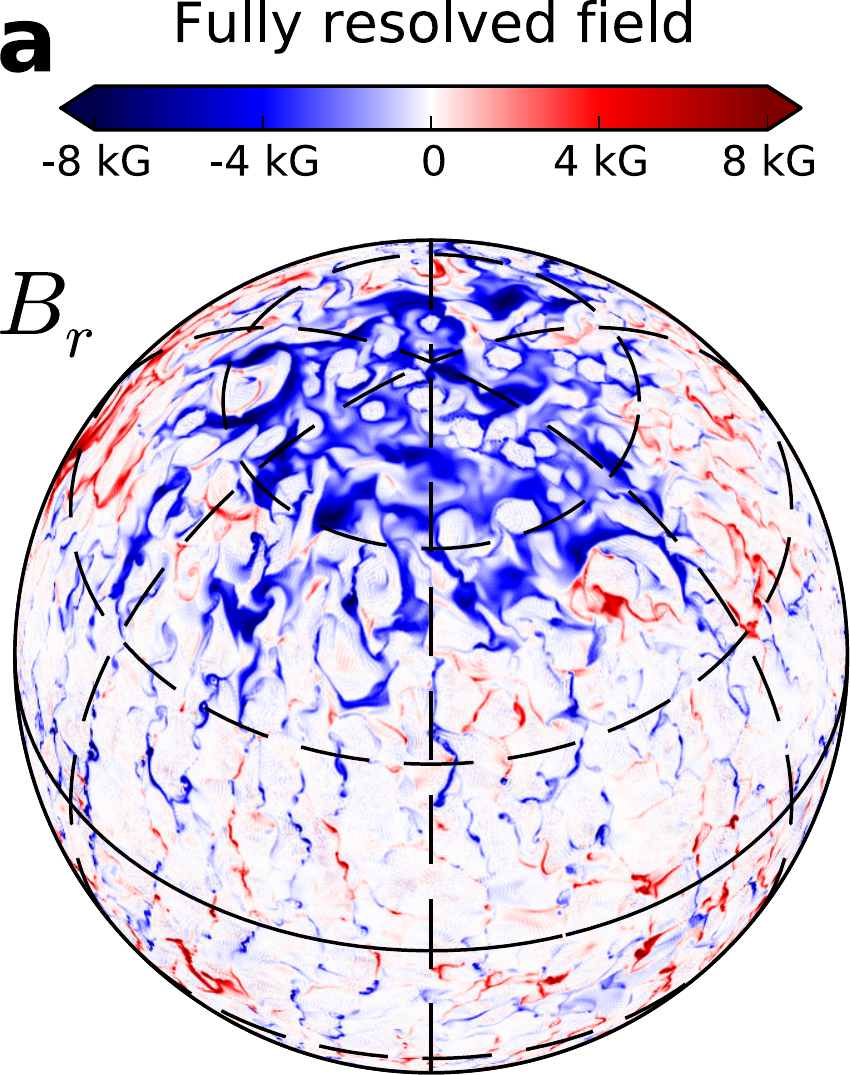}\hspace*{0.6cm}
\includegraphics[width=0.4\textwidth]{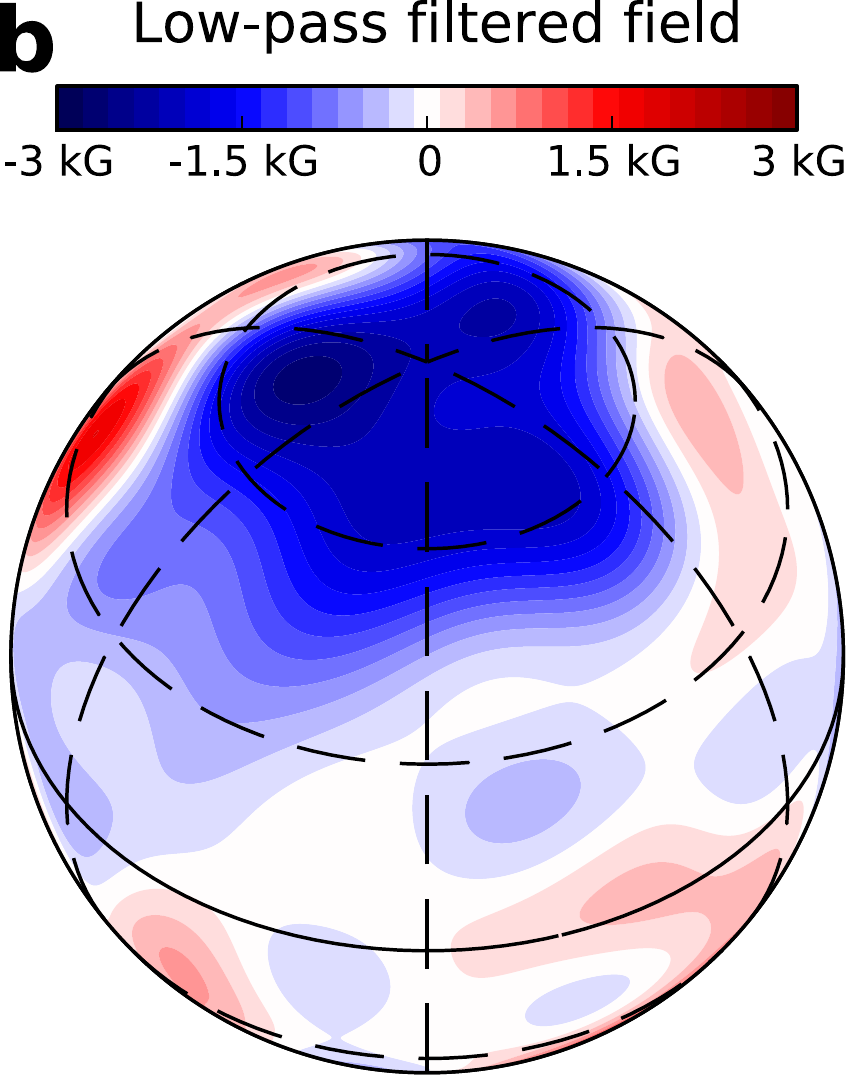}
\caption{Near-surface distribution of the radial magnetic field component in three-dimensional dynamo simulation of a fully convective star. Panel (a) shows the field structure at full resolution. Panel (b) illustrates the low-pass filtered surface field map, as would be observed by ZDI. 
Image reproduced with permission from \citet{yadav:2015}, copyright by AAS.}
%% Credit: \citeauthor{yadav:2015}, 2015, ApJ, 813, L31.}
\label{fig:mhd}
\end{figure}

Another class of theoretical dynamo models developed for fully convective M dwarfs attempts to derive the full spatial power spectrum of magnetic field through computationally expensive direct three-dimensional \cla{MHD} numerical simulations \citep{dobler:2006,browning:2008,gastine:2013,yadav:2015,yadav:2016}. Among these calculations, the series of models by \citet{gastine:2013} and \citet{yadav:2015,yadav:2016}, based on an anelastic treatment of the dynamo action in spherical shells, appears to be the most advanced in terms of numerical methods and underlying physics, and most useful for comparison with observational results. Specifically, low-resolution calculations by \citet{gastine:2013} demonstrated bistability of dynamo at low Rossby numbers, which results in solutions with persistent dipole-dominated and multipolar large-scale fields, very similar to the observed dichotomy of the global magnetic fields in late-M dwarfs. Subsequently, \citet{yadav:2015} presented high-resolution simulation of magnetic field of the former type for an M dwarf with a 20~d rotation period. A snapshot of the radial magnetic field map obtained by these authors is shown in Fig.~\ref{fig:mhd}. This theoretical field structure contains both a large-scale component, formed by a network of magnetic spots, and a uniformly distributed small-scale field that carries most of the magnetic energy. Owing to its limited resolution, ZDI retrieves only a few low-order spherical harmonic components of this complex, \cla{dynamic} field (Fig.~\ref{fig:mhd}b), yielding a stationary, dipole-like global field configuration with a strength of $\sim$\,500 G. This is much smaller than the average total field strength of 2.3~kG for this model. This difference is in line with the systematic discrepancy between the outcomes of ZDI and Zeeman broadening measurements of M-dwarf field strengths. \cla{However, these observational results pertain mostly to stars rotating much faster than $P_{\rm rot}=20$~d assumed by \citet{yadav:2015}. It cannot be excluded that simulations results will change significantly once this modelling framework is extended to the limit of full dynamo saturation corresponding to the fastest rotators.}

\cla{Sensitivity of the three-dimensional dynamo simulations to the stellar rotation period was demonstrated by the work of} \citet{yadav:2016}. \cla{They} found that \cla{an MHD simulation carried out} for a much slower rotating fully convective M dwarf (loosely modelled after Proxima Cen with its 83~d rotation period) \cla{resulted} in a qualitatively different cyclic dynamo state characterised by periodic changes of the average field strength and global field configuration. Although the length of this cycle -- nine years -- is compatible with the seven-year activity cycle of Proxima Cen suggested by indirect activity proxies \citep{wargelin:2017}, a long-term spectropolarimetric monitoring of this star is necessary to test the prediction of changing polarity of the global magnetic field. At the same time, substantial mean field strength changes anticipated by this theoretical model can probably be already validated using archival collections of the high-resolution spectra of this star.

\cla{A broader perspective on the physics of convective dynamos in stars and sub-stellar objects was offered by the work of \citet{christensen:2009}. They suggested an empirical scaling law, inspired by geodynamo simulations, that accounts for the magnetism of widely different objects, ranging from terrestrial planets to low-mass stars, provided their dynamos are operating in the saturated regime. This scaling law is based on the general idea that the mean internal magnetic field strength is set by the energy flux available in convection zones. Characteristic field intensities predicted by this scaling principle are reasonably consistent with the kG-strength fields observed for fast-rotating M dwarfs and T~Tauri stars if one assumes that their internal fields are a factor of 3.5 stronger than the measured surface fields.}

It should be remarked that none of the dynamo models discussed above has a sufficient resolution to be directly used for the line formation calculations in stellar atmospheres. Instead, local magnetohydrodynamic modelling of the surface layers of M dwarfs has been carried out by a different family of radiation-hydro codes \citep{beeck:2015,beeck:2015a,wedemeyer:2016}. These calculations consider in detail the interaction between convection and magnetic field in a stellar atmosphere and chromosphere, but adopt unstructured seed magnetic field as an arbitrary initial parameter and do not treat the dynamo problem self-consistently. Mean magnetic fluxes of up to 500~G considered by these calculations are well below the average field strengths typical of active M dwarfs. Thus, these local three-dimensional MHD models inform us about dynamics of atmospheres of inactive low-mass stars but provide little insight into the structure of active M-dwarf atmospheres permeated by kG-strength fields.

\subsection{Magnetic stellar structure models}

\citet{mullan:2001} demonstrated that inclusion of magnetic field in theoretical one-dimensional interior structure calculations for low-mass stars increases the stellar radius and reduces the effective temperature compared to non-magnetic model of the same mass. These effects, resulting from the modification of parameterised treatment of the convective energy transport in stellar structure models, are potentially capable of explaining the radius inflation observed for M dwarfs in eclipsing binaries \citep{ribas:2006,torres:2013}. Initial models by \citet{mullan:2001} and \citet{macdonald:2014} called for 1--100 MG interior fields in M dwarfs, which are untenable for reasons of stability \citep{browning:2016} and are inconsistent with the results of three-dimensional numerical simulations \citep{browning:2008,yadav:2015}. Revised magneto-convection models by \citet{macdonald:2017} were able to reproduce inflated radii of M dwarfs with the interior fields capped at $\sim$\,10~kG and the average surface vertical fields, $B_{\rm vert}$, of typically 500~G. The models developed for the late-M components of the GJ~65 system \citep{macdonald:2018} and for TRAPPIST-1 \citep{mullan:2018} took into account finite magnetic diffusivity, arriving at an estimate of $B_{\rm vert}$ of about 1.5--2 and 1.5--1.7~kG, respectively.

An independent set of theoretical stellar interior models incorporating magnetic field was developed by \citet{feiden:2013,feiden:2014}. Their treatment of the suppression of convection by magnetic field differs in certain details and assumptions from the approach followed by \citet{mullan:2001}, resulting in a systematically stronger surface magnetic field necessary to reproduce masses and radii of well-studied binary systems. For example, \citet{feiden:2013} predicted $\langle B f\rangle=3.6$--4.3~kG for the partially convective components of the eclipsing binaries YY~Gem and CU~Cnc. For the fully convective components of CM~Dra, \citet{feiden:2014} anticipated $\langle B f\rangle\approx5$~kG.

Attempts to confront the surface magnetic fields invoked by these theoretical stellar structure calculations with the observed M-dwarf magnetic field strengths are hampered by ambiguity of the relationship between the model $B_{\rm vert}$ and empirical \bi\ deduced by the Zeeman broadening and intensification analyses. The latter measurements cannot constrain the local field inclination. Therefore, one can plausibly have relations ranging from $B_{\rm vert}\approx\langle B \rangle$ for predominantly vertical small-scale fields to $\sqrt{3} B_{\rm vert}\approx\langle B \rangle$ for an isotropic field vector orientation. Taking this uncertainty into account, \citet{kochukhov:2019a} concluded that  $\langle B f\rangle=4.0$--4.5~kG predicted by \citet{feiden:2013} for YY~Gem is in much better agreement with the observed field strength of 3.2--3.4~kG in comparison to $\sqrt{3} B_{\rm vert}<1$~kG required by \citet{macdonald:2017}. On the other hand, \citet{macdonald:2018} argued that their $\sqrt{3} B_{\rm vert}=2.9$--3.6~kG surface field estimate is marginally consistent with \bi\,=\,4.5--5.8~kG reported for GJ~65A and B by \citet{shulyak:2017}. A new precise direct measurement of the surface field strength of TRAPPIST-1, for which \citet{mullan:2018} predicted a relatively high and easily observable photospheric field strength $\sqrt{3} B_{\rm vert}=2.5$--2.9~kG, will serve as a crucial test of this theoretical modelling framework.

\subsection{Future research directions}

A major progress in observational characterisation of M-dwarf magnetic fields is expected from opening the near-infrared spectral domain to high-resolution spectropolarimetry. This development is associated with the commissioning of SPIRou at CFHT and the upgraded CRIRES at VLT. These instruments will enable studies of many more M dwarfs at higher precision thanks to a several magnitude gain in brightness of these stars at near-infrared wavelengths compared to the optical. The forthcoming near-infrared spectropolarimetric data will make it possible to detect and monitor global magnetic fields of slowly rotating, inactive M dwarfs. A multitude of near-infrared molecular and atomic diagnostic lines are suitable for effective Zeeman broadening measurements. The $\lambda^2$ dependence of the Zeeman effect will enable measurements of weaker total fields of inactive stars and will allow one to infer more reliable field strength distributions for active M dwarfs. Very high $S/N$ ratio spectra of low-mass stars, particularly data that could be obtained with CRIRES fed by the 8-m VLT, will make it possible to study polarisation of individual spectral lines, thereby lifting many of the assumptions and simplifications associated with LSD modelling. These data will facilitate reconstruction of more robust and physically more realistic ZDI maps, addressing the issue of uncomfortably large local field strengths resulting from the application of global field filling factor approach in tomographic mapping of low-mass stars.

Advances in observational magnetic field studies of M dwarfs have to be matched by improvements of theoretical modelling of the interiors and atmospheres of fully convective stars. The global 3D MHD dynamo simulations by \citet{gastine:2013} and \citet{yadav:2015,yadav:2016} have reached a remarkable degree of sophistication and realism, succeeding in reproducing several observed characteristics of global and small-scale fields. However, detailed simulation predictions were published for two sets of stellar parameters only. Given the diversity of simulation outcomes as a function of the stellar rotation rate and seed magnetic field, a systematic mapping of the entire parameter range relevant for M dwarfs with these simulations is required for meaningful interpretation of observations. The global 3D simulations have to be coupled to detailed local MHD modelling of M dwarf atmospheres, such as the one carried out by \citet{beeck:2015} and \citet{wedemeyer:2016}. Although global 3D stellar models with a resolution sufficient to study atmospheric layers of dwarf stars are unlikely to become available in the near future, it should be possible to run local box-in-a-star calculations with the input magnetic field corresponding to different patches of global star-in-a-box models. In this context, understanding the response of cool-star atmospheres to multi-kG magnetic fields is badly sought after by spectral modelling studies. Both ZDI and Zeeman broadening analyses may be affected by hitherto unrecognised systematic biases stemming from historic reliance of these studies on standard non-magnetic, single-component, plane-parallel model atmospheres.

In summary, the methods of observational characterisation of M-dwarf magnetic fields and theoretical models explaining the origin and evolution of these fields are reaching maturity. The synergy of modern theoretical and observational approaches to the M-dwarf magnetism problem is unparalleled in stellar physics, leading to a far greater clarity of the properties of surface magnetic fields and their relation to the underlying dynamo physics compared to any other class of active late-type stars. This development firmly establishes magnetism as a compulsory ingredient of essentially any modern study of low-mass stars and their planetary systems.

\begin{acknowledgements}
I thank my long-term collaborator and friend Denis Shulyak for many years of inspiring work on magnetic fields of M-dwarf stars. I also thank colleagues in Uppsala and elsewhere, Alexis Lavail, Ulrike Heiter, Terese Olander, Nikolai Piskunov, Gregg Wade, who are actively contributing to this research. Financial support by several funding agencies, including the Swedish Research Council, the Knut and Alice Wallenberg Foundation, and the Swedish National Space Agency, is gratefully acknowledged. In preparation of this review I made extensive use of the SIMBAD database and the VizieR catalogue access tool, both operated at CDS, Strasbourg, France, and of NASA's Astrophysics Data System Abstract Service.
\end{acknowledgements}

\newpage

\begin{appendix}
%%%%%%%%%%%%%%%%%%%%%%%%%%%%%%%%%%%%%%%%%%%%%%%%
% Table 1 with Stokes I magnetic fields measurements
%%%%%%%%%%%%%%%%%%%%%%%%%%%%%%%%%%%%%%%%%%%%%%%%
\section{Appendix 1: Summary of M-dwarf magnetic field strength measurements using Zeeman broadening and intensification}
{\small
\begin{center}

\begin{ThreePartTable}
  \begin{TableNotes}  %%% This should come before longtable and it is not tablenotes but TableNotes
  \small
  \item References: 1, \citet{shulyak:2011}; 2, \citet{shulyak:2017}; 3, \citet{shulyak:2019}; 4, \citet{kochukhov:2017c}; 5, \citet{kochukhov:2001}; 6, \citet{kochukhov:2009b}; 7, \citet{kochukhov:2019a}; 8, \citet{johns-krull:2000}; 9, \citet{afram:2009}; 10, \citet{shulyak:2014}; 11, \citet{saar:1985}; 12, \citet{saar:1994a}; 13, \citet{phan-bao:2009}; 14, \citet{saar:1996}; 15, \citet{johns-krull:1996}.
  \end{TableNotes}

\begin{longtable}{llcl}

\caption{(continued)} \\
\hline\noalign{\smallskip}
GJ, Name & \bi\ (kG) & Method & References  \\
\noalign{\smallskip}\hline\noalign{\smallskip}
\endhead

\noalign{\smallskip}\hline
\endfoot

\noalign{\smallskip}\hline
\insertTableNotes
\endlastfoot

\caption{Average magnetic field strengths of M dwarfs derived from spectrum synthesis modelling of the Stokes $I$ spectra. The column ``Method'' indicates parameterisation of the magnetic field strength distribution (1: a single field strength value, 2: two-component model, 3: three and more field strength components). \label{tab:stokesI}} \\
\hline\noalign{\smallskip}
GJ, Name & \bi\ (kG) & Method & References  \\
\noalign{\smallskip}\hline\noalign{\smallskip}
\endfirsthead

GJ 2005A           & $2.0      $ & 1 & 1 \\
\noalign{\smallskip}
GJ 49              & $0.8\pm0.2$ & 3 & 2 \\
\noalign{\smallskip}
GJ 51, V388 Cas    & $6.1\pm0.2$ & 3 & 2 \\
                   & $4.8\pm0.3$ & 3 & 3 \\
\noalign{\smallskip}
Barta 161 12       & $5.8\pm1.0$ & 3 & 3 \\
\noalign{\smallskip}
GJ 65A, BL Cet     & $5.2\pm0.5$ & 2 & 4 \\
                   & $4.5\pm1.0$ & 2 & 2 \\
\noalign{\smallskip}
GJ 65B, UV Cet     & $6.7\pm0.6$ & 2 & 4 \\
                   & $5.8\pm1.0$ & 2 & 2 \\
\noalign{\smallskip}
GJ 1049            & $2.5      $ & 2 & 5 \\
                   & $2.9      $ & 3 & 6 \\
\noalign{\smallskip}
GJ 3136            & $4.9\pm1.0$ & 3 & 3 \\
\noalign{\smallskip}
G 80-21            & $3.2\pm0.1$ & 3 & 3 \\
\noalign{\smallskip}
GJ 3263A           & $0.56     $ & 1 & 1 \\    
\noalign{\smallskip}
GJ 3263B           & $0.5      $ & 1 & 1 \\
\noalign{\smallskip}
RX J0447.2+2038    & $5.7\pm1.0$ & 3 & 3 \\
\noalign{\smallskip}
GJ 182, V1005 Ori  & $2.6\pm0.6$ & 3 & 2 \\
\noalign{\smallskip}
GJ 3322A           & $2.75     $ & 1 & 1 \\    
\noalign{\smallskip}
LP 717-36          & $1.75     $ & 1 & 1 \\
\noalign{\smallskip}
GJ 208, V2689 Ori  & $1.2\pm0.3$ & 3 & 3 \\
\noalign{\smallskip}
GJ 3379            & $2.5\pm1.0$ & 3 & 3 \\
\noalign{\smallskip}
GJ 234A, V577 Mon  & $2.75     $ & 1 & 1 \\    
\noalign{\smallskip}
GJ278Ca, YY Gem A  & $3.4\pm0.3$ & 2 & 7 \\
\noalign{\smallskip}
GJ278Cb, YY Gem B  & $3.2\pm0.3$ & 2 & 7 \\
\noalign{\smallskip}
GJ 285, YZ CMi     & $3.3      $ & 3 & 8 \\
                   & $4.5      $ & 3 & 6 \\
                   & $3.4\pm0.2$ & 1 & 9 \\
                   & $3.6\pm0.1$ & 3 & 10 \\
                   & $4.8\pm0.2$ & 3 & 2 \\
                   & $4.6\pm0.3$ & 3 & 3 \\
\noalign{\smallskip}
GJ 1111, DX Cnc    & $3.2\pm0.5$ & 3 & 2 \\
                   & $3.3\pm0.6$ & 3 & 3 \\
\noalign{\smallskip}
G 161-071          & $5.3\pm1.0$ & 3 & 3 \\
\noalign{\smallskip}
GJ 388, AD Leo     & $2.8\pm0.3$ & 2 & 11 \\
                   & $2.6      $ & 2 & 12 \\
                   & $3.3      $ & 3 & 8 \\
                   & $3.2      $ & 3 & 6 \\
                   & $3.3\pm0.1$ & 1 & 9 \\
                   & $2.9\pm0.2$ & 3 & 10 \\
                   & $3.1\pm0.2$ & 3 & 2 \\
\noalign{\smallskip}
GJ 398, RY Sex     & $4.3      $ & 3 & 6 \\
\noalign{\smallskip}
GJ 3622            & $1.4\pm0.2$ & 3 & 2 \\
\noalign{\smallskip}
GJ 406, CN Leo     & $2.3\pm0.3$ & 1 & 9 \\
                   & $2.3\pm0.3$ & 3 & 10 \\
\noalign{\smallskip}
GJ 410, DS Leo     & $0.9\pm0.3$ & 3 & 2 \\
\noalign{\smallskip}
GJ 412B, WX UMa    & $7.3\pm0.3$ & 3 & 2 \\
\noalign{\smallskip}
RX J1215.6+5239    & $4.5\pm1.0$ & 3 & 3 \\
\noalign{\smallskip}
GJ 1156, GL Vir    & $3.3\pm0.4$ & 3 & 2 \\
                   & $3.6\pm0.7$ & 3 & 3 \\
\noalign{\smallskip}
LP 218-8           & $2.8\pm0.2$ & 1 & 9 \\
\noalign{\smallskip}
GJ 490B            & $3.2\pm0.4$ & 2 & 13 \\
\noalign{\smallskip}
GJ 494A, DT Vir    & $1.5      $ & 2 & 14 \\
                   & $2.6\pm0.1$ & 3 & 2 \\
\noalign{\smallskip}
RX J1417.3+4525    & $5.2\pm0.2$ & 3 & 3 \\
\noalign{\smallskip}
GJ 569A, CE Boo    & $1.8\pm0.1$ & 3 & 2 \\
\noalign{\smallskip}
GJ 3877            & $2.2\pm0.2$ & 1 & 9 \\
\noalign{\smallskip}
GJ 9520, OT Ser    & $2.7\pm0.1$ & 3 & 2 \\
                   & $3.2\pm0.2$ & 3 & 3 \\
\noalign{\smallskip}
G 256-25           & $3.4\pm1.0$ & 3 & 3 \\
\noalign{\smallskip}
GJ 3959            & $4.1\pm0.1$ & 3 & 3 \\
\noalign{\smallskip}
GJ 644 C, VB 8     & $2.8\pm0.1$ & 1 & 9 \\
                   & $2.8\pm0.4$ & 3 & 3 \\
\noalign{\smallskip}
GJ 1207            & $3.3\pm0.4$ & 3 & 3 \\
\noalign{\smallskip}
V1274 Her          & $6.9\pm1.0$ & 3 & 3 \\
\noalign{\smallskip}
G 227-22           & $4.3\pm0.9$ & 3 & 3 \\
\noalign{\smallskip}
GJ 1224            & $3.0\pm0.1$ & 1 & 9 \\
\noalign{\smallskip}
GJ 4053            & $2.1\pm1.6$ & 3 & 3 \\
\noalign{\smallskip}
GJ 729, V1216 Sgr  & $2.6\pm0.3$ & 2 & 15 \\
                   & $2.0      $ & 3 & 8 \\
                   & $2.5\pm0.2$ & 1 & 9 \\
                   & $2.3\pm0.2$ & 3 & 10 \\
                   & $2.2\pm0.8$ & 3 & 3 \\
\noalign{\smallskip}
GJ 752B, VB 10     & $2.3\pm0.2$ & 3 & 3 \\    
\noalign{\smallskip}
GJ 1243            & $3.2\pm1.0$ & 3 & 3 \\
\noalign{\smallskip}
GJ 1245B           & $1.9\pm0.5$ & 1 & 9 \\
                   & $3.4\pm0.4$ & 3 & 2 \\
\noalign{\smallskip}
SCR J2009-0113     & $3.2\pm0.3$ & 3 & 3 \\
\noalign{\smallskip}
GJ 803, AU Mic     & $2.3      $ & 2 & 12 \\      
\noalign{\smallskip}
GJ 4247, V374 Peg  & $5.3\pm1.0$ & 2 & 2 \\
                   & $4.4\pm1.0$ & 3 & 3 \\
\noalign{\smallskip}
GJ 852A, FG Aqr    & $3.0      $ & 1 & 1 \\
\noalign{\smallskip}
GJ 852B            & $1.5      $ & 1 & 1 \\
\noalign{\smallskip}
GJ 873, EV Lac     & $3.7      $ & 2 & 12 \\
                   & $3.8\pm0.5$ & 2 & 15 \\
                   & $3.9      $ & 3 & 8 \\
                   & $3.7\pm0.2$ & 1 & 9 \\
                   & $4.2\pm0.3$ & 3 & 2 \\
                   & $4.1\pm0.2$ & 3 & 3 \\
\noalign{\smallskip}
GJ 9799, GT Peg    & $3.4\pm0.6$ & 3 & 3 \\
\noalign{\smallskip}
GJ 896A, EQ Peg A  & $3.6\pm0.3$ & 3 & 2 \\
\noalign{\smallskip}
GJ 896B, EQ Peg B  & $4.2\pm1.0$ & 2 & 2 \\
\noalign{\smallskip}
GJ 4368A, LHS4022A & $0.5      $ & 1 & 1 \\
\noalign{\smallskip}
GJ 4368B, LHS4022B & $0.75     $ & 1 & 1 \\
\noalign{\smallskip}
RX J2354.8+3831    & $4.6\pm0.4$ & 3 & 3 \\

\end{longtable}
\end{ThreePartTable}
\end{center}
}

%%%%%%%%%%%%%%%%%%%%%%%%%%%%%%%%%%%%%%%%%%%%%%%%
% Table 2 with Stokes V ZDI analysis results
%%%%%%%%%%%%%%%%%%%%%%%%%%%%%%%%%%%%%%%%%%%%%%%%
\section{Appendix 2: Summary of Zeeman Doppler imaging results for M dwarfs}
{\small
\begin{center}

\begin{ThreePartTable}
  \begin{TableNotes}  %%% This should come before longtable and it is not tablenotes but TableNotes
  \small
  \item References: 1, \citet{donati:2008}; 2, \citet{morin:2010}; 3, \citet{kochukhov:2017c}; 4, \citet{hebrard:2016}; 5, \citet{moutou:2017}; 6, \citet{kochukhov:2019a}; 7, \citet{morin:2008}; 8, \citet{lavail:2018}; 9, \citet{phan-bao:2009}; 10, \citet{morin:2008a}
  \end{TableNotes}

\begin{longtable}{llllll}

\caption{(continued)} \\
\hline\noalign{\smallskip}
%Name & \bv\ (kG) & $E_{\rm pol}$ (\%) & $E_{\ell=1}$ (\%) & $E_{|m|<\ell/2}$ (\%) & Reference  \\
GJ, Name & \bv\ (kG) & \multicolumn{3}{c}{Energy fraction (\%)} & References  \\
         &           & Pol. & Dip. & Axis.           &            \\
\noalign{\smallskip}\hline\noalign{\smallskip}
\endhead

\noalign{\smallskip}\hline
\endfoot

\noalign{\smallskip}\hline
\insertTableNotes
\endlastfoot

\caption{Parameters of the large-scale magnetic field geometries of M dwarfs derived from ZDI analysis of the Stokes $V$ spectra. Magnetic energy fractions are given for the poloidal, dipolar, and axisymmetric field components. \label{tab:stokesV}} \\
\hline\noalign{\smallskip}
%Name & \bv\ (kG) & $E_{\rm pol}$ (\%) & $E_{\ell=1}$ (\%) & $E_{|m|<\ell/2}$ (\%) & Reference  \\
GJ, Name & \bv\ (kG) & \multicolumn{3}{c}{Energy fraction (\%)} & References  \\
         &           & Pol. & Dip. & Axis.           &            \\
\noalign{\smallskip}\hline\noalign{\smallskip}
\endfirsthead

GJ 49                 & 0.027  & 48  &    71   &   67   & 1  \\
\noalign{\smallskip}
GJ 51, V388 Cas       & 1.61   & 99  &    96   &   91   & 2  \\
                      & 1.58   & 99  &    92   &   77   & 2  \\ 
                      & 1.65   & 97  &    92   &   89   & 2  \\
\noalign{\smallskip}
GJ 65A, BL Cet        & 0.338  & 89  &    92   &   56   & 3  \\
\noalign{\smallskip}
GJ 65B, UV Cet        & 1.340  & 93  &    70   &   89   & 3  \\
\noalign{\smallskip}
GJ 182, V1005 Ori     & 0.172  & 32  &    48   &   17   & 1  \\ 
\noalign{\smallskip}
GJ 205                & 0.020  & 99  &    90   &   73   & 4  \\
\noalign{\smallskip}
GJ 251                & 0.027  & 99  &    100  &   88   & 5  \\
\noalign{\smallskip}
GJ278Ca, YY Gem A     & 0.260  & 71  &    52   &   58   & 6  \\
\noalign{\smallskip}
GJ278Cb, YY Gem B     & 0.205  & 72  &    46   &   45   & 6  \\ 
\noalign{\smallskip}
GJ 285, YZ CMi        & 0.56   & 92  &    69   &   61   & 7  \\
                      & 0.55   & 97  &    72   &   86   & 7  \\
\noalign{\smallskip}
GJ 1111, DX Cnc       & 0.11   & 93  &    69   &   77   & 2  \\ 
                      & 0.08   & 73  &    31   &   49   & 2  \\
                      & 0.08   & 62  &    42   &   70   & 2  \\
\noalign{\smallskip}
GJ 358                & 0.130  & 97  &    98   &   85   & 4  \\
\noalign{\smallskip}
GJ 388, AD Leo        & 0.19   & 99  &    56   &   97   & 7  \\ 
                      & 0.18   & 95  &    63   &   88   & 7  \\
                      & 0.33   & 94  &    89   &   93   & 8  \\
                      & 0.30   & 91  &    94   &   97   & 8  \\
\noalign{\smallskip}
GJ 3622               & 0.05   & 96  &    90   &   73   & 2  \\ 
                      & 0.06   & 93  &    84   &   80   & 2  \\
\noalign{\smallskip}
GJ 410, DS Leo        & 0.101  & 18  &    52   &   58   & 1  \\
                      & 0.087  & 20  &    52   &   16   & 1  \\ 
                      & 0.065  & 25  &    88   &   11   & 4  \\
\noalign{\smallskip}
GJ 412B, WX UMa       & 0.89   & 97  &    66   &   92   & 2  \\
                      & 0.94   & 97  &    71   &   92   & 2  \\
                      & 1.03   & 97  &    69   &   83   & 2  \\ 
                      & 1.06   & 96  &    89   &   95   & 2  \\
\noalign{\smallskip}
GJ 1156, GL Vir       & 0.05   & 88  &    30   &   6    & 2  \\
                      & 0.11   & 83  &    41   &   20   & 2  \\
                      & 0.10   & 94  &    54   &   2    & 2  \\ 
\noalign{\smallskip}
GJ 479                & 0.065  & 37  &    74   &   29   & 4  \\
\noalign{\smallskip}
GJ 490B               & 0.680  & 99  &    95   &   95   & 9  \\
\noalign{\smallskip}
GJ 494A, DT Vir       & 0.145  & 38  &    64   &   12   & 1  \\ 
                      & 0.149  & 53  &    10   &   20   & 1  \\
\noalign{\smallskip}
GJ 569A, CE Boo       & 0.103  & 95  &    87   &   96   & 1  \\
\noalign{\smallskip}
GJ 9520, OT Ser       & 0.136  & 80  &    47   &   86   & 1  \\
                      & 0.123  & 67  &    33   &   66   & 1  \\ 
\noalign{\smallskip}
GJ 1245B              & 0.17   & 80  &    45   &   15   & 2  \\
                      & 0.18   & 84  &    46   &   52   & 2  \\
                      & 0.06   & 85  &    33   &   20   & 2  \\
\noalign{\smallskip}
GJ 793                & 0.200  & 64  &    44   &   82   & 5  \\ 
\noalign{\smallskip}
GJ 4247, V374 Peg     & 0.78   & 96  &    72   &   76   & 10 \\
                      & 0.64   & 96  &    70   &   77   & 10 \\
\noalign{\smallskip}
GJ 846                & 0.045  & 27  &    69   &   68   & 4  \\ 
                      & 0.030  & 63  &    52   &   86   & 4  \\
\noalign{\smallskip}
GJ 873, EV Lac        & 0.57   & 87  &    60   &   36   & 7  \\
                      & 0.49   & 98  &    75   &   31   & 7  \\
\noalign{\smallskip}
GJ 896A, EQ Peg A     & 0.48   & 85  &    70   &   70   & 7  \\ 
\noalign{\smallskip}
GJ 896B, EQ Peg B     & 0.45   & 97  &    79   &   94   & 7  \\
\noalign{\smallskip}
GJ 1289               & 0.275  & 99  &    98   &   90   & 5  \\
                                                                                  
\end{longtable}                                                                    
\end{ThreePartTable}
\end{center}
}

\newpage

%%%%%%%%%%%%%%%%%%%%%%%%%%%%%%%%%%%%%%%%%%%%%%%%
% Table 3 with spectral types and rotation periods
%%%%%%%%%%%%%%%%%%%%%%%%%%%%%%%%%%%%%%%%%%%%%%%%
\section{Appendix 3: Spectral types and rotation periods of M dwarfs with magnetic field measurements}
{\small
\begin{center}
\begin{longtable}{llcl}

\caption{(continued)} \\
\hline\noalign{\smallskip}
2MASS & GJ, Name & Sp. type & $P_{\rm rot}$ (d) \\
\noalign{\smallskip}\hline\noalign{\smallskip}
\endhead

\noalign{\smallskip}\hline
\endfoot

\noalign{\smallskip}\hline
\endlastfoot

\caption{Spectral types and rotation periods of stars listed in Tables~\ref{tab:stokesI} and \ref{tab:stokesV}. Asterisk indicates new rotation periods derived from TESS light curves. \label{tab:params}} \\
\hline\noalign{\smallskip}
2MASS & GJ, Name & Sp. type & $P_{\rm rot}$ (d) \\
\noalign{\smallskip}\hline\noalign{\smallskip}
\endfirsthead

J00244419-2708242    & GJ 2005A           & M5.5 &  0.94$^\ast$ \\
J01023895+6220422    & GJ 49              & M1.5 & 18.6         \\
J01031971+6221557    & GJ 51, V388 Cas    & M5.0 &  1.06        \\ 
J01351393-0712517    & Barta 161 12       & M4.0 &  0.70        \\
J01390120-1757026(A) & GJ 65A, BL Cet     & M5.5 &  0.24        \\
J01390120-1757026(B) & GJ 65B, UV Cet     & M6.0 &  0.23        \\
J02390117-5811138    & GJ 1049            & M0.0 &  2.63$^\ast$ \\
J02085359+4926565    & GJ 3136            & M3.5 &  0.75        \\
J03472333-0158195    & G 80-21            & M3.0 &  3.88        \\
J04072048-2429129(A) & GJ 3263A           & M3.5 &  $<$6.4      \\
J04072048-2429129(B) & GJ 3263B           & M4.0 &  $<$4.8      \\
J04471225+2038109    & RX J0447.2+2038    & M5.0 &  0.34        \\
J04593483+0147007    & GJ 182, V1005 Ori  & M0.0 &  4.35        \\
J05015881+0958587    & GJ 3322A           & M4.0 &  1.22        \\
J05254166-0909123    & LP 717-36          & M3.5 &  1.09$^\ast$ \\
J05312734-0340356    & GJ 205             & M1.0 & 33.63        \\
J05363099+1119401    & GJ 208, V2689 Ori  & M0.0 & 12.04        \\
J06000351+0242236    & GJ 3379            & M4.0 &  1.81        \\
J06292339-0248499    & GJ 234A, V577 Mon  & M4.5 &  1.58$^\ast$ \\
J06544902+3316058    & GJ 251             & M3.5 &  90          \\
J07343745+3152102(A) & GJ278Ca, YY Gem A  & M0.5 &  0.81        \\
J07343745+3152102(B) & GJ278Cb, YY Gem B  & M0.5 &  0.81        \\
J07444018+0333089    & GJ 285, YZ CMi     & M4.5 &  2.78        \\
J08294949+2646348    & GJ 1111, DX Cnc    & M6.5 &  0.46        \\
J09394631-4104029    & GJ 358             & M2.0 & 25.37        \\
J09445422-1220544    & G 161-071          & M5.0 &  0.31        \\
J10193634+1952122    & GJ 388, AD Leo     & M3.5 &  2.24        \\
J10360120+0507128    & GJ 398, RY Sex     & M4.0 &  $<$4.5      \\
J10481258-1120082    & GJ 3622            & M6.5 &   1.5        \\
J10562886+0700527    & GJ 406, CN Leo     & M5.5 &  $<$2.5      \\
J11023832+2158017    & GJ 410, DS Leo     & M1.0 &  14.0        \\
J11053133+4331170    & GJ 412B, WX UMa    & M6.0 &  0.78        \\
J12153937+5239088    & RX J1215.6+5239    & M4.0 & $<$0.35      \\
J12185939+1107338    & GJ 1156, GL Vir    & M5.0 &  0.49        \\
J12375231-5200055    & GJ 479             & M2.0 & 24.04        \\
J12531240+4034038    & LP 218-8           & M7.5 & $<$0.71      \\
J12573935+3513194    & GJ 490B            & M4.0 &  0.54        \\
J13004666+1222325    & GJ 494A, DT Vir    & M2.0 &  2.85        \\
J14172209+4525461    & RX J1417.3+4525    & M5.0 &  0.36$^\ast$ \\
J14542923+1606039    & GJ 569A, CE Boo    & M2.0 &  14.7        \\
J14563831-2809473    & GJ 3877            & M7.0 &  2.01$^\ast$ \\
J15215291+2058394    & GJ 9520, OT Ser    & M1.5 &  3.37        \\
J15495517+7939517    & G 256-25           & M5.0 &  0.19        \\
J16311879+4051516    & GJ 3959            & M5.0 &  0.51        \\
J16553529-0823401    & GJ 644 C, VB 8     & M7.0 &  $<$1.0      \\
J16570570-0420559    & GJ 1207            & M3.5 &  1.21        \\
J17335314+1655129    & V1274 Her          & M5.5 &  0.27        \\
J18021660+6415445    & G 227-22           & M5.0 &  0.28        \\
J18073292-1557464    & GJ 1224            & M4.5 &  3.9         \\
J18185725+6611332    & GJ 4053            & M4.5 &  0.52        \\
J18494929-2350101    & GJ 729, V1216 Sgr  & M3.5 &  2.87        \\
J19165762+0509021    & GJ 752B, VB 10     & M8.0 &  1.00        \\
J19510930+4628598    & GJ 1243            & M4.0 &  0.59        \\
J19535508+4424550    & GJ 1245B           & M5.5 &  0.71        \\
J20091824-0113377    & SCR J2009-0113     & M5.0 &  $<$2.0      \\
J20303207+6526586    & GJ 793             & M3.0 &  22          \\
J20450949-3120266    & GJ 803, AU Mic     & M1.0 &  4.86        \\
J22011310+2818248    & GJ 4247, V374 Peg  & M4.0 &  0.45        \\
J22021026+0124006    & GJ 846             & M0.0 & 10.73        \\
J22171899-0848122    & GJ 852A, FG Aqr    & M4.0 &  $<$2.6      \\
J22171870-0848186    & GJ 852B            & M4.5 &  $<$3.3      \\
J22464980+4420030    & GJ 873, EV Lac     & M4.0 &  4.37        \\
J22515348+3145153    & GJ 9799, GT Peg    & M3.0 &  1.64        \\
J23315208+1956142    & GJ 896A, EQ Peg A  & M3.5 &  1.06        \\
J23315244+1956138    & GJ 896B, EQ Peg B  & M4.5 &  0.40        \\
J23430628+3632132    & GJ 1289            & M4.5 &  54          \\
J23503619+0956537(A) & GJ 4368A, LHS4022A & M4.0 &  $<$4.4      \\
J23503619+0956537(B) & GJ 4368B, LHS4022B & M5.5 &  $<$2.7      \\
J23545147+3831363    & RX J2354.8+3831    & M4.0 &  4.76        \\
                                                              
\end{longtable}                                               
\end{center}
}

\end{appendix}

%\bibliographystyle{spbasic}      % basic style, author-year citations
%\bibliographystyle{spbasic1}      % basic style, author-year citations, suppress eprint output
%\bibliographystyle{spmpsci}      % mathematics and physical sciences
%\bibliographystyle{spphys}       % APS-like style for physics
%\bibliography{astro_papers}   % name your BibTeX data base

\end{document}